\documentclass[twocolumn,twoside]{article}

\usepackage{a4}
\usepackage{amssymb}
\usepackage{amsmath}
\usepackage[numbers,sort&compress]{natbib}
\usepackage{graphicx}

\begin{document}

\title{{\bf Magnetohydrodynamic models of astrophysical jets}\\
{\normalsize\sc in memory of vitaly lazarevich ginzburg}}

\date{{\normalsize\textit{P N Lebedev Physical Institute, Russian Academy of Sciences,\\
Leninskii prosp. 53, 119991, Moscow, Russian Federation,}}\\[1ex]
{\small \textit{Usp.\ Fiz.\ Nauk} \textbf{180}, 
1241--1278 (2010)
[in Russian]\\
English translation: \textit{Physics -- Uspekhi}, \textbf{53}, 1199-1233 (2010)}
\\{\small Translated by K A Postnov; edited by A Radzig}
}

\author{V.S.Beskin}

\maketitle

\begin{abstract}             
In this review, analytical results obtained for a wide class of stationary 
axisymmetric flows in the vicinity of compact astrophysical objects are analyzed, 
with an emphasis on quantitative predictions for specific sources. Recent years 
have witnessed a great increase in understanding the formation and properties of 
astrophysical jets. This is due not only to new observations but also to advances 
in analytical theory which has produced fairly simple relations, and to what can 
undoubtedly be called a breakthrough in numerical simulation which has enabled 
confirmation of theoretical predictions. Of course, we are still very far from 
fully understanding the physical processes occurring in compact sources. Nevertheless, 
the progress made raises hopes for near-future test observations that can give insight 
into the physical processes occurring in active astrophysical objects.
\end{abstract}

\setcounter{secnumdepth}{3}
\setcounter{tocdepth}{2}

\tableofcontents

\section{Introduction} 

This review was not specially written for the {\it Uspekhi Fizicheskikh Nauk 
(Physics-Uspekhi)} issue devoted to the memory of V L Ginzburg. Nevertheless, 
I would like to hope that the spirit of this review is close to that of other 
papers in this issue. I was happy to work closely with Vitaly Lazarevich
for more than 30 years, starting from my graduate student days, and he definitely 
played a significant role in my scientific development. So, this review, hopefully, 
bears a fraction of the soul of Vitaly Lazarevich.

Vitaly Lazarevich was a passionate person. Astrophysics, undoubtedly, fascinated 
him most. But the scale of his personality was such that this passion did not 
separate, but instead united, people. So, it is not surprising that at the Lebedev 
%Physical 
Institute the astrophysical seminar headed by Vitaly Lazarevich for more 
than several decades continues, and scientists from many institutes participate 
in its work. The Department of Physics and Astrophysics Problems at Moscow Institute 
of Physics and Technology (MIPT), which Vitaly Lazarevich founded in 1968 and headed 
until recently, continues to be one of the leading institutions in teaching young 
astrophysicists. Numerous pupils of V L Ginzburg and pupils of his pupils working 
in the leading astrophysical centers of the world keep his unique trademark in their 
studies.

The astrophysical heritage of V L Ginzburg is enormous. He obtained fundamental 
results in the theory of propagation of electromagnetic waves in cosmic plasma, 
in the theory of the origin of cosmic rays, and in the theory of neutron stars 
and black holes. In all cases, a simple model allowing the understanding of the 
essence of physical process in observed astrophysical sources laid the basis of 
the theory. The present review, hopefully, was written in the same spirit.

Astronomy, as follows from the very appellation, is the science that stemmed 
from the observations of stars. During hundreds of years the people observed 
stars in the sky and gained insight into the laws of Nature. The stars appeared 
to be always unchanged and existing for an infinite amount of time. After the 
appearance of spectral analysis, the first astrophysical observations and, later, 
the theory of radiation generally confirmed this point of view. The lifetimes of 
most stars turned out to be comparable to the age of the Universe. Thus, in the 1950s, 
when radio astronomy began, stars emitting thermal radiation seemed to be the main 
objects for studies. In radio astronomy, the brightness temperature remains even 
now the basic characteristic of radiation intensity.

However, the first radio astronomical observations, and especially observations in 
the X-ray and gamma-ray ranges, which started in the middle of the 1970s, discovered 
numerous nonthermal sources in the Universe. These objects are sufficiently compact 
(i.e., the spatial resolution of the existing detectors is insufficient to determine their 
internal structure) and, in addition, are highly variable. In active galactic nuclei 
the variability timescale (months or sometimes even days) is small according to the 
cosmic timescale, with the variability timescale of radio pulsars and sources of 
gamma-ray bursts being the fractions of a second, which is small even to Earth's 
measures. The activity, i.e., high variability on timescales $\tau \sim R/c$, as well 
as the generation of nonthermal radiation indicates that in most cases we are dealing 
with relativistic objects, namely, with objects in which matter moves with velocities 
close to the speed of light.

Jet eruptions represent one of the visible appearances of the activity of compact 
astrophysical objects. We shall briefly discuss their properties in Section 2. They 
are observed in both relativistic objects (such as active galactic nuclei and 
microquasars) and in young stars where the motion of matter is definitely nonrelativistic. 
This means that we are dealing with some universal and extremely efficient mechanism of 
energy release. Therefore, the key theoretical problems include the question of the 
energy source of the activity of compact objects, the understanding of their energy 
release mechanism, and the collimation of matter outflows. We shall postpone the detailed 
discussion of arguments against alternative models until the next section, and here we 
only remind the main arguments favoring the magnetohydrody-namic model of activity of 
compact sources, which is accepted by most astrophysicists.

The model of the unipolar inductor, i.e., the source of direct current, lies at the 
heart of the magnetohydrodynamic approach. As we shall show in Section 2, conditions 
for the existence of such a 'central engine' are satisfied in all the compact sources 
discussed below. Indeed, all compact sources are assumed to harbor a rapidly spinning 
central body (black hole, neutron star, or young star) and some regular magnetic field, 
which leads to the emergence of strong induction electric fields. The electric fields, 
in turn, lead to the appearance of longitudinal electric currents and effective particle 
acceleration. The collimation mechanism in this model is related to the well-known property 
of mutual attraction of parallel currents.

The first studies of the electromagnetic model of compact sources (namely, radio pulsars) 
were carried out as early as the end of the 1960s [1-4]. It was evidenced that there are 
objects in the Universe in which electrodynamical processes can play the decisive role in 
the energy release. Then, in 1976 R Blandford [5] and R Lovelace [6] independently suggested 
that the same mechanism can also operate in active galactic nuclei. In the same year, 
G S Bisnovatyi-Kogan, Yu P Popov, and A A Samokhin proposed a magnetorotational mechanism 
of the supernova explosion [7] (i.e., the model of an essentially nonstationary phenomenon), 
in which jet eruptions can also be formed [8]. This model has remained the leading one 
for nearly 40 years. However, only recently have some key properties become clear. This 
is related both to advances in the theory which have at last formulated sufficiently simple 
analytical relations, and to the breakthrough in numerical simulations which confirmed 
theoretical predictions.

The reader can find the detailed introduction to the analytical theory in the author's 
monograph [9] (see also the review in {\it Physics-Uspekhi} in 1997 [10]). However, first, the 
monograph was devoted to the basics of the theory, and qualitative predictions for specific 
astrophysical sources were discussed only very briefly. Second, the monograph clearly could 
not include the results of numerical calculations carried out in the last five years since 
its publication. This is the main reason for writing the present review. In addition, here 
we shall correct formulas from the monograph in which misprints were found.

Of course, we are still far away from the full understanding of the essence of physical 
processes proceeding in compact sources. In fact, now we have only agreement between theory 
and numerical modeling. All results have been obtained applying ideal one-liquid 
magneto\-hyd\-ro\-dynamics, though by different methods (the theory is based on stationary 
equations, while numerically the time relaxation problem is solved). In particular, it 
is not yet clear which of the main physical characteristics of the central engine (such as 
the mass of the central body or its rotation velocity) should fully determine the observed 
energy release. Nevertheless, the progress achieved over recent years raises hopes for test 
observations already in the nearest future, which can give insight into physical processes 
occurring in active astrophysical sources.

\section{Jets}

\subsection{Active galactic nuclei}

The main properties of the central engine in active galactic nuclei, which are 
presently accepted by most astrophysicists, can be summarized as follows [11,12]. 
In the center of the host galaxy there is a supermassive black hole (its mass reaches 
$10^6$--$10^9 \, M_{\odot}$, where \mbox{$M_{\odot} \approx 2 \times 10^{33}$ g}
is the mass of the Sun), onto which accretion of the surrounding matter occurs [13]. 
Only in this case it is possible to explain the very high efficiency of the energy 
release and the compactness of the central engine. The energy source of activity of 
galactic nuclei can be related to both the rotational energy of the black hole, viz.
\begin{equation}
E_{\rm tot} = \frac{J_{\rm r}\Omega_{\rm H}^2}{2} \approx
10^{62}\left(\frac{M}{10^9M_{\odot}}\right)\left(\frac{\Omega_{\rm H} r_{\rm g}}{c}\right)^2
\,{\rm erg},
\label{k1}
\end{equation}
and the energy of the accreting matter. Here \mbox{$r_{\rm g} = 2 GM/c^2$} 
is the radius of the black hole, $J_{\rm r}$ is the moment of inertia, $\Omega_{\rm H}$ and $M$ are 
the angular velocity and the mass of the black hole, respectively, and $c$ is the speed 
of light. The existence of supermassive objects is also supported by the fact that the 
Eddington luminosity
\begin{equation}
L_{\rm Edd} \approx 10^{47}\left(\frac{M}{10^{9}M_{\odot}}\right)
\, {\rm erg} \, {\rm s}^{-1},
\label{k2}
\end{equation}
(i.e., the luminosity at which the gravitational force acting on the accreting matter is 
balanced by the radiation pressure force) is close to the characteristic luminosity of 
active galactic nuclei [14]. Moreover, the duration of the active phase
$\tau_{\rm D} = E_{\rm tot}/L_{\rm Edd}$ estimated using formulas (1) and (2) is on the 
order of $10^7$ years, which is also in agreement with observations.

Further, it is usually assumed that the accretion of matter proceeds through a disc [15]. 
Thus, the preferential direction --- the axis of rotation --- emerges naturally in space, along 
which the formation of jets is possible. As a black hole itself cannot have the self-magnetic 
field (the so-called 'no-hair theorem'), the generation of a large-scale magnetic field in the 
vicinity of the black hole is believed to occur in the accretion disc [16-18].

\begin{figure}
\begin{center}
\includegraphics[width=\columnwidth]{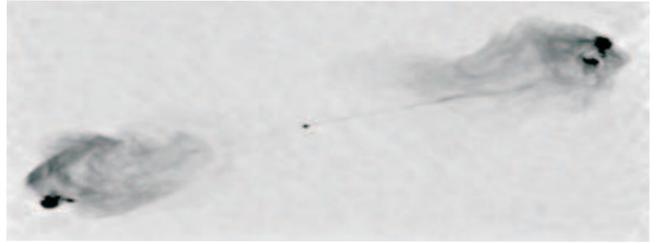}
\end{center}
\caption{Radio image (5 GHz) of active regions and jet eruptions 
        from the nucleus of the Cygnus A galaxy [20]. The distance between 
        bright spots is about 80 kpc, which is 9-10 orders of magnitude greater 
        than the size of the central black hole.
   }
\label{fig1_01}
\end{figure}

According to the modern concept, massive central objects are present in most galaxies and 
remain active only if a sufficient amount of matter falls on them. This restricts their 
active lifetime. Unfortunately, as stated above, the angular resolution of modern detectors 
does not allow us to directly observe plasma flow on the scales comparable to the black hole 
size \mbox{$r_{\rm g} \approx 3 \times 10^{14}(M/10^{9} \, M_{\odot})$ cm.} Therefore, we have to 
judge the activity of galactic nuclei only using indirect evidence, by observing flows on 
much larger scales.

Let us remember that the diffuse radio emission around active galaxies is observed from 
regions located at distances of tens or even a hundred kiloparsecs from their nuclei. 
Very shortly after the discovery of these regions at the beginning of the 1960s, this 
emission was associated with collimated plasma ejections (jets) flowing out the galactic 
nuclei [12]. It is precisely these jets that transport matter and energy from the active 
nuclei to those regions (Fig. 1). Observations show that the jets can be accelerated 
and collimated very close to a galactic nucleus. For example, in the case of the nearest 
active galaxy M87 the formation of the jet occurs within a radius of $60 \, r_{\rm g}$
from the nucleus [19]. In recent years, the internal structure close to the jet base was 
resolved in several sources, where the jet tranverse dimension usually does not exceed several 
parsecs [21, 22] (Fig. 2).

\begin{figure}
\begin{center}
\includegraphics[width=0.9\columnwidth]{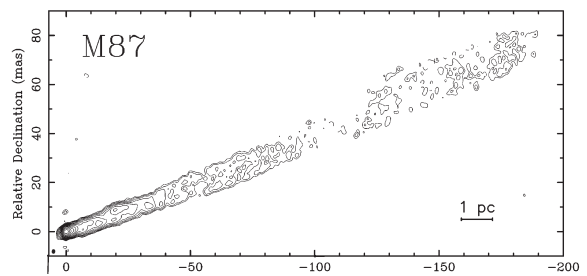}
\caption{Radio image of the jet eruption from the galaxy M87 near the 
        central engine [22]. The jet transverse size is about 1 pc.
}
\label{fig1_02}
\end{center}
\end{figure}

The matter in jets from active galactic nuclei has a very high energy --- the bulk Lorentz 
factor of a jet is at least a few unities. For example, this motion is directly observed 
in the M87 galaxy, with the bulk Lorentz factor of the outflow being 
$\gamma \approx 6$ [23]. In many cases, the matter continues moving with relativistic 
velocities up to huge distances from the nucleus before noticeable braking due to interaction 
with the ambient intergalactic medium. Another peculiar feature of jets is their high degree 
of collimation within a cone characterized by an opening angle of only several degrees.

Unfortunately, observations do not yet allow reliable estimations of the energy and mass 
fluxes in jets from active galactic nuclei, of the magnitude of the magnetic field both close 
to the black hole and in the jet itself, or of the composition of jet eruptions. The spectrum 
of radiation from galactic jets (in contrast, for example, to the spectrum of jets from young 
stars) does not exhibit any spectral features of moving matter, i.e., neither atomic (ionic) 
lines nor the electron-positron pair annihilation line are observed. To this regard, there are 
arguments both in favor [24] and against [25] the leading role of electron-positron plasma, 
so it is now impossible to say exactly which mechanism of energy transfer to the jet actually 
operates.

Where a physical nature of the galactic nucleous activity is concerned, several mechanisms 
of particle acceleration and jet collimation have been proposed, but so far there is no 
definite answer as to which of them are actually realized. It is possible that different 
mechanisms operate in different sources, or, just the opposite, all mechanisms are realized 
simultaneously.

{\it Gas-dynamic acceleration}. The acceleration and collimation of a jet can be related to the 
presence of an ambient medium with high pressure which decreases with distance from the 
center [26, 27]. Such a medium could play the role of an external wall collimating the 
outflow. The pressure of the external hot medium can, in principle, be estimated from X-ray 
observations [28]. This mechanism possibly explains how weak jets in our Galaxy and in some 
Seyfert galaxies (i.e., low-active galaxies) are formed. On the other hand, the observed 
pressure of the hot matter around the most powerful jets from active galactic nuclei is not 
sufficiently high, and there must be an alternative mechanism of plasma confinement.

{\it Acceleration by radiation}. As the photon density near the central source can be very high, 
the radiation-driven mechanism of jet matter acceleration by radiation pressure was proposed 
[29, 30]. In this model, it is assumed that the inner parts of the disc can serve as a nozzle 
directing matter outflows accelerated by the radiation pressure. However, this mechanism also 
meets some difficulties. For example, there is no correlation between the jet power and 
the luminosity of the source --- many sources with very powerful jets are low-luminous sources 
[31]. Another difficulty comes from the fact that, starting from the sufficiently low particle 
energies $\gamma \approx 3$, the radiation field more effectively brakes particles than accelerates them [32]. 
This contradicts observations of 'superluminal' sources in which the energy of plasma particles 
is much higher. In addition, if the jet was formed in a system with a thin accretion disc 
emitting radiation more or less isotropically, additional mechanisms for the jet collimation 
should be invoked. A modification of this model involving the formation of a funnel in a thick 
accretion disc can explain the initial jet collimation, but there are indications that such a 
structure is unstable [31].

{\it Magnetohydrodynamic mechanism}. As noted above, most researchers favor the 
magneto\-hyd\-ro\-dynamic model of jet formation. The magnetohydrodynamic (MHD) model was 
successfully utilized to describe many processes in active nuclei, and, in particular, in 
connection with the problem of the origin and stability of jets, as well as to explain the 
energetics of processes proceeding near the central black hole. The magnetic field here is 
the natural link between the central engine and the jet. Moreover, in this model it is easy 
to understand why the jet matter can predominantly consist of electron-positron plasma. As 
was shown in Refs [33-35], it can be generated on the magnetic field lines threading the 
black hole horizon.

In the simplest version, the picture is as follows: the regular magnetic field generated in 
the disc links the spinning central engine (the disc and the black hole) with infinity. The 
plasma outflow occurs along the magnetic field lines; the electromagnetic energy flux is also 
directed along the magnetic field lines. The longitudinal electric current flowing along the
jet forms a toroidal magnetic field, and the magnetic field pressure associated with this 
toroidal component can collimate the jet.

It should be noted, however, that in a real astrophysical system the total current flowing 
from the central engine should vanish, so the Ampere force in the current closure region will, 
on the contrary, decollimate the flow (the antiparallel currents repulse). Therefore, an
external medium (for example, a subrelativistic wind outflow from the accretion disc) is 
necessary to collimate the jet. In addition, the question as to whether it is possible to 
consider a black hole immersed in the external magnetic field as a unipolar inductor turned 
out to be also rather nontrivial. It required almost 30 years of studies after the paper by 
Blandford and Znajek [33], which laid the basis of the theory in 1977, before the needed 
clarity was reached in this question. We shall discuss these points in more detail in Section 3.

\subsection {Microquasars}

Microquasars comprise galactic objects in which the jet formation is due to accretion onto a 
compact relativistic object (neutron star or black hole). In other words, all microquasars 
reside in sufficiently close binary systems in which the effective flow of matter from the 
star companion occurs. The rate of matter inflow in such systems is larger than can be swallowed 
by the central object. As a result, some accreting matter that carries, in particular, an 
excessive angular momentum is expelled from the system in the form of jets. Observations of
 microquasars show that jets are related to thick accretion discs. In other words, no jets 
 are known for systems with thin discs. The reason for that is unclear: either a thin disc 
 insufficiently collimates the outflow, or the magnetic field generated by the thin disc is 
 not strong enough.

Microquasars represent a small population of objects, including only around ten sources [36], 
with only half of them demonstrating noticeable relativistic jets \mbox{($v > 0.9 \, c$).} 
The characteristic longitudinal size of jets is usually \mbox{$0.1$ pc}, with the jet spread 
angle being within several degrees (Fig. 3). The total energetics are about $10^{37}$ erg s$^{-1}$. 
Due to the relativistic velocity of the bulk motion of matter in the jets, some sources 
demonstrate the superluminal motion effect, with the apparent angular velocity being several 
orders of magnitude larger (due to a relative proximity of these objects) than that observed 
in jets from active galactic nuclei.

\begin{figure}
\begin{center}
\includegraphics[width=\columnwidth]{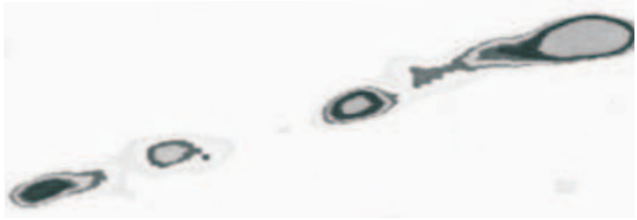}
\caption{Radio image of the jet from the microquasar iE1740.7-2942. The characteristic 
        length of the jet measures $0.1$ pc [38].
}
\label{fig1_03}
\end{center}
\end{figure}

Historically, the first revealed object of this class was the famous source SS433 [37] in 
which, however, the gas ejection velocity in jets is only $0.26 \, c$. Such a velocity can be easily 
explained by the radiation pressure from highly heated internal regions of the accretion disc. 
As for relativistic jets, the first source was discovered only in 1994 [38]. Since the appearance 
of near-light velocities due to radiation or gas pressure is problematic, it has not been ruled 
out that, to explain them, an electrodynamic model similar to that used in explaining the origin 
and collimation of extragalactic jets should be invoked again. This model is also supported by 
the fact that in all but one microquasar (SS433) no emission lines from jets are observed. This 
indirectly points to the electron-positron composition of matter in jets [36]. Finally, it 
should be noted that in most microquasars the jet is separated in individual blobs at large 
distances from the central engine, which is thought to be due to a long duty cycle of the work 
of the central engine.

\subsection {Sources of cosmological gamma-ray bursts}

As regards the sources of cosmological gamma-ray bursts, there are indirect, although 
sufficiently reliable, arguments in favor of the presence of jets related exactly to 
relativistic strongly magnetized outflows, which we shall discuss in this review. It 
is well known that the discovery of the optical afterglow [39], as well as afterglows 
in other spectral ranges, which allowed the measurement of the distance to these sources 
from the observed redshifts of the host galaxies, put serious constraints on their energetics 
[40]. If the observed gamma-ray radiation were emitted isotropically, the total energy 
release for the typical distance to these sources of several gigaparsecs would reach 
$10^{54}$ erg. However, we do not know at present processes with such huge energy liberation. 
On the other hand, the small duration of the burst \mbox{($\sim 10 \,$ s)} restricts the size of the 
emission region, which, in turn, does not allow us to explain the observed nonthermal 
gamma-ray spectra, since the optical depth in the source proves to be very high [41].

If it is assumed that gamma-rays are emitted within a narrow cone angle 
$\vartheta \sim 1^{\circ}$, the observed energy can be reduced to $10^{51}$ erg, which 
is already to an order of magnitude of the energy release during supernova explosions. 
On the other hand, the observed optically thin nonthermal gamma-ray spectra immediately 
imply the presence of ultrarelativistic outflows with bulk Lorentz factors of $\sim 100$--$300$. 
Only in this case can the compactness problem of the source be resolved, since the 
estimated size of the emitting region also increases as the square of the bulk Lorentz 
factor (i.e., by $10^{4}$--$10^{5}$ times), and the optical depth, which is proportional 
to the density multiplied by the size of the region, decreases respectively by 
$10^{8}$--$10^{10}$ times.

However, the ultrarelativistic character of the outflow, in turn, puts constraints on the 
particle composition in the jet, since the presence of a significant fraction of baryons 
with such energy in the outflow would contradict the total energy release in the gamma-ray 
burst. Therefore, the contribution of protons must be smaller than $10^{-2}$ of the total 
number of particles, so that only electron-positron jets should be considered. The 
existence of jets is also evidenced by the presence of the characteristic bend of the 
light curve of the afterglow, when the power law index $\alpha$ in the radiation intensity 
dependence on time, $W_{\rm tot} \propto t^{-\alpha}$, changes from $\alpha \approx 1.1$ to 
$\alpha \approx 2.0$ after a span of about a few days following the burst. This effect is 
related to the cessation of relativistic contraction of the radiation cone in the direction 
of particles' motion toward an observer. Incidentally, this model allowed independent 
confirmation of the jet spread angle $\vartheta \sim 1^{\circ}$ and the bulk Lorentz factor 
$\gamma \sim 100$--$300$ [42].

The nature of the central engine giving rise to strongly magnetized jets can be usually 
related to the collision of two neutron stars [43, 44] or of a neutron star and a black 
hole [45], or, most likely, to the collapse of the massive core of an unusual supernova 
[46, 47]. However, in most models a rapidly spinning solar-mass black hole ultimately 
emerges, which loses its rotation energy via the Blandford-Znajek process [45, 48 50]. 
Indeed, as we have seen, this process easily provides a natural explanation for both the 
low baryonic load of the jet and the large bulk Lorentz factors of jet particles. In other 
words, the model again is constructed similarly to the scheme proposed for active galactic 
nuclei. In particular, the key processes here also include the magnetic field generation 
in the plasma around the black hole, the interaction of the black hole with the accretion 
disc via magnetic field lines, and the generation of particles in the magnetosphere. To 
explain the observed energy release, it is necessary to assume that the magnetic field 
near the black hole must be as high as $10^{14}$ or even \mbox{$10^{15}$ G.} The generation of 
such a high field is thought to be possible in nonstationary processes like the supernova 
core collapse or binary neutron star coalescence [51, 52].

\subsection  {Radio pulsars}

The discovery of radio pulsars at the end of the 1960s, which are the sources of pulsating 
cosmic radio emission with the characteristic period $P \sim 1$ s [53], is definitely one 
of the major astrophysical discoveries of the 20th century. Indeed, for the first time a 
new class of cosmic sources related to neutron stars, whose existence was theoretically 
predicted away back in the 1930s [54], was discovered. Neutron stars (mass of about 
\mbox{$1.2$--$1.4$ $M_{\odot}$,} and radius $R$ of only $10$--$15$ km) must result from the 
catastrophic compression (collapse) of usual massive stars at the late stage of their 
evolution or, for example, of white dwarfs whose mass exceeds the Chandrasekhar mass 
limit of $1.4$ $M_{\odot}$ due to accretion from the companion star. It is this formation 
mechanism that provides the simplest explanation for both small spin periods $P$ (the 
smallest known spin period $P = 1.39$ ms) and superstrong magnetic fields with 
\mbox{$B_0 \sim 10^{12}$ G} [1,2].

Interestingly, the basic physical processes determining the observed activity of radio 
pulsars were understood almost immediately after their discovery. For example, it became 
clear that highly regular pulsations of observed radio emission are related to the rotation 
of neutron stars. Next, radio pulsars are powered by the rotational energy of the neutron 
star, and the mechanism of energy release is related to their superstrong magnetic field 
with $B_0 \sim 10 ^{12}$ G. Indeed, energy losses estimated using the simple magnetodipole 
formula [44] are as follows:
\begin{equation}
W_{\rm tot} = -J_{\rm r}\Omega\dot\Omega \approx
\frac{1}{6}\frac{B_0^2\Omega^4R^6}{c^3}\sin^2\chi,
\label{wmd}
\end{equation}
where $J_{\rm r} \sim MR^2$ is the moment of inertia of the neutron star, $\chi$  
is the angle between the magnetic dipole axis and the spin axis, and  $\Omega = 2\pi/P$ 
is the angular velocity of the neutron star rotation. For most pulsars, energy losses range 
from  $10^{31}$--$10^{34}$ erg s$^{-1}$. These energy losses exactly correspond to the 
observed spin-down rate ${\rm d}P/{\rm d}t \sim 10^{-15}$, or to the spin-down time
$\tau_{\rm D} = P/\dot P \sim$ $1$--$10$ mln years. Let us keep in mind that the fraction 
of radio emission amounts to only $10^{-4}$--$10^{-6}$ of total energy losses. For most 
pulsars this corresponds to \mbox{$10^{26}$--$10^{28}$ erg s$^{-1}$,} which is 5--7 orders of 
magnitude less than the luminosity of the Sun.

As shown in Refs [56, 57], the actual energy losses cannot be due to magnetodipole radiation 
because the plasma that fills the magnetosphere will fully screen the low-frequency radiation 
from the neutron star. However, energy losses can be caused by longitudinal electric currents 
circulating in the magnetosphere and looped across the surface of the central engine. As a 
result, in this case, too, the main energy release near the neutron star is related to the 
electromagnetic energy flux (the Poynting vector flux), and the total energy losses can be 
again estimated using formula (3).

Most radio pulsars constitute single neutron stars. Of the 1880 pulsars known by the middle 
of 2010, only 140 were members of binary systems. However, in all these cases it is reliably
known that there is no somewhat appreciable mass transfer from the companion star to the 
neutron star. Since, as already stressed, the radio luminosities of pulsars are low, the 
modern sensitivity of detectors allows observations of pulsars only up to distances of 
$3$--$5$ kpc, which is smaller than the distance to the galactic center. Therefore, we can 
observe only a small fraction of all 'active' pulsars. The total number of neutron stars 
in our Galaxy must be around \mbox{$10^{8}$--$10^{9}$.} Such a big number of extinguished neutron 
stars can be naturally related to the small duration of their active life, as discussed above.

The jets are only observed in Crab and Vela radio pulsars [58, 59], which is not surprising, 
since, in contrast to the compact objects considered above, the pulsar magnetosphere is not 
axisymmetric. On the other hand, only axisymmetric configurations were actually 
considered until recently in the theory of pulsar wind. Based on these studies, the main 
features of strongly magnetized winds were understood. Nevertheless, even in this approximation 
for smooth flows, it has thus far been impossible to construct a self-consistent model which 
jointly describes the energy transfer from the neutron star surface to infinity and includes 
effective particle acceleration, i.e., an almost complete transformation of the electromagnetic 
field energy into the energy of the plasma flowing out. Because of this, different models are 
actively being discussed at present, which, to various degrees, propose going beyond the 
framework of the 'classical' scheme (see, for example, Refs [60-62]).

Indeed, observations show that most energy far from the neutron star must be carried by 
relativistic particles. For example, the analysis of the emission from the Crab Nebula 
in the shock region located at a distance of $\sim 10^{17}$ cm from the pulsar in the 
region of interaction of the pulsar wind with the supernova remnant definitely shows 
that the total flux $W_{\rm em}$ of the electromagnetic energy in this region is no more 
than $\sim 10^{-3}$ of the particle energy flux $W_{\rm part}$ [63]. Thus, the Poynting 
vector flux in the asymptotically remote region must be completely converted into the 
outgoing plasma flux. The presently known axisymmetric numerical models of jets 
from radio pulsars [64-66] were constructed exactly under this assumption.

However, the transformation apparently occurs already much closer to the neutron star, 
namely at distances comparable to the size of the light cylinder. This is evidenced by 
the detection of variable optical emission from companions in some close binary systems 
involving radio pulsars [67]. This variable optical emission with a period equal exactly 
to the orbital period of the binary can be naturally related to the heating of the companion's 
part facing the radio pulsar. It was found that the energy reradiated by the companion star 
almost matches the total energy emitted by the radio pulsar into the corresponding solid angle. 
Clearly, this fact cannot be understood either in the magnetodipole radiation model or by 
assuming a Poynting-dominated strongly magnetized outflow, since the transformation coefficient 
of a low-frequency electromagnetic wave cannot be close to unity. Only if a significant fraction 
of the energy is related to the relativistic particle flux can the heating of the star's surface 
be effective enough. Therefore, the so-called $\sigma$-problem --- the question as to how the energy 
is transferred from the electromagnetic field to particles in the pulsar wind --- remains one of 
big puzzles in modern astrophysics.

\subsection{Young stars}

Jets from young stars were indirectly discovered at the beginning of the 1950s, when 
G Herbig and G Haro [68, 69] discovered a new class of extended diffuse objects usually 
existing in pairs and, as became clear later, connected by thin jets with young 
rapidly rotating stars [70]. The formation of such jets can naturally be related to the 
need of removing most effectively the excessive angular momentum that prevents the formation 
of a star. As we see, the situation here is quite similar to that with active galactic nuclei, 
where first a diversity of different types of sources (quasars, Seyfert galaxies, and radio 
galaxies) were discovered, and only later on did it become clear that the activity of all
these sources has a similar nature. Moreover, the similarity of the observational features 
suggests that the physical mechanism of jet formation from young stars can also be similar 
to that from active galactic nuclei. And this is despite the fact that physical conditions 
near a young star (mass of order $3$--$10$ $M_{\odot}$, and total energy release ranging 
from $10^{31}$ to $10^{36}$ erg s$^{-1}$ are dramatically different from those in the centers 
of active galactic nuclei. One of the main differences here is the nonrelativistic character 
of gas outflow from young stars.

\begin{figure}
\begin{center}
\includegraphics[width=\columnwidth]{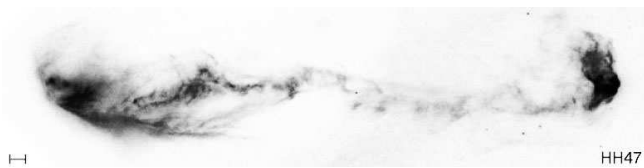}
\caption{Optical image of jets from the system HH47 (see, for example, Ref. [70]). 
        The scale corresponds to 1000 a.u.
}
\label{fig1_04}
\end{center}
\end{figure}

Presently, more than 250 Herbig-Haro objects are known [71]. As shown in Fig. 4, they 
represent bright condensations with an angular size of several seconds of arc (linear 
size of order $500$--$1000$ a.u.), usually surrounded by a bright diffuse envelope. 
Their spectra mainly show emission lines of hydrogen and some other low-excitation 
elements. A shock wave propagating with velocities $40$--$200$ km s$^{-1}$ through 
a gas with a density of $\sim 10^2$ cm$^{-3}$  is apparently the main source of excitation [70].

As in the case of radio galaxies, the activity of Herbig-Haro objects is dictated by 
collimated outflows which are well seen in forbidden lines. Nearly $60\%$ of the objects 
demonstrate both jets, while in other cases the receding jet is blocked by the accretion 
disc. The extent of the optical jets is of order $0.01$--$2$ pc, and their velocity 
reaches $600$ km s$^{-1}$. The gas density in the jets is estimated to be $10$--$100$ cm$^{-3}$, 
and the mass outflow rate comes to $10^{-9}$--$10^{-10}$ $M_{\odot}$ yr$^{-1}$. The degree of 
collimation of the jets (the ratio of the observed length to the width) can be as high as 30. 
The total jet opening angle is in the range of $5$--$10^{\circ}$. In addition to highly 
elongated jets, molecular outflows with a much smaller collimation degree are observed near 
young stars. Their size may run to \mbox{$0.04$--$4$ pc,} and the velocity of gas motion does not 
exceed \mbox{$5$--$100$ km s$^{-1}$.} Here we should stress that this velocity is much higher than 
the speed of sound in an outflow with a temperature of only $10$--$90$ K. The total mass of 
the ejected gas is estimated to be \mbox{$0.1$--$200$ $M_{\odot}$,} and the total kinetic energy 
stored in the molecular outflows can reach $10^{43}$ and even $10^{47}$ erg. The direct 
observation of rotation of the jets is the most important recent discovery. The characteristic 
velocities at an axial distance of $20$--$30$ a.u. range from 3--10 km s$^{-1}$ [72, 73]. 
There is also direct evidence of the spiral structure of the magnetic field in the jets [74]. 
All these facts unambiguously support the MHD model.

As in jets from microquasars, a strong instability frequently develops in collimated outflows 
from young stars at large distances from the central engine (see Fig. 4), so that the outflow 
is split into separate blobs. On the other hand, as seen from Fig. 5, the flow near the base 
of the jet can be considered sufficiently regular.

\begin{figure}[t]
\begin{center}
\includegraphics[width=0.8\columnwidth]{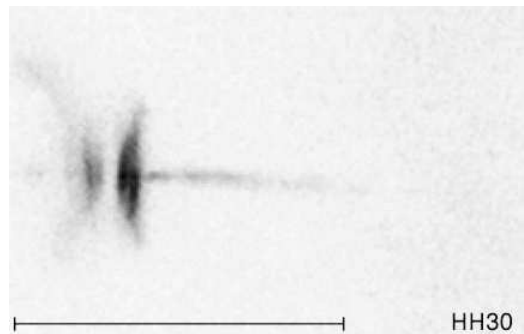}
\caption{Formation of a jet from a young star in the system HH 30 [70]. 
        The accretion disc is clearly seen. Here also the scale corresponds to 1000 a.u.
}
\label{fig1_05}
\end{center}
\end{figure}

As for the physical nature of collimated jet formation, this question is still far from solved. 
It is only clear that the power of the central engine is always sufficient to accelerate the 
outflowing gas; however, the mechanism of energy transformation remains unclear. We stress that 
in contrast to relativistic galactic objects (for example, microquasars), where the formation 
of jets is possibly caused by supercritical accretion, the luminosity in young stars never 
approaches the Eddington limit. On the other hand, it is clear that the key role in the collimated 
outflow formation is just played by accretion discs which undoubtedly exist around young stars. 
This is supported by the direct correlation between the power of the gas flux and the mass of the 
disc, estimated from its luminosity, as well as some other correlations [75, 76]. The parameters 
of the discs can be very different. For example, their masses range from $0.1$--$100$ 
$M_{\odot}$, while the outer radii can vary from 10 a.u. to 0.1 pc.

It is important that, in contrast to discs around relativistic objects (neutron stars and 
black holes), the gas temperature in discs around young stars is only \mbox{$20$--$100$ K.} As a 
result, as in the case with active galactic nuclei, neither the radiation pressure force 
nor gas pressure can explain the high velocities observed in the collimated outflows [71]. 
Therefore, to explain the jet formation and particle acceleration, models in which the magnetic 
field plays the key role and effectively mediates the interaction between the accretion disc 
and the jet were invoked once again. Because the real structure of the magnetic field in the 
proximity of a young star is presently unknown, here, too, both models in which the magnetic 
field of the star itself has a dominant role [77] and models in which the magnetic field of 
the disc plays the decisive role [75, 78] have been proposed. It is seen that here we meet 
the same problems regarding the structure of the initial magnetic field as in the study of 
the  black hole magnetosphere.

\section{Basics of the MHD approach}

\subsection{The key idea --- unipolar inductor}
 
As already said, the notion of a unipolar inductor is the main physical idea that 
underlies the MHD theory of compact objects. Referring to Fig. 6, a rotating magnetized 
ball can serve as the battery that determines the energy release from the central engine. 
Indeed, assuming the high conductivity of the ball, the freezing-in condition of the magnetic 
field, viz.
\begin{equation}
{\bf E}_{\rm in}
+ \frac{{\bf\Omega} \times {\bf r}}{c}\times {\bf B}_{\rm in} = 0
\label{d1}
\end{equation}
(i.e., simply the condition that the electric field in the rotating reference frame 
vanishes), leads to the appearance of the potential difference $\delta U$ between points 
$a$ and $b$. To an order of magnitude, this potential difference can be 
\begin{equation}
\delta U \sim E R_{0} \sim \frac{\Omega R_{0}^2}{c} B,
\label{d1'}
\end{equation}
where $R_{0}$ is the transverse size of the working area. As a result, the total energy 
release $W_{\rm tot}$ on the external load ${\cal R}$ will be given as
\begin{equation}
W_{\rm tot} = I \delta U,
\label{IU}
\end{equation}
where the electric current $I = \delta U/{\cal R}$. Here, however, several conditions 
should be met. First, the electric circuit must touch the ball at different latitudes, 
i.e., at points with different electric potentials. Second, the electric circuit should 
rotate with an angular velocity $\Omega$ different from that of the magnetized ball. 
The current flowing along a wire tightly welded on the ball will be absent.

\begin{figure}
\begin{center}
\includegraphics[width=0.8\columnwidth]{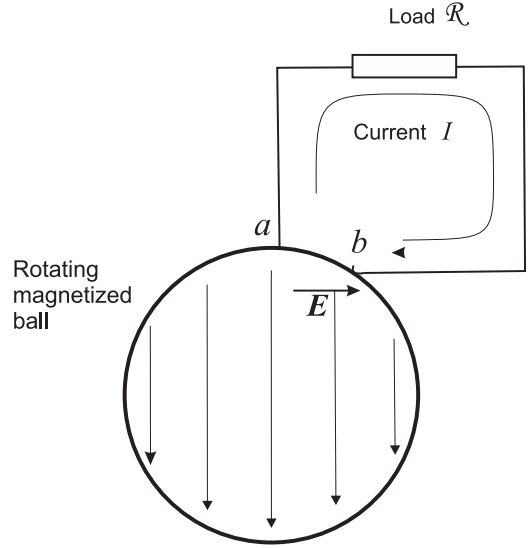}
\end{center}
\caption{The unipolar inductor as the source of a direct current. 
        Inside the magnetized ball, the electric current flows against the 
        electric field direction.
   }
\label{fig2_01}
\end{figure}

We stress that the energy source [electromotive force (EMF)] in the unipolar inductor is 
due to the kinetic energy of rotation. Indeed, as seen from Fig. 6, charges inside the ball 
move against the direction of the electric field. This becomes possible due to the force 
by which the lattice acts on charges carried along the wire, which violate the freezing-in 
condition inside the ball. Conversely, the Ampere force acting from the side of the surface 
electric current on the ball's material brakes its rotation. Therefore, the principle of 
work of the unipolar inductor (or, as it is sometimes called, the unipolar Faraday generator) 
is not the Faraday effect as such (where the EMF induced in a current loop depends on the 
variation of the magnetic flux), since the flux through the circuit remains constant. Notice 
that the reverse situation is also possible: if one applies a potential difference to a 
magnetized ball (i.e., if one replaces the load in Fig. 6 by the voltage source), the ball 
starts rotating. On the site http://fiziks.org.ua/samyj-prostoj-v-mire-elektrodvigatel/, 
which is devoted to laboratory studies in secondary school, one can find a video illustrating 
the work of such a device.

As we have understood, for the central engine to operate it is necessary to have:
\begin{itemize}
\item
rotating body;
\item              
regular magnetic field, and
\item
well-conducting wire.
\end{itemize}
Then the current, and hence the energy losses, will be determined by the value of the 
external resistance ${\cal R}$. Let us see now whether these conditions can be met in 
compact astrophysical objects.

As we have seen, a central rotating body in active astrophysical sources is undoubtedly 
present. For example, the spin periods of young stars are about several days (the inner 
parts of accretion discs rotate even faster). The spin periods of most radio pulsars are 
close to 1 s; however, they can be as small as a few milliseconds, which is already close 
to the limiting speed of rotation ($\Omega R/c \sim 0.1$). The rotational velocities of 
black holes in active galactic nuclei, to tell the truth, are unknown, but we can suppose 
that due to disc accretion (it is in this way that the millisecond-period pulsars are 
thought to have been spun up) their spin parameter $\Omega_{\rm H}R/c = a/2M$ (see the 
Appendix) can also be sufficiently large. For example, the estimate of the black hole 
rotational velocity in the nucleus of Seyfert galaxy MCG 06-30-15, as inferred from the 
iron 6.4-keV line profile distortion, yields $a/M = 0.989_{-0.002}^{+0.009}$ [79] (see 
also Ref [80]). As a result, the kinetic energy of rotation ${\cal E}_{\rm kin} = J_r \Omega^2/2$ 
stored in the central engine turns out to be quite sufficient to explain the energy source 
of activity of compact objects.

There are no particular problems with a regular magnetic field, either. In young stars, the 
proper magnetic field $B_{0}$ is measured directly and can be as high as \mbox{$10^{3}$ G [71].} 
At present, there are no direct observations of magnetic fields in radio pulsars, but they can 
be measured in X-ray (accreting) pulsars, which are also neutron stars [11]. Therefore, 
nobody now doubts that the magnetic field of a neutron star can reach $10^{12}$ G, and even 
extend up to $10^{15}$ G in magnetars [81]. The situation is worse with the magnetic fields 
of supermassive black holes. As is well known, a black hole cannot have a proper magnetic 
field, but the field can be generated in the surrounding accretion discs [82]. Unfortunately, 
so far there is no self-consistent theory of such generation, so we have to apply the estimate 
$B_0 \sim B_{\rm Edd}$, where
\begin{equation}
B_{\rm Edd} \approx 10^4 \left(\frac{M}{10^9M_{\odot}}\right)^{-1/2} {\rm G}.
\label{bedd}
\end{equation}
Let us keep in mind that such an estimate comes from the simple assumption that the energy 
density of the magnetic field is comparable to the total energy density in the accreting 
plasma yielding the Eddington luminosity (2). Clearly, estimate (7) represents rather an 
upper limit of the magnetic field near the black hole. In particular, it 
does not take into account the contribution from the thermal pressure, which can be significant 
in gamma-ray burst sources.

Finally, the problem of the 'electric wiring' can also be easily solved at first glance. Due 
to the presence of a strong magnetic field, in all cases the Larmor radius of particles 
$r_{\rm L} = m c v/eB$ is always much smaller than the size $R$ of the central engine. Therefore, 
one can consider with good accuracy that the electric current flows along the direction of the 
regular magnetic field. However, here we meet the problem of current closing, since particles 
in the region of the load must move across the magnetic field. We shall necessarily discuss 
this point below.

As an example, Fig. 7 illustrates how the braking occurs in an axisymmetric magnetosphere
of radio pulsars. Clearly, the total current flowing out of the pulsar surface must vanish; 
thus, there must necessarily be a reverse current in the magnetosphere to compensate for the 
loss of charges from the neutron star. As a result, currents ${\bf J}_{\rm s}$ closing the 
longitudinal currents in the magnetosphere must flow over the pulsar surface. The ponderomotive 
action of these currents must brake the rotation of the radio pulsars [3, 56].

\begin{figure}
\begin{center}
\includegraphics[width=0.95\columnwidth]{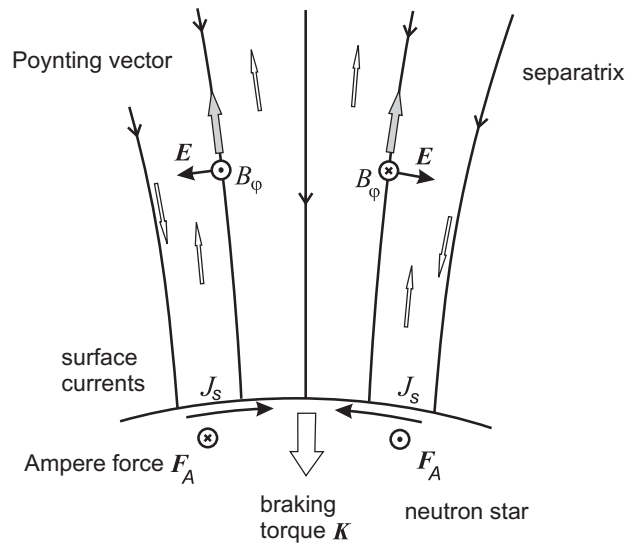}
\end{center}
\caption{The structure of electric currents (contour arrows) near the polar 
        caps of a neutron star. The Ampere force related to the surface current 
        ${\bf J}_{\rm s}$, produces the torque ${\bf K}$ braking the neutron star 
        rotation. Above the acceleration region, the energy flux is predominantly 
        transported by the Poynting vector (hatched arrows).
   }
\label{fig2_07}
\end{figure}

Thus, the problem of the magnitude of potential difference is solved quite easily. But the 
problem of the load that determines the current $I$ and, hence, the energy 
losses, proved much more difficult. A long way had to be covered in order to solve it, and 
this, essentially, will be discussed in this review. Nevertheless, we shall go somewhat 
ahead and give here the preliminary estimates confirming the applicability of the discussed 
mechanism. As shown below, a good estimate of the electric current density is given by the 
expression
\begin{equation}
j_{\rm GJ} = \rho_{\rm GJ}c,
\label{jGJ}
\end{equation}
where 
\begin{equation}
\rho_{\rm GJ} = -\frac{{\bf \Omega} \cdot {\bf B}}{2\pi c}
\label{GJ}
\end{equation}
is the electric charge density that is needed for the electric field in the rotating 
reference frame to vanish. Formula (9) can be easily derived from relation (4). It 
was first applied to the neutron star magnetosphere in the pioneering paper by P Goldreich 
and W H Julian [3], so the charge density (9) is usually called the Goldreich density.

\begin{table*}[ht]
\caption{ Parameters of the central engine: AGN --- active galactic nucleus, 
GRB --- gamma-ray burst, $\mu$QSO --- microquasar, PSR --- radio pulsar, 
msPSR --- millisecond radio pulsar, and YSO --- young stellar object.}
\vspace{0.3cm}
\centering
\begin{tabular}{|l|c|c|c|c|c|c|}
\hline
                        & AGN & GRB & $\mu$QSO & PSR & msPSR & YSO  \\
\hline
Mass $M$ in $M_{\odot}$& $10^6$--$10^9$ & $\sim 10$ & $\sim 10$ & $\approx 1.4 $ & $\approx 1.4 $ & $\sim 10$ \\
\hline
Radius $R$, cm        & $10^{11}$--$10^{14}$ & $\sim 10^6$ & $\sim 10^6$ & $\sim 10^6$ & $\sim 10^6$ & $\sim 10^{11}$ \\
\hline
Working radius $R_0$   & $\sim R$ & $\sim R$ & $\sim R$ & $(\Omega R/c)^{1/2} R$ & $(\Omega R/c)^{1/2} R$  & $\sim R$ \\
\hline
Period $P$             & $10$--$10^3$ \, s & $\sim 1$ \, ms & $\sim 1$ \, ms & $\sim 1$ \, s & $1.39$--$10$ \, ms  & $1$--$10$ \, d \\
\hline
$\Omega R/c$           & $\geq 0.1$ & $\geq 0.1$ & $\geq 0.1$ &$ \sim 10^{-4}$ &$\sim 10^{-1}$ & $\sim 10^{-5}$\\
\hline
Magnetic field $B_0$, G & $10^3$--$10^4$ & $\sim 10^{15}$ & $\sim 10^{10}$ & $\sim 10^{12}$& $\sim 10^{8}$ & $\sim 10^3$ \\
\hline
Energy storage ${\cal E}_{\rm kin}$, erg  & $10^{58}$--$10^{61}$ & $\sim 10^{52}$ & $\sim 10^{52}$ & $10^{44}$--$10^{46}$ & $\sim 10^{51}$ & $\sim 10^{44}$ \\
\hline
Dimensionless current $i_0$& 1 & 1 & 1 & 1 & 1 &  $ \sim c/v_{\rm in}$ \\ 
\hline
Power $W_{\rm tot}$, erg s$^{-1}$& $10^{42}$--$10^{45}$ & $10^{51}$--$10^{52}$ & $\sim 10^{38}$ & $10^{31}$--$10^{34}$ & $10^{34}$--$10^{35}$ &  $\sim 10^{35}$ \\
\hline
Lifetime $\tau_{\rm D}$, yr & $\sim 10^{7}$ & $\sim 10^{-6}$ & $\sim 10^{4}$ & $10^{6}$--$10^{7}$ & $10^{8}$--$10^{9}$&  $\sim 10^{4}$ \\
\hline
Current $I$, CGSE & $10^{26}$--$10^{28}$ & $\sim 10^{31}$ & $\sim 10^{25}$ & $10^{21}$--$10^{22}$ & $\sim 10^{22}$&  $ \sim 10^{25}$ \\
\hline
\end{tabular}
\label{table2_01} 
\end{table*}

Clearly, the total electric current circulating in the magnetosphere of the central 
engine can be conveniently written out in the form
\begin{equation}
I_{\rm tot} = i_{0} I_{\rm GJ}.
\label{iq0}
\end{equation}
Here $i_0$ is the dimensionless current, and  \mbox{$I_{\rm GJ}= \pi R_0^2 c \rho_{\rm GJ}$,} 
i.e., for the case $\rho_{\rm GJ} \approx$ const we obtain
\begin{equation}
I_{\rm GJ}=\frac{\Omega B_{0}R_0^2}{2}.
\label{d46}
\end{equation}
Finally, $R_0$ is again the size of the working area on the central engine surface. 
For black holes we can set $R_0 \approx R=r_{\rm g}$, and for neutron stars (radio 
pulsars) it must be on the order of the radius of a polar cap from which magnetic field 
lines can go beyond the light cylinder $R_{\rm L} = c/\Omega$. Indeed, inside the 
closed magnetosphere, by virtue of the remarkable Ferraro isorotation law, the plasma 
starts rotating with the star as a solid body, and, hence, this region cannot work as 
a unipolar inductor. The working area will include only the region of open field lines, 
inside which the plasma rotational velocity can be different from that of the star. 
As a result, for the dipole magnetic field we obtain
\begin{equation}
R_{0} \approx R\left(\frac{\Omega R}{c}\right)^{1/2}.
\label{r0}
\end{equation}

For relativistic strongly magnetized wind it is natural to assume that
\begin{equation}
i_0 \approx 1,
\label{i_0r}
\end{equation}
which corresponds to a free plasma outflow with the velocity $c$. As we shall see, this 
estimate is indeed correct. Therefore, the total energy losses can be estimated as
\begin{equation}
W_{\rm tot} \approx \left(\frac{\Omega R_0}{c}\right)^{2} B_{0}^{2} R_{0}^{2} c.
\label{i_0rnew}
\end{equation}
In consequence of this, as shown in Table 1, the unipolar inductor model allows us 
to explain both the total energy release $W_{\rm tot}$ and the time of activity of compact 
sources, $\tau_{\rm D} = {\cal E}_{\rm kin}/W_{\rm tot}$. As mentioned above, for radio 
pulsars estimate (14) with account for relation (12) coincides to within an order of 
magnitude with the magnetodipole losses (3).

For nonrelativistic outflows, estimate (13) is incorrect, and, as a detailed analysis 
shows, $i_0 \gg 1$ [9]. For a sufficiently rapid rotation with $\Omega > \Omega_{\rm cr}$, 
where
\begin{equation}
\Omega_{\rm cr} = \frac{v_{\rm in}}{R_{0}}\left(
\frac{4\pi\rho_{\rm in}v_{ \rm in}^2}{B_{0}^2}\right)^{1/2} \sim 10^{-6} \, {\rm s}^{-1},
\label{ocr}
\end{equation}
we have
\begin{equation}
i_0 \approx \frac{c}{v_{\rm in}}\left(
\frac{\Omega_{\rm F}}{\Omega_{\rm cr}}\right)^{-2/3},
\label{i00}
\end{equation}
and the dimensionless current for slow rotation limit takes the form
\begin{equation}
i_0 \approx \frac{c}{v_{\rm in}}.
\label{i_0n}
\end{equation} 
Here $\rho_{\rm in}$ is the density of the outflowing matter near the surface of the star, 
and $v_{\rm in}$ is the characteristic velocity of the outflow along the jet axis. As a 
result, the total energy losses for rapidly rotating stars can be expressed through the 
directly observed quantities:
\begin{equation}
W_{\rm tot} \approx \Omega^{4/3} \Psi_{\rm tot}^{4/3} {\dot M}^{1/3},
\label{exta1}
\end{equation}
i.e., through the total magnetic flux $\Psi_{\rm tot} = \pi R_0^2 B_0$, the rotational 
angular velocity $\Omega$, and the mass loss rate in the jet ${\dot M}$. For the parameters 
typical in young stars we have
\begin{eqnarray}
W_{\rm tot} \sim 10^{36} \, 
\left(\frac{P}{10^{6} \, {\rm s}}\right)^{4/3}
\left(\frac{B_{\rm in}}{10^{3} \, {\rm G}}\right)^{4/3} \nonumber \\
\left(\frac{R_{\rm in}}{10^{11} \, {\rm cm}}\right)^{8/3}
\left(\frac{\dot M}{10^{-9}\, M_{\odot} \, {\rm yr}^{-1}}\right)^{1/3} \, {\rm erg} \,{\rm s}^{-1}.
\label{exta3}
\end{eqnarray}
It is seen that this value is indeed close to the energy losses from young stellar objects. 
Thus, the unipolar inductor model allows us to explain the main jet characteristics for nonrelativistic 
sources, too.

Interestingly, the knowledge of the total energy losses $W_{\rm tot}$ immediately allows the total
longitudinal electric current the circulating in the magnetosphere to be estimated. Indeed, by 
comparing expressions (11) and (14), we straightforwardly obtain
\begin{equation}
I \approx i_0 c^{1/2}W_{\rm tot}^{1/2}.
\label{i_0}
\end{equation}
The characteristic amplitudes of currents are also collated in Table 1.

\subsection{Grad-Shafranov equation method}

The Grad-Shafranov equation method lies at the heart of the analytical theory which, 
in our opinion, is able to quite successfully describe the main properties of active 
compact astrophysical sources. Simply speaking, this approach describes axisymmetric 
stationary flows in the framework of ideal magnetohydrodynamics. This approximation 
is based on the assumption of a high conductivity of the plasma that fills the 
magnetosphere of the central engine (the high energy release guarantees a high degree 
of ionization of matter, and the effective production of electron-positron pairs in 
the vicinity of black holes). Moreover, most of the sources discussed above (except 
for radio pulsars) can be considered to a good approximation as axisymmetric 
and stationary.

\begin{figure}
\begin{center}
\includegraphics[width=0.6\columnwidth]{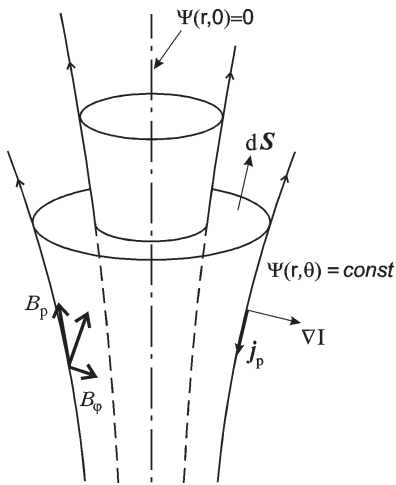}
\end{center}
\caption{Axisymmetric magnetic surfaces $\Psi(r,\theta) = $ const.
   }
\label{fig2_02}
\end{figure}

The attractiveness of this approach is related to the fact that there are quite a lot of 
integrals of motion in stationary ideal magnetohydrodynamics, i.e., quantities which are 
conserved along particle trajectories. This immediately provides us with important 
information without complicated calculations. Indeed, to determine the height a throwing 
stone reaches it is not necessary to solve equations of its motion: it is sufficient to 
apply the energy conservation law.

In the axisymmetric case, as illustrated in Fig. 8, the magnetic field vectors must 
lie on the magnetic surfaces which can be easily parametrized using the magnetic flux 
function $\Psi(r,\theta)$ that determines the magnetic field
\begin{equation}
{\bf B} = \frac{{\bf\nabla}\Psi \times {\bf e}_{\varphi}}{2\pi\varpi}
-\frac{2I}{c\varpi}{\bf e}_{\varphi}.
\label{d30}
\end{equation}
Here $\varpi = r\sin\theta$ is the distance from the rotational axis, and the 
numerical coefficient in the first term is chosen such that the function 
$\Psi(r,\theta)$ indeed coincides with the magnetic flux passing through 
a circle $r$, $\theta$, \mbox{$0<\varphi<2\pi$}. As for the quantity $I(r,\theta)$, 
it represents the total electric current flowing through the same circle.
It is easy to check that the following important properties are satisfied.
\begin{enumerate}
\item
At all times ${\rm d}\Psi = {\bf B} \cdot {\rm d}{\bf S}$ (${\rm d}{\bf S}$
is the surface element). Therefore, the 
function $\Psi(r,\theta)$ indeed bears the sense of the magnetic flux.
\item
Since the poloidal part of the magnetic field in formula (21) can 
be written out as $(2 \pi)^{-1} \nabla \Psi \times \nabla \varphi$, 
the condition $\nabla \cdot {\bf B} = 0$ is automatically 
satisfied. Thus, three components of the magnetic field are completely     
determined by two scalar functions $\Psi(r,\theta)$ and $I(r,\theta)$.
\item
For the same reason it is clear that the condition  ${\bf B} \cdot \nabla \Psi = 0$ 
will be satisfied for axisymmetric flows. Therefore, the lines $\Psi(r,\theta) =$ const 
define the form of magnetic surfaces. As a result, the integrals of motion should 
depend on only one scalar function, $\Psi(r,\theta)$.
\end{enumerate}

Let us now understand which integrals of motion appear in the case of axisymmetric 
stationary flows. Incidentally, the very structure of the Grad-Shafranov equation 
method will become clear. For simplicity, let us consider first a purely hydrodynamic 
flow. In this case, in analogy with relation (21), it is necessary to introduce the 
function $\Phi(r,\theta)$ of hydrodynamic flux defined as
\begin{equation}
\rho{\bf v}_{\rm p}
= \frac{\nabla \Phi \times {\bf e}_{\varphi}}{2\pi r\sin\theta},
\label{b1}
\end{equation}
where hereinafter the subscript 'p' will correspond to the poloidal [i.e., lying in the 
($r, \theta$) plane] components of vectors. 

In hydrodynamics, there are five scalar equations (the mass continuity equation, 
three components of the momentum Euler equation, and the energy conservation equation) 
for five unknown quantities --- three velocity components, and two thermodynamic functions. 
However, due to the axial symmetry, stationarity, and ideality of the flow three of five 
equations can be represented in the form $({\bf v}\nabla){\cal I}^{(i)} = 0$, which means 
that integrals ${\cal I}^{(i)}$ must be constant on the surfaces $\Phi(r,\theta) =$ const. 
As is well known, these integrals include the energy (Bernoulli integral) $E_{\rm n}$, 
the specific angular momentum $L_{\rm n}$, and the entropy $s$:
\begin{eqnarray}
E_{\rm n} & = & E_{\rm n}(\Phi) = \frac{v^2}{2} + w(\rho, s) + \varphi_{\rm g},
\label{b2} \\
L_{\rm n} & = & L_{\rm n}(\Phi) = v_{\varphi}r\sin\theta,
\label{b6} \\
s & = & s(\Phi).
\label{b3}
\end{eqnarray}
Here, $\varphi_{\rm g}$ is the gravitational potential, $w(\rho,s)$ is the specific enthalpy, 
and the subscript 'n' corresponds to nonrelativistic quantities, while Bernoulli integral 
$E_{\rm n}(\Phi)$ corresponds to the projection of the Euler equation (i.e., the equation 
of the force balance) onto the direction along the poloidal velocity ${\bf v}_{\rm p}$, and 
the angular momentum $L_{\rm n}(\Phi)$ represents the projection onto the unit vector 
${\bf e}_{\varphi}$. The remaining two first-order equations can be reduced to one second-order 
equation for the flux function $\Phi(r,\theta)$. It is clear that this equation will describe 
the force balance in the direction perpendicular to surfaces $\Phi(r,\theta) =$ const. In the 
compact form, it can be written out as
\begin{eqnarray}
-\varpi^2 \nabla_{k}
\left(\frac{1}{\varpi^{2}\rho}\nabla^{k}\Phi\right)
-4\pi^{2}\rho L_{\rm n}\frac{{\rm d}L_{\rm n}}{{\rm d}\Phi} \nonumber \\
+4\pi^{2}\varpi^{2}\rho \frac{{\rm d}E_{\rm n}}{{\rm d}\Phi}
-4\pi^2\varpi^{2}\rho \frac{T}{m_{\rm p}}\frac{{\rm d}s}{{\rm d}\Phi}=0.
\label{gscomp}
\end{eqnarray}  

It should be noted that we deliberately defined the enthalpy  $w$ as a function of 
density $\rho$ and entropy $s$. The point is that Bernoulli equation with the use 
of definition (22) can be recast into the form
\begin{equation}
E_{\rm n} = \frac{(\nabla\Phi)^2}{8\pi^2\varpi^2\rho^2}
+ \frac{1}{2}\,\frac{L_{\rm n}^2}{\varpi^2} + w(\rho,s) + \varphi_{\rm g}.
\label{b17}
\end{equation}
One can see that Bernoulli equation written in this form, in addition to the integrals 
of motion and the flux function $\Phi(r,\theta)$, contains only the density $\rho$. Consequently, 
it indirectly defines the density $\rho$ though the flux function $\Psi$ and integrals of motion:
\begin{equation}
\rho = \rho(\nabla\Phi;E_{\rm n},L_{\rm n},s;r,\theta).
\label{b18'new}
\end{equation}
This implies that after substituting Eqn (28) into Eqn (26), the latter will contain only 
one unknown flux function $\Phi(r,\theta)$ and three integrals of motion depending on it.

Such an equation can also be written out in the framework of ideal magnetohydrodynamics, 
too. In this case, however, not three but five integrals of motion exist. Two additional 
integrals come from the freezing-in condition
\begin{equation}
{\bf E} + {\bf v} \times {\bf B}/c = 0. 
\label{frozen}
\end{equation}
Indeed, from condition (29) follows that the electric field is perpendicular to the
magnetic field. In the axisymmetric case this means that magnetic surfaces 
\mbox{$\Psi(r,\theta) =$ const} will be equipotential ones. This condition can be conveniently 
rewritten in the form
\begin{equation}
\Omega_{\rm F} = \Omega_{\rm F}(\Psi),
\label{k41'}
\end{equation}
where the scalar quantity  $\Omega_{\rm F}$ determines the electric field according 
to the definition
\begin{equation}
{\bf E} = -\frac{\Omega_{\rm F}}{2\pi c}{\bf\nabla}\Psi.
\label{d34}
\end{equation}
This is related to the fact that:
\begin{itemize}
\item
in the axisymmetric case ($\partial/\partial t = 0$) Maxwell equation 
$\nabla \times {\bf E} = 0$  leads to the condition $E_{\varphi} = 0$,
\item
the freezing-in condition yields  $E_{\parallel} = 0$,
\item
the definition [31] together with Maxwell equation $\nabla \times {\bf E} = 0$ leads to the condition
$\nabla\Omega_{\rm F} \times \nabla \Psi = 0$, where relation (30) comes from.
\end{itemize}
Function $\Omega_{\rm F}$ introduced in this way bears the meaning of the angular 
velocity of particles (more precisely, the motion of particles is the sum of rotation with 
the angular velocity $\Omega_{\rm F}$ and sliding along the magnetic field). Condition (30) 
represents the Ferraro isorotation law [83], according to which the angular velocity of 
particle motion relative to the magnetic field must be constant on axisymmetric magnetic 
surfaces.

On the other hand, the freezing-in plasma condition implies that plasma velocity vectors ${\bf v}$ 
have also to lie on the magnetic surfaces, i.e., the flux of matter does not intersect the 
boundaries of the magnetic surfaces. This means that the particle flux function $\Phi(r,\theta)$ 
must be a function of the magnetic flux $\Psi(r,\theta)$. This fact allows us to introduce one
more integral of motion
\begin{equation}
\eta_{\rm n}(\Psi) = \frac{{\rm d}\Phi}{{\rm d}\Psi},
\label{p5}
\end{equation}
which, as evidenced by the foregoing, bears the sense of the ratio of the particle flux to the 
magnetic field flux. Correspondingly, the poloidal velocity of matter can be written as
\begin{equation}
{\bf v}_{\rm p} = \frac{\eta_{\rm n}}{\rho}{\bf B}_{\rm p}.
\label{p2'extra}
\end{equation}

As for the energy and angular momentum integrals (which in the nonrelativistic 
case are usually considered as functions of the particle flux), they now take the form
\begin{eqnarray}
  E_{\rm n}(\Phi) & = & \frac{\Omega_{\rm F} I}{2\pi \eta_{\rm n} c}
  +\frac{v^2}{2} + w + \varphi_{\rm g},
  \label{5a}  \\
   L_{\rm n}(\Phi) & = & \frac{I}{2\pi \eta_{\rm n} c}+ v_{\varphi}r\sin\theta,
    \label{6a}  
\end{eqnarray}
respectively. The entropy $s(\Psi)$ is ones again the one more (fifth) invariant. It is clear 
that both particles and electromagnetic field contribute to the energy and angular momentum, 
and, as can be easily checked, the term $\Omega_{\rm F} I/2\pi \eta_{\rm n} c$
corresponds simply to the Poynting vector flux.

Next, Bernoulli equation (34) can now be rewritten in the form
\begin{eqnarray}
  \frac{{\cal M}^4}{64\pi^4\eta_{\rm n}^2}\left(\nabla\Psi\right)^2
    = 2\varpi^{2} (E_{\rm n} - w -\varphi_{\rm g}) \nonumber \\
  -\frac{(\Omega_{\rm F}\varpi^2- L_{\rm n}{\cal M}^2)^2}
{(1-{\cal M}^2)^2}
      -2\varpi^2\Omega_{\rm F}\frac{L_{\rm n}-\Omega_{\rm F}\varpi^2}
{1-{\cal M}^2},
       \label{11a}
        \end{eqnarray}   
where 
\begin{equation}
 {\cal M}^{2}=\frac{4\pi\eta_{\rm n}^{2}}{\rho}.
  \label{10a}
   \end{equation}
The quantity ${\cal M}^{2}$  is the square of the Mach number of the poloidal 
velocity $v_{\rm p}$  relative to the poloidal component of the Alfv\'en 
velocity:
\begin{equation}
v_{\rm Ap} = \frac{B_{\rm p}}{\sqrt{4\pi \rho}},
\label{alfA}
\end{equation}
i.e., ${\cal M}^2=v_{\rm p}^2/v_{\rm Ap}^2$. It should be recalled that the specific 
enthalpy $w$ in equation (36) must be considered as a function of entropy $s$, as well 
as of the Mach number ${\cal M}^2$  and the integral $\eta_{\rm n}$. The corresponding 
relationship has the form
\begin{equation}
\nabla w = c_{\rm s}^2 \left(2\frac{\nabla\eta_{\rm n}}{\eta_{\rm n}}
-\frac{\nabla {\cal M}^{2}}{{\cal M}^{2}}\right)
+\left[\frac{1}{\rho}\left(\frac{\partial P}{\partial s}\right)_{n}
+\frac{T}{m_{\rm p}}\right]\nabla s.
\label{12aa}
\end{equation}
In consequence, as in the hydrodynamic limit, Bernoulli equation allows one to 
determine, albeit indirectly, the quantity ${\cal M}^{2}$ via the magnetic flux 
$\Psi(r,\theta)$ and five integrals of motion:
\begin{equation}
{\cal M}^2 = {\cal M}^2(\nabla\Psi;E_{\rm n},L_{\rm n},s,\eta_{\rm n},\Omega_{\rm F};r,\theta).
\label{b18'qqq}
\end{equation}
As for the projection of the force balance onto the direction perpendicular to the magnetic 
surfaces, it can be written in the form{\footnote{Unfortunately, in monograph [9] 
terms $-w$ $- \varphi_{\rm g}$ in Eqn (4.102) were discarded.}}
\begin{eqnarray}
&&\frac{1}{16 \pi^3 \rho}
\nabla_{k}\left(\frac{1-{\cal M}^2}{\varpi^{2}}
    \nabla^{k}\Psi\right)
+\frac{{\rm d}E_{\rm n}}{{\rm d}\Psi}
 \nonumber \\
&&+\frac{\Omega_{\rm F}\varpi^2 - L_{\rm n}}{1-{\cal M}^2} \,
\frac{{\rm d}\Omega_{\rm F}}{{\rm d}\Psi}
+\frac{1}{\varpi^2} \,
\frac{{\cal M}^2 L_{\rm n} - \Omega_{\rm F}\varpi^2}{1-{\cal M}^2}
\frac{{\rm d}L_{\rm n}}{{\rm d}\Psi}
\nonumber \\
&&+\left[2(E_{\rm n} -w - \varphi_{\rm g}) + 
\frac{\Omega_{\rm F}^2\varpi^4 -2\Omega_{\rm F}L_{\rm n}\varpi^2
+ {\cal M}^2 L_{\rm n}^2}{\varpi^2(1-{\cal M}^2)}\right]  
\nonumber \\
&&\times\frac{1}{\eta_{\rm n}}\frac{{\rm d}\eta_{\rm n}}{{\rm d}\Psi} 
-\frac{T}{m_{\rm p}} \, \frac{{\rm d}s}{{\rm d}\Psi} = 0.
\label{19b}
\end{eqnarray}

Now the structure of the treatment considered here becomes clear. Equation (41) 
jointly with Bernoulli equation (36) determines the value of the magnetic flux 
$\Psi(r,\theta)$. Then, again using Bernoulli equation, one can determine 
the value of the Mach number ${\cal M}$ at each point. After that it turned out 
that the number of integrals of motion is enough for determining all other 
quantities from simple algebraic equations. For example, one arrives at [84] 
\begin{eqnarray}
 \frac{I}{2\pi} & = & c\eta_{\rm n}
\frac{L_{\rm n}-\Omega_{\rm F}\varpi^2}{1-{\cal M}^2},
\label{Inrel} \\
   v_{\varphi} & = & \frac{1}{\varpi}\frac{\Omega_{\rm F}\varpi^2
    -L_{\rm n}{\cal M}^2}{1-{\cal M}^2},
    \label{9a}
     \end{eqnarray}
and, respectively, $\rho = 4 \pi \eta_{\rm n}^2/{\cal M}^2$. This is essentially the main 
attractiveness of the approach we discuss. Sometimes, as we shall see, the key properties 
can be directly obtained from algebraic relations. In other words, by making sufficiently 
reasonable assumptions about the structure of the flow [i.e., about flux function $\Psi(r,\theta)$], 
it is possible not to solve equation (41) at all and to analyze only algebraic, albeit indirect, 
relations.

The full version of the nonrelativistic equation containing all five invariants was first 
formulated by L S Solov'ev in 1963 in the third volume of the {\it Reviews of Plasma Physics} [85]. 
Being virtually unknown for astrophysicists, this equation was later reformulated anew several 
times [86-88]. For this reason, in particular, to date there has been no unique system of 
notations, so sometimes it is difficult to compare the results of different studies. In the 
literature, equations of this type are commonly called Grad-Shafranov equations, which were 
formulated at the end of the 1950s in relation to controlled thermonuclear fusion [89, 90], 
although the hydrodynamic version of this equation was known even earlier (see, for example, 
Ref. [91]). Similar equations going back to the classical Tricomi equation, were discussed 
as early as the beginning of the twentieth century in the context of transonic hydrodynamic 
flows [92, 93].

For simplicity, we have written out above equations only for the nonrelativistic case. 
However, it was not too difficult to obtain the corresponding equations both for the 
relativistic case [94] and for flows in the vicinities of nonrotating [95] and rotating 
[96, 97] black holes, since the Kerr metric is axisymmetric and stationary. It is 
these relativistic equations that we shall discuss below. In the main text we shall try 
to formulate sufficiently simple asymptotic expressions by focusing on the qualitative 
description of the flow properties. Sufficiently lengthy full equations are presented 
in the Appendix. Here, we shall restrict ourselves by writing out additionally the 
integrals of motion for relativistic flows in flat space.

Clearly, magnetic surfaces remain equipotential in the relativistic case, too. Thus, the 
angular frequency $\Omega_{\rm F}$ in definition (31) remains the integral of motion. As 
for the integrals of energy $E$ and the $z$-component of the angular momentum $L$, now 
they should be written out as{\footnote{To avoid misunderstanding, from now on the electric 
field {\bf E} will always be boldfaced.}}
\begin{eqnarray}
 & & E = E(\Psi) = \frac{\Omega_{\rm F}I}{2\pi}
+ \gamma \mu \eta c^2,
\label{p31} \\
 & & L=L(\Psi) = \frac{I}{2\pi} + \mu\eta \varpi u_{\varphi} c.
\label{p32}
\end{eqnarray}
Here ${\bf u}$ is the spatial part of the four-velocity vector 
\mbox{($\gamma = \sqrt{u^2 + 1}$} is the Lorentz factor), and
\begin{equation}
\mu %=\frac{\rho_{\rm m}+P}{n} 
\approx m_{\rm p}c^2 + m_{\rm p}w + \dots
\label{c17}
\end{equation}
is the relativistic enthalpy including the rest mass of particles. Finally, the relativistic 
integral of motion $\eta$ is now determined from the condition
\begin{equation}
{\bf u}_{\rm p} = \frac{\eta}{n}{\bf B}_{\rm p},
\label{p2'}
\end{equation}
where $n$ is the particle number density, and hereinafter all thermodynamic functions 
will be defined in the co-moving reference frame. Thus, for relativistic flows where
$|{\bf u}_{\rm p}| \approx \gamma$ we simply have
\begin{equation}
 \eta = \frac{n^{({\rm lab})}}{{B}_{\rm p}}.
\label{extraeta}
\end{equation}
It should be noted that the relativistic and nonrelativistic integrals of motion have 
different dimensions, since the relativistic integrals are normalized not on the unit 
matter flux ${\rm d}\Phi$ but on the unit magnetic flux ${\rm d}\Psi$.

It is evident that again both the energy flux and the angular momentum flux include the 
contributions from the electromagnetic field and particles, with the electromagnetic 
contribution [accurate up to an additional factor $\eta(\Psi)$] fully coinciding with 
that obtained in the nonrelativistic limit. In the general magnetohydrodynamic case, 
the total energy and angular momentum losses, $W_{\rm tot}$ and $K_{\rm tot}$, will be 
determined by the relationships
\begin{eqnarray}
 & & W_{\rm tot} = \frac{1}{c} \int_{0}^{\Psi_{\rm max}}E(\Psi){\rm d}\Psi,
\label{p32a} \\
 & & \nonumber  \\
 & & K_{\rm tot} = \frac{1}{c} \int_{0}^{\Psi_{\rm max}}L(\Psi){\rm d}\Psi,
\label{p32b}
\end{eqnarray}
respectively.

\subsection{Supersonic flows}

\vspace*{.2cm}

{\bf 3.3.1. The model.} In order to clearly understand the main features of the model considered here, it is 
convenient from the very beginning to analyze a sufficiently simple geometry of magnetic 
surfaces, and to formulate basic parameters characterizing the flow. Figure 9 demonstrates 
the simplestsplit monopole model of the magnetized wind [33], which during many years served 
as the 'hydrogen atom' for all researchers who studied the nature of the activity of galactic 
nuclei, gamma-ray bursts, and microquasars. Notably, most analytical results were obtained 
exactly for such flows.

It is assumed in the framework of this model that the 'central engine' involves a compact 
object (neutron star or black hole) and an accretion disc which separates the converging 
and diverging magnetic field fluxes. The accretion disc here is nedeed both to separate 
the oppositely directed magnetic field fluxes and, in the case of a black-hole magnetosphere, 
to generate the regular poloidal magnetic field (it will be produced by the toroidal currents 
flowing in the disc). In the absence of the accretion disc, a black hole, as is well known, 
cannot have the proper magnetic field (the so-called 'no-hair theorem' [18]). In addition, 
poloidal currents will also flow in the disc, closing the bulk currents flowing in the 
magnetosphere. Notice that a similar configuration with a disc separating magnetic field 
fluxes far from the neutron star also emerges in many models of radio pulsars [98 104], 
because it is natural to assume that at large distances the flow becomes quasispherical.

\begin{figure}
\begin{center}
\includegraphics[width=0.9\columnwidth]{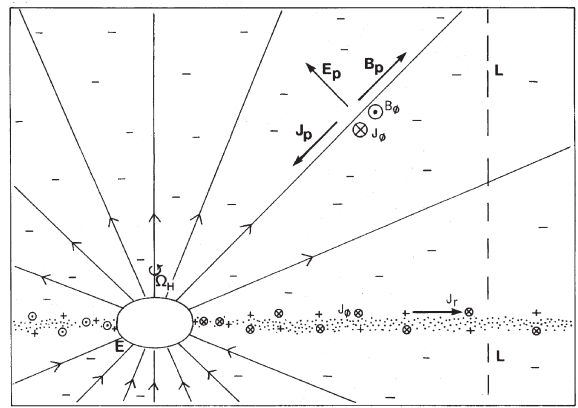}
\end{center}
\caption{The structure of electromagnetic fields for the split monopole magnetic field near a 
slowly rotating black hole [33]. Currents flowing in the highly conducting disc in the 
equatorial plane provide both the poloidal magnetic field jump and the closure of bulk 
currents flowing out the upper and bottom hemispheres.
   }
\label{fig2_03}
\end{figure}

Next, it is important that the crossed fields $E_{\theta}$ and $B_{\varphi}$ form the 
electromagnetic energy flux (the Poynting vector flux), which is directed along the 
magnetic surfaces. It should be stressed that this energy is transferred at the zero 
frequency, so the electromagnetic field that carries the energy is not an electromagnetic 
wave in the usual sense. The electromagnetic energy flux therefore appears only due to 
the longitudinal current generating the toroidal magnetic field; in the absence of 
particles, such energy release becomes impossible. The plasma also moves along the 
magnetic surfaces, so the sum of their energy fluxes is the integral of motion. On 
the other hand, the longitudinal current $I$ is not the integral of motion, so the MHD 
approximation we are considering allows, in principle, describing the current closure 
phenomenon. However, the magnetic surfaces here remain equipotential. Therefore, such 
flows can carry high electric voltages over large distances from the central engine. 
This should always be borne in mind when discussing the interaction of a magnetized wind 
with the surrounding medium. For example, this effect must be taken into account in close 
binary systems with radio pulsars.

As for the main dimensionless quantities characterizing the flow, they include the 
magnetization parameter $\sigma$, the particle production multiplicity $\lambda$, and 
the compactness parameter $l_{a}$. The {\it magnetization parameter} $\sigma$ shows by 
how much the electromagnetic energy flux near the central engine can exceed the particle 
energy flux. Thus, as seen from the definition of the energy integral (44), it can be 
more conveniently defined for relativistic flows as
\begin{equation}
\sigma = \left(\frac{E}{\mu\eta c^2}\right)_{\rm max},
\label{sigma}
\end{equation}
where the maximal value is chosen for all magnetic surfaces. As a result, the value 
of $\sigma$ corresponds to the maximal Lorentz factor of the plasma that can be 
reached in the case where all the electromagnetic field energy is transferred to 
the particles. In other words, $\sigma$ is the maximum Lorentz factor that can be 
achieved in the magnetized wind. Of course, here the mean hydrodynamic energy of 
the plasma flowing out is assumed. In particular, for the split monopole magnetic 
field (for which this quantity was first introduced by F C Michel [4] in 1969),
we obtain
\begin{equation}
\sigma = \frac{\Omega^2\Psi_{\rm tot}}{8\pi^2c^2 \mu\eta}.
\label{b12*}
\end{equation}
Correspondingly, for the nonrelativistic flow it is convenient to use the quantity
\begin{equation}
\sigma_{\rm n} = \frac{\Omega^2\Psi_{\rm tot}}{8\pi^2{v_{\rm in}}^3\eta_{\rm n}},
\label{b12*bis}
\end{equation}
where $v_{\rm in}$ is the velocity of matter flow along the jet axis. It is easy to 
check that the strong magnetization condition $\Omega > \Omega_{\rm cr}$  (15) coincides 
with the condition \mbox{$\sigma_{\rm n} > 1$}.

Recall that we are mainly interested in strongly magnetized flows, i.e., flows in which 
the main energy flux near the central engine is due to the Poynting vector flux
$\Omega_{\rm F}I/2\pi$.. In the opposite case, the flow will be only slightly different 
from the hydrodynamic outflow. Using definition (51), this condition can be rewritten as
\begin{equation}
\gamma_{\rm in} \ll \sigma,
\label{mdom}
\end{equation}
where $\gamma_{\rm in}$ is the injection Lorentz factor. As we shall see, the magnetization 
parameter $\sigma$ is the key parameter determining the basic features of the flow.

Next, to find the ejected plasma density, it is convenient to introduce the dimensionless 
particle production {\it multiplicity} $\lambda$ 
\begin{equation}
\lambda = \frac{n^{({\rm lab})}}{n_{\rm GJ}},
\label{lam}
\end{equation}
where $n_{\rm GJ}=|\rho_{\rm GJ}|/e$. Such a definition is connected with the fact that, 
as we shall show below, both in pulsar magnetospheres and in black hole magnetospheres 
the densest is the secondary electron-positron plasma generated either due to conversion 
of hard gamma-quanta in the magnetic field or due to gamma-ray collisions with thermal 
photons [33, 105]. However, hard gamma-quanta must be emitted in both cases by primary 
particles whose density is supposed to be close to the Goldreich density. The convenience 
of the particle production multiplicity $\lambda$ also stems from the fact that the 
magnetization parameter $\sigma$ can be rewritten with its help in the form
\begin{equation}
\sigma = \frac{e\Omega \Psi_{\rm tot}}{4\lambda m_{\rm e}c^3} 
\sim \frac{1}{\lambda}\left(\frac{W_{\rm tot}}{W_{\rm A}}\right)^{1/2},
\label{newsigma}
\end{equation}
where $W_{\rm A} = m_{\rm e}^{2}c^{5}/e^{2} \approx 10^{17}$ erg \, s$^{-1}$.
Thus, the knowledge of two of three quantities $W_{\rm tot}$, $\sigma$, and $\lambda$,
allows the determination of the third one.

Finally, the {\it compactness parameter}
\begin{equation}
l_{a} = \frac{\sigma_{\rm T} L_{\rm tot}}{m_{\rm e}c^3 R}
\label{compact}
\end{equation}
is in fact the optical depth for Thomson cross section $\sigma_{\rm T}$ at a distance $R$ from a 
source with the total luminosity $L_{\rm tot}$. Below, it will be important for us 
that the parameter $l_{a}$  provide an upper limit of particle energy in the acceleration 
region. On the other hand, a large $l_{a}$  is necessary for effective particle production.

\vspace*{.2cm}
{\bf 3.3.2. Singular surfaces.}
Singular surfaces represent the most important structural element of flows. As we shall see, 
it is the analysis of the conditions of the smooth passing of a flow through singular surfaces 
that allows sometimes rather general relationships to be obtained without solving the 
Grad-Shafranov equation itself. Notice from the very beginning that, for simplicity, below 
we shall only analyze the case of cold flows. The point here is that, at large distances 
thermal effects are insignificant for the polytropic index $\Gamma > 1$ (pressure 
$P \propto n^{\Gamma}$). This conclusion can readily be obtained from both the Grad-Shafranov 
equation itself and Bernoulli equation. Indeed, from the analysis, for instance, of 
nonrelativistic equations (36) and (41) it follows that both the enthalpy  
$w = c_{\rm s}^2/(\Gamma - 1) \propto n^{\Gamma -1}$ in Eqn (36) and the temperature 
$T \propto n^{\Gamma - 1}$ in Eqn (41) decrease with the distance from the compact source, 
since for any divergent outflow the particle number density $n \rightarrow 0$ for 
$r \rightarrow \infty$. Therefore, the contribution from a final temperature (enthalpy, 
entropy) compared to the total energy $E$ and its derivative ${\rm d}E/{\rm d}\Psi$ can 
be neglected at large distances. Thus, it becomes clear why in the analysis of relativistic 
flows the final temperature effects (and, in particular, critical conditions on the slow 
magnetosonic surface) are usually neglected. On the other hand, the pressure can be significant 
for cylindrical flows, i.e., for flows in which the density does not decrease with distance 
from compact object [9,106].

The first natural scale that emerges in the theory of relativistic winds is the {\it light 
cylinder}
\begin{equation}
R_{\rm L} = \frac{c}{\Omega},
\label{d14}
\end{equation}
i.e., the axial distance at which solid-body rotation together with the central object 
becomes impossible. It is easy to show that the light cylinder is the scale where:
\begin{enumerate}
\item
the magnitude of the electric field becomes comparable to that of the poloidal 
magnetic field;
\item
toroidal electric currents flowing in the magnetosphere start perturbing the poloidal 
manetic field of the central engine;
\item
the magnitude of the toroidal magnetic field produced by the longitudinal 
Goldreich current becomes comparable with that of the poloidal magnetic field.
\end{enumerate}
It follows from the first statement above and definitions (21) and (31) that beyond the 
light cylinder the electric field becomes stronger than the poloidal magnetic field. 
In particular, the poloidal magnetic field will decrease as $r^{-2}$ and the electric
field as $r^{-1}$ for the spit monopole outflow shown in Fig. 9. On the other hand, 
freezing-in condition (29) requires that the magnetic field be stronger than the electric 
field. This can be possible only when a strong enough longitudinal electric current is 
flowing in the magnetosphere, because the toroidal magnetic field also decreases as $r^{-1}$ 
according to Eqn (21).

We can therefore conclude that the question as to whether or not a smooth relativistic 
MHD outflow \mbox{($|{\bf E}| < |{\bf B}|$)} exists beyond the light cylinder 
is also directly related to the question of the magnitude of the longitudinal current 
circulating in the magnetosphere of a compact object. Then, for currents below some 
critical value, a so-called {\it light surface} is bound to appear in the magnetosphere, 
on which the electric field matches the magnetic field ($|{\bf E}| = |{\bf B}|$), and 
hence the approximation considered here itself becomes invalid. Calculations [107, 108] 
showed that the closure of currents occurs near this surface in the region with the 
thickness $\delta r \sim R_{\rm L}/\lambda$, and particles are effectively accelerated 
there up to energies of $\gamma \sim \sigma$. 

If the longitudinal currents are sufficiently high, the smooth MHD outflow can exist 
beyond the light cylinder as well. The electric field there will be almost equal to the 
magnetic field. Indeed, as directly follows from the relativistic Bernoulli equation 
($A.12$), in the limit $\varpi \gg R_{\rm L}$ we obtain simply{\footnote{This expression 
corrects Eqn (4.144) from monograph [9].}}
\begin{equation}
{\bf B}^2 - |{\bf E}|^2 = \frac{B_{\varphi}^2}{\gamma^2}.
\label{w10k}
\end{equation}
Since, as can be easily checked, $B_{\rm p} \approx B_{\varphi}/x_r$, where 
$x_r = \Omega \varpi/c$, we can always apply the estimate
\begin{equation}
B_{\varphi}^2 - |{\bf E}|^2 \leq
\, \frac{B_{\varphi}^2}{\gamma^2}.
\label{w10k'}
\end{equation}
As a result, the radial drift motion in the crossed electromagnetic fields dominates 
in a strongly magnetized relativistic outflow beyond the light cylinder. Indeed, as 
can be easily verified, the Lorentz factor entering into Eqn (59) satisfies the 
condition $\gamma^{-2} = 1 - U_{\rm dr}^2$, where
\begin{equation}
{\bf U}_{\rm dr} = c \frac{{\bf E} \times {\bf B}}{B^2}.
\label{drdr}
\end{equation}
In other words, the velocity parallel to the magnetic field does not contribute at all 
to the value of the Lorentz factor [109]. 

The {\it fast magnetosonic surface} is another important surface of the magnetized flows. 
It is fully equivalent to the sonic surface in the ideal hydrodynamics. Indeed, 
Bernoulli equation (27) is well known to have a singularity on the sonic surface. 
For example, the logarithmic derivative of density determined from Eqn (27) is 
written for a spherically symmetric flow in the form
\begin{equation}
\eta_1 = \frac{r}{\rho}\frac{{\rm d}\rho}{{\rm d}r}=
\frac{{\displaystyle{2v^2-\frac{GM}{r}}}}{c_{\rm s}^2-v^2}=
\frac{{\displaystyle{2-\frac{GM}{rv^2}}}}
{{\displaystyle{-1+\frac{c_{\rm s}^2}{v^2}}}}=\frac{N}{D}.
\label{a18}
\end{equation}
It is obvious that derivative (62) has a singularity when the velocity of matter 
equals the speed of sound: \mbox{$v = c_{\rm s} = c_{*}$ ($D=0$).} This means that in 
order to cross the sonic surface $r=r_{*}$ smoothly, the additional condition
\begin{equation}
N(r_{*}) = 2 - \frac{GM}{r_{*}c_{*}^2} = 0
\label{a19}
\end{equation}
must be satisfied. As a result, the additional critical condition (63) fixes the 
accretion (ejection) rate of matter [14].

The fast magnetosonic surface plays a similar role. But now it determines not the 
accretion or ejection rate, but the magnitude of the longitudinal current $I$ 
(more precisely, the integral $L$). It is this critical condition on this surface 
that shows us that in the relativistic case the longitudinal current $I$ near the 
central engine must be close to the Goldreich current $I_{\rm GJ}$.  On the other 
hand, as we have already noted, for a nonrelativistic outflow $i_0 \gg 1$, and the 
conditions of the smooth crossing of singular surfaces lead to relations (15)--(17) 
used above. Their derivation, however, is rather cumbersome, and we shall omit it 
here. It should only be emphasized that they can be obtained directly from an analysis 
of Bernoulli equation. The point is that the sonic surface is the $X$-point on 
the (distance $r$-velocity $v$) plane. That is, it is the point of crossing  
the roots of the algebraic Bernoulli equation. The condition of coincidence of roots 
of the algebraic equation puts certain bounds on the coefficients of the equation 
itself, which enables the magnitude of the longitudinal current to be estimated. 
Notice that expressions for the current formulated above were exactly obtained in 
Refs [110,111] for the simplest split monopole geometry shown in Fig. 9.

In a similar way, the following theorem can be proved: {\it In a relativistic outflow near 
the outer fast magnetosonic surface, the energy of the particles reaches the values
\begin{eqnarray}
\gamma & = & \left(\frac{E}{\mu\eta c^2}\right)^{1/3} \sim \sigma^{1/3}, \quad
 \gamma_{\rm in} \ll \sigma^{1/3},  \\
\gamma & = & \gamma_{\rm in}, \hspace*{3cm} \gamma_{\rm in} \gg \sigma^{1/3}. 
\end{eqnarray}
Thus, the fraction of energy carried by particles in the vicinity of the fast magnetosonic 
surface is a small fraction ($\sim \sigma^{-2/3}$) of the electromagnetic energy flux for 
strongly magnetized outflows $\sigma \gg \gamma_{\rm in}$. The surface itself is located 
at the distance of
\begin{eqnarray}
r_{\rm F}  & \approx & \left(\frac{E}{\mu\eta c^2}\right)^{1/3}R_{\rm L} 
\sim \sigma^{1/3}R_{\rm L}, \hspace*{.3cm} \gamma_{\rm in} \ll \sigma^{1/3}, 
\label{q05} \\
r_{\rm F}  & \approx &  \left(\frac{E}{\mu\eta\gamma_{\rm in}c^2}\right)^{1/2}R_{\rm L}
\sim \left(\frac{\sigma}{\gamma_{\rm in}}\right)^{1/2}R_{\rm L}, \label{q03} \\ 
&&\hspace*{4cm} \gamma_{\rm in} \gg \sigma^{1/3}
\nonumber
\end{eqnarray}
(the first relation holds true not too close to the rotational axis)}.
Interestingly, expression (66) is valid for both relativistic and nonrelativistic 
flows since it does not include in fact the speed of light $c$.

Finally, the singularity at $A = 1 - {\cal M}^2 = 0$ that appeared in nonrelativistic 
equations (42), (43) suggests that the {\it Alfv\'enic surface} must also play an important 
role in the structure of magnetized flows. Its location for nonrelativistic flows 
can easily be estimated from the numerator of relation (42):
\begin{equation}
\varpi_{\rm A}^2 = \frac{L_{\rm n}}{\Omega_{\rm F}}.
\label{w18}
\end{equation}
In this case, the Alfv\'enic surface turns out to be located close to the fast 
magnetosonic surface. 

As to a relativistic flow (and a flat space), the corresponding condition should be written differently:
\begin{equation}
A = 1 -\frac{\Omega_{\rm F}^{2}\varpi^{2}}{c^2}-{\cal M}^{2}.
\label{p39}
\end{equation}
On the other hand, it is easy to check that the parameter $q = W_{\rm part}/W_{\rm em}$
(i.e., the particle-to-electromagnetic energy flux ratio) can be presented in the form
\begin{equation}
q = \frac{{\cal M}^2c^2}{\Omega_{\rm F}^2\varpi^2}.
\label{defg}
\end{equation}
Thus, in the region where the energy flux is Poynting-dominated ($q \ll 1$), the following 
condition should be satisfied:
\begin{equation}
{\cal M}^2 \ll \frac{\Omega_{\rm F}^2\varpi^2}{c^2}.
\label{w10c}
\end{equation}
Consequently, the Alfv\'enic surface for such flows is located near the light cylinder. 
Hence (except for the polar region where both the Alfv\'enic and fast magnetosonic surfaces are
close to each other), the fast magnetosonic surface for strongly magnetized flows is located 
$\sigma^{1/3}$ times further from the central engine as compared with the Alfv\'enic surface.

Let us remember that the Alfv\'enic surface in the relativistic case determines the scale on 
which the toroidal magnetic field becomes comparable in magnitude to the poloidal field. 
It is easy to check that for rapid rotation $\Omega > \Omega_{\rm cr}$ (15) a similar 
situation holds for nonrelativistic flows, too. As regards the electric field, it is 
always weaker than the magnetic field in the nonrelativistic case. Notably, that is why 
the light surface cannot appear in the nonrelativistic case.

The Alfv\'enic surface represents a higher-order singularity as against the fast magnetosonic 
surface. Therefore, relations (42) and (43) do not put any constraint on the integrals of 
motion and only determine the location of the Alfv\'enic surface and the magnetic field 
structure. In this event, however, particles can intersect the Alfv\'enic surface only in 
one direction. For example, when the central source loses its rotational energy, crossing 
the Alfv\'enic surface is possible only in the direction outward from the compact object. 
When the energy flux is directed toward the central engine (for instance, if it is spun 
up by the accreting material), the flow must also be directed toward the central engine. 
Certainly, this statement is invalid in the region of an accretion disc where viscosity 
cannot be neglected.

This statement can easily be proved in the relativistic case by recalling that the motion 
of particles is the sum of the drift motion in the crossed fields and the motion along the 
magnetic field. The condition $v < c$ that limits the longitudinal velocity puts bounds on 
the radial velocity of matter. This is related to the fact that the drift velocity itself 
becomes close to the speed of light on the Alfv\'enic surface. However, accretion of matter 
with positive energy release cannot be realized in the nonrelativistic case, either. In this 
event, the interaction of the supersonic accretion flow with the rotating magnetized central 
body would take place. A shock wave is known to be formed in such an interaction [112].

\begin{figure}
\begin{center}
\includegraphics[width=0.9\columnwidth]{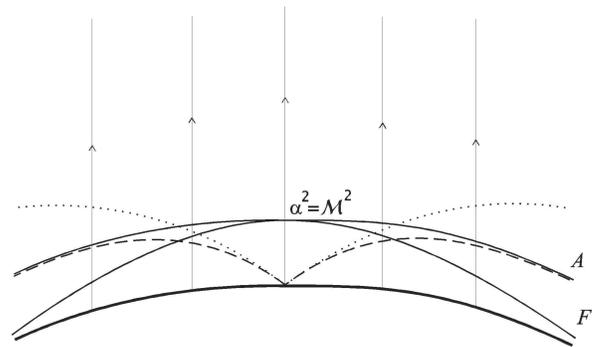}
\end{center}
\caption{The location of the Alfv\'enic ($A$) and fast magnetosonic ($F$) surfaces near the 
black hole horizon. The dashed line shows the Alfv\'enic surface in the force-free 
approximation (i.e., the 'light cylinder'), while the dotted line indicates the 
ergosphere surface. Here, $\alpha$ is the gravitational redshift (see the Appendix).
   }
\label{fig2_04}
\end{figure}

In conclusion, we should comment on the features of a black hole magnetosphere. As seen 
from exact expressions for the Alfv\'enic $A$ ($A.13$) and sonic $D$ ($A.20$) factors presented 
in the Appendix, a second family of singular surfaces inevitably emerges near the black hole 
horizon. Here, as shown in Fig. 10, the infalling matter, as in the case of the outflow, must 
first cross the Alfv\'enic surface and only then intersect the fast magnetosonic surface (recall 
that thermal effects are not discussed here). This is related to the fact that the strong 
gravitational field of the black hole forces the matter to approach the event horizon.

The appearance of the second family of singular surfaces leads to new important properties. 
First of all, the matter can intersect the inner Alfv\'enic surface only in the direction 
towards the black hole horizon. But this means that if the central engine loses rotational 
energy (and hence particles can cross the outer Alfv\'enic surface only in the direction away 
from the compact object), the plasma is bound to be generated in the magnetic field lines 
'anchored' to the black hole horizon. Only in this case can electric currents appear in the 
black hole magnetosphere, which are necessary, as we have seen, to explain the observed 
energy release. In turn, the appearance of one more critical condition on the inner fast 
magnetosonic surface proved to be sufficient to determine the angular velocity $\Omega_{\rm F}$.
Here, $\Omega_{\rm F}$ indeed must be close to $\Omega_{\rm H}/2$ 
[$\Omega_{\rm H} = \omega(r_{\rm g})$ is the angular velocity of the black hole; 
see the Appendix], as was understood as early as the Blandford and Znajek paper [33]. 
This problem has been possible to solve exactly for slow rotation in the split monopole 
magnetic field [113].

\vspace*{.2cm}

{\bf 3.3.3. The problem setting.}
Before considering the main results which were obtained using the analytical theory, it is 
necessary to discuss the formulation of the problem. The point is that we shall primarily 
be interested in transonic flows, i.e., those which are subsonic near the compact object and 
supersonic in the wind region. Indeed, as we shall see, the distance from a central 
engine to singular surfaces in all compact objects are much smaller than even the transverse 
size of the collimated outflows. The difficulty here is that the direct problem setup itself 
in the framework of the Grad-Shafranov equation method turns out to be nontrivial. For example, 
the second-order equation describing the two-dimensional flow in the hydrodynamic limit, when 
only three integrals of motion are available, requires four boundary conditions to be imposed 
in the transonic regime. The fifth condition is the critical condition on the sonic surface. 
This means that on some surface, for example, two thermodynamic functions and two velocity 
components must be specified. We stress that here only flows depending on two variables are 
considered. As discussed in detail in monograph [9], the well-known spherically symmetric 
flows (the Bondi accretion, the Parker ejection) are degenerate, since the structure of the 
flow itself is specified in them. In the general case of spherical accretion, a nonstationary 
solution with a shock wave appears (see, for example, Ref. [114]).

However, to determine Bernoulli integral, which is naturally needed in solving the 
equilibrium equation, we should specify all three components of the velocity, which is 
impossible since the third velocity component itself should be found from the solution. 
In the general case, one should set
\begin{equation}
b = 2 + i - s'
\label{a35}
\end{equation}
boundary conditions, where $i$ is the number of invariants, and $s'$ is the number of 
singular surfaces (and for the magnetic field lines threading the black hole horizon 
this number is doubled, i.e., separately for ejecting and accreting plasma). Such an 
internal inconsistency of this approach in the general case does not allow us to solve 
direct problems, namely, to determine the structure of the flow in some region using the
given physical parameters on its boundary. Therefore, it is not astonishing that most 
researchers primarily interested in astrophysical applications already in the middle 
of the 1990s started addressing a totally different class of equations, namely those 
covering time relaxation problems, which can only be solved numerically [79, 115-119]. 
However, only in the last several years has significant progress here been achieved 
[103, 109, 120-127], which, among other things, has confirmed many analytical results 
obtained before.

It should be noted that this problem does not appear in both subsonic and supersonic 
cases. For these flows, all necessary integrals of motion must be determined from the 
boundary conditions. In particular, the boundary conditions will determine in the subsonic 
case the longitudinal current $I$, too. This exactly corresponds to the 
unipolar inductor model, where the current (and hence the energy losses) is determined 
by the external load. Unfortunately, this ideology has also spread into the theory of 
magnetized winds. This is related to the fact that many results were obtained in the 
1970s-1980s using the force-free approximation [i.e., when $\sigma \rightarrow \infty$ and 
masses of particles can be neglected]. In this approximation, the Grad-Shafranov equation 
becomes elliptical and, hence, the flow structure must depend on conditions at the external 
boundary. But the theory of pulsar magnetospheres so far has been constructed 
in the force-free approximation. The class of subsonic flows also includes the so-called 
'magnetic tower' [128, 129]. As this question is highly important, we shall discuss 
it below in more detail.

Correspondingly, the Blandford-Znajek model was also constructed in the force-free 
approximation, which required the boundary condition to be set on the black hole horizon. 
Since the electromagnetic wave (like other material bodies) can propagate near the horizon 
only normal to the horizon toward the black hole, the boundary condition in fact 
is equivalent to the Leontovich boundary condition in radio physics [130]. However, it 
is usually obtained by requiring the finiteness of fields in the freely falling 
observer's frame of reference, which yields $B_{\varphi}(r_{\rm g}) =  -E_{\theta}(r_{\rm g})$. 
This condition, as is well known, can be rewritten in the form of the Ohm law for the 
formally introduced 'surface current' [131]
\begin{equation}
{\bf J}(r_{\rm g}) = \frac{c}{4\pi}{\bf E}(r_{\rm g}),
\label{k50}
\end{equation}
which corresponds to the universal 'internal' resistance of the battery, 
${\cal R} = 4\pi/c =$ 377 $\Omega$. In another form, this boundary condition 
can be rewritten as
\begin{equation}
4 \pi I(\Psi)
=[\Omega_{\rm H}-\Omega_{\rm F}(\Psi)]\sin\theta
%\frac{r_{\rm g}^2+a^2}{r_{\rm g}^2+a^2\cos^2\theta}
\left(\frac{{\rm d}\Psi}{{\rm d}\theta}\right).
\label{k48}
\end{equation}
Here, we have utilized definitions ($A.7$) and ($A.8$) and, for simplicity, written the 
equality $B_{\varphi}(r_{\rm g}) =  -E_{\theta}(r_{\rm g})$ for slow rotation. Thus, it 
is not surprising that in the framework of this approximation the mechanism of energy 
loss by a black hole was connected, in analogy with the unipolar inductor, with the Ampere 
force acting on the black hole horizon from the side of the surface current [131].

Only much later was it understood that the force-free approximation gives inaccurate and 
sometimes erroneous results. This is related to the fact that in the force-free 
approximation, i.e., when particles are assumed to be massless, the flow always remains 
subsonic. Under this assumption, the fast magnetosonic surface on which the poloidal 
velocity of particles matches the fast magnetosonic wave velocity goes formally to 
infinity (and the inner surface, oppositely, coincides with the black hole horizon). 
But we have seen that it is the conditions on the fast magnetosonic surface that fix 
the longitudinal current circulating in the magnetosphere. In addition, the very 
necessity of establishing the boundary condition on the black hole horizon, i.e., 
in the causally disconnected region, shows that the physical interpretation given 
above does not relate to the reality [132].

\begin{figure}
\begin{center}
\includegraphics[width=0.6\columnwidth]{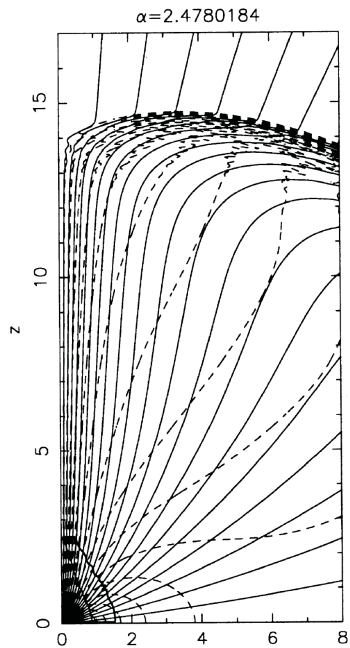}
\end{center}
\caption{The structure of a magnetic field behind a switching-on wave propagating with velocity 
$c$ from a compact object [120]. Inside the switching-on wave, the flow rapidly becomes 
stationary, in correspondence with the analytical solution.
   }
\label{fig2_05}
\end{figure}

To make this point more clear and, in particular, to understand the boundedness of stationary 
solutions (and hence the analytical method itself) in studies of the current closure, it is 
useful to consider the results obtained in paper [120] in which the problem was formulated 
as follows. There is a magnetized ball at rest which at the moment $t = 0$ starts 
rotating with angular velocity $\Omega$. As a result, the switching-on wave starts propagating 
from the ball with velocity $c$, so that the magnetic field remains unperturbed beyond it and 
electric currents are absent, while inside the switching-on wave (and this is a very important 
result) the solution rapidly approaches the stationary transonic regime, which is in full 
agreement with the analytical solution. Thus, the assumption of the stationary solution for 
longitudal currents flowing actually along magnetic surfaces is confirmed.

As for the current closure, no current closure as such happens at all in the ideal case 
where the outflow occurs in a vacuum. This is related to the fact that in the switching-on 
wave the flow is time-dependent (Fig. 11), so there ${\rm div} {\bf j} \neq 0$
(S S Komissarov, private communication). In reality, the current closure will take place on 
a shock wave which must necessarily emerge in the region where the supersonic switching-on 
wave collides with the surrounding medium. In any case, however, the ambient medium for 
transonic flows cannot influence the magnitude of the longitudinal current for $r < r_{\rm F}$ and, 
hence, affect the central engine energy release. As soon as the switching-on wave crosses 
the singular surfaces (they are also shown in the left lower corner of Fig. 11), the 
longitudinal current flowing in the magnetosphere stops depending on time. For this reason, 
one can indeed consider in the framework of the stationary approximation that the electric 
current closure occurs at infinity, as is usually assumed. Thus, transonic flows are 
significantly different from subsonic ones, when the electric current circulating in the
magnetosphere is determined by the conductivity of the boundary of the region occupied by plasma 
(see, for example, Ref. [133]).

Accounting for the nonzero mass of particles allows us to clarify the situation with the 
'boundary condition on the horizon', and thus with the mechanism of energy release by a 
black hole. Indeed, as shown in Fig. 10, the fast magnetosonic surface for nonzero masses 
of particles is located above the black hole horizon. Therefore, the critical condition 
should also be set here, which is definitely located in the region casually connected to 
the outer space. The black hole horizon will be located in the region of the supersonic 
flow and, hence, cannot affect the properties of the flow. As a result, the additional 
critical condition must also be kept in the force-free limit $m_{\rm p} \rightarrow 0$, when 
the fast magnetosonic surface, as noted above, formally coincides with the event horizon. 
It is not then surprising that this limit of the critical condition on the fast magnetosonic 
surface exactly coincides with condition (74) [10]. Thus, the boundary condition on the 
horizon (74), which was necessarily used in the force-free approximation, represents a relict 
of the critical condition on the fast magnetosonic surface.

Correspondingly, it also becomes clear how to interpret the infinite time retardation near 
the black hole horizon. Indeed, the time it takes for the plasma to reach the horizon must 
be infinite from the point of view of the remote observer. As the time of existence of the 
black hole (the surrounding plasma, etc.) is finite, the remote observer will register a 
'switching-on wave' corresponding to the very initial stages of the existence of the central 
engine over the black hole surface. This was, in fact, one more argument in support of the 
fact that it is incorrect to set any boundary condition on the black hole horizon [132]. 
However, as seen by the example of the outflow (see Fig. 11), to form the current system 
that fully determines the central engine power it is sufficient to wait a finite time until 
the plasma intersects the fast magnetosonic surface. This also implies, inter alia, that 
after a finite time the switching-on wave will turn out to be in the supersonic region of 
the flow and, hence, cannot affect the magnetosphere structure. This conclusion was 
numerically confirmed by Komissarov [134].

If this is indeed the case, however, another source of EMF should be found, since the black 
hole itself now can no longer be the source of extraneous forces and, consequently, serve as 
a battery. It turned out that the appearance of EMF in the black hole magnetosphere is related 
to the Lense-Thirring (the frame-dragging) effect, which appears due to rotation of the 
black hole. Indeed, according to general relativity, the space itself in the vicinity of a 
Kerr black hole starts rotating with angular velocity $\omega$ ($A.4$). Only in the reference 
frame rotating with this angular velocity will an observer not register a precession of 
gyroscopes. On the contrary, noninertial forces will appear in the laboratory frame at rest 
with respect to remote observers and they can be detected. It is the frame-dragging effect
that leads to the appearance of the electromotive force in the black hole magnetosphere.
 
\begin{figure}
\begin{center}
\includegraphics[width=\columnwidth]{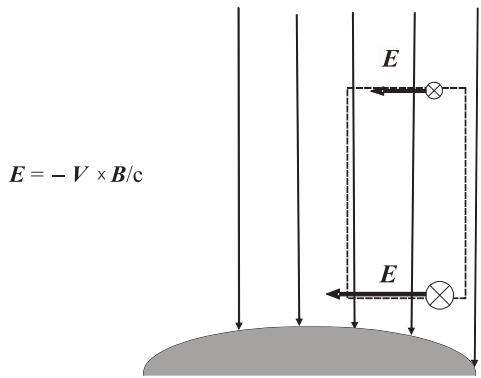}
\end{center}
\caption{The appearance of the electromotive force in a circuit at rest 
relative to a rotating black hole immersed in an external magnetic field. 
The electric field directions are shown for observers at rest.
   }
\label{fig2_06}
\end{figure}

Indeed, as in the case of any body moving in the magnetic field, the 'motion of space' relative 
to the observer at rest will produce the electric field ${\bf E} = -{\bf V} \times {\bf B}/c$, 
where now ${\bf V} = {\bf \omega} \times {\bf r}$ is the velocity of the body at rest 
relative to the preferred reference frame. Here, it is important that the angular velocity 
$\omega$  be different at various distances from the black hole. Therefore, 
as shown in Fig. 12, the circulation of the electric field in a circuit will be nonzero 
even if the electric circuit is at rest relative to the black hole, i.e., when the magnetic 
flux through the contour remains constant (of course, this state is possible only above the 
ergosphere). It is the motion of space through the circuit that generates the electromotive 
force. We see here that the 'electric battery' will be located above the black hole horizon, 
and possibly outside the ergosphere.

Finally, as already stressed, for positive energy flux from the central engine the electric 
currents in the black hole magnetosphere can flow only in the case where the plasma generation 
mechanism is operating above the horizon. Electron-positron pair production in the collision 
of hard gamma-ray quanta emitted from the accretion disc surface was already discussed above. 
Here, we should note that, as in the Penrose effect, one particle should fall into the black 
hole, and another particle should escape to infinity. It should be recalled that the Penrose 
effect has its origins in the remarkable property of rotating black holes --- the relativistic 
mass defect can exceed $100\%$ inside the ergosphere 
\mbox{$r_{\rm g} < r < r_{\rm e} = M + \sqrt{M^2 - a^2 \cos^2\theta}$}
($G = c = 1$) and, hence, above all the horizon surface [14]. Therefore, it becomes evident 
that the Blandford-Znajek effect is in fact the electromagnetic realization of the Penrose 
process. The difference is that it concerns not charged particles themselves but the 
electromagnetic field they induce. In other words, the spin-down of the black hole is not 
due to electric currents flowing over the horizon but due to the negative 
electromagnetic energy flux falling onto the black hol. It is this role of 
the ergosphere inside which the relativistic energy of any material bodies (including 
the electromagnetic field) can become negative that is great. Such interpretation seems 
now to be the most likely, and most researchers involved in these studies tend to accept 
this interpretation (see, for example, book [135]).

Let us summarize. We have shown that the critical conditions on the fast magnetosonic 
surface mostly determine the energy release of the central engine. This surface serves 
as a valve that determines the magnitude of the longitudinal current circulating in the 
magnetosphere. Like the usual sonic surface in hydrodynamics, it separates subsonic and 
supersonic parts of the flow. As a result, the longitudinal current for transonic flows 
must be determined not by the external conditions but by the condition of smoothly crossing 
the singular surfaces. Even more complicated is the situation in a black hole magnetosphere, 
where both the current and the angular velocity $\Omega_{\rm F}$ should be deduced from the 
critical conditions on the singular surfaces. In consequence of this, it is these critical 
conditions on the singular surfaces that will determine the central engine power.

\section{Theoretical predictions}

\subsection{Collimation}

{\bf 4.1.1 The force balance across the flow.} 
Let now analyze the results of the analytical theory. First of all, we should try to 
formulate some general properties of a magnetized outflow, which must show their worth 
at large distances from the central object. As already noted, we are primarily interested 
in transonic flows, in which the flow at large distances is supersonic. Moreover, we shall 
unconditionally assume that the solution can be continued to infinity. This is possible, 
as we have seen, only if the longitudinal electric current is sufficiently large. As we have 
already stressed, thermal effects can almost always be neglected. As a consequence, Bernoulli 
equation becomes a fourth-order algebraic equation with respect to ${\cal M}^2$, which in 
many cases allows us to write out rather simple analytical asymptotic solutions.

For simplicity, let us start from the nonrelativistic case. By analyzing the leading 
terms in the Grad-Shafranov equation (41), it is possible to show that the force balance 
equation can be written in the form [136]
\begin{equation}
 \frac{\rho v_{\parallel}^2}{R_{\rm c}}
= \frac{1}{c}j_{\parallel}B_{\varphi},
\label{t-f-n}
\end{equation}
where $R_{\rm c}$ is the radius of the magnetic field line in the poloidal plane. 
In other words, the equilibrium must be established due to the balance between the 
centrifugal force  $\rho v_{\parallel}^2/R_{\rm c}$ and the Ampere force related to 
the longitudinal electric current $j_{\parallel}$. It is evident that the Ampere 
force, depending on the sign of the longitudinal current, can both collimate and 
decollimate the flow. In this case, the collimation must occur close to the jet axis, 
while the decollimation occurs at its periphery. Exactly this behavior was obtained 
both analytically [111] and numerically [120] (Fig. 13). It should be stressed that 
here we do not consider how strongly the magnetic surfaces can be bent and only 
investigate their form.

\begin{figure}
\begin{center}
\includegraphics[width=0.8\columnwidth]{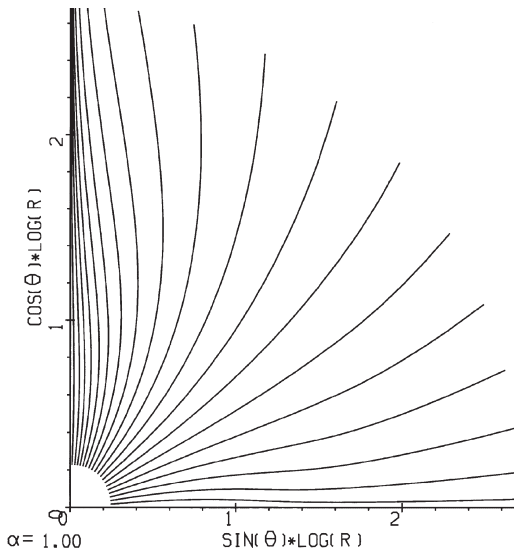}
\end{center}
\caption{The structure of magnetic surfaces for a nonrelativistic plasma outflow 
in the split monopole magnetic field [120]. The decollimation in the region of the 
reverse bulk current near the central engine is clearly seen. However, the 
redistribution of longitudinal currents occurs at far distances, so that the 
current flowing out is concentrated near the rotational axis, and the reverse 
current flows near the equatorial plane.
   }
\label{fig3_01}
\end{figure}

A similar picture, however, cannot be realized up to very large distances from the 
central engine. 'Natura abhorret vacuum', and the diverging magnetic surfaces inevitably 
must start collimating. It turned out that this is possible exactly because the 
longitudinal current $I$ is not the integral of motion, and hence the electric 
current, unlike particles, can intersect the magnetic surfaces. Indeed, it is easy 
to check that, for an almost radial flow at large distances, the right-hand side of 
equation (75) should decrease as $r^{-3}$, while the numerator on the left-hand side 
decreases as $r^{-2}$. As a result, for equality (75) to hold, the radius of curvature 
of the magnetic surfaces, $R_{\rm c}$, should increase as $r$. But such a behavior cannot be 
realized at mathematical infinity [137], so for $r \rightarrow \infty$ only the right-hand 
side of equation (75) is leading.

In the long run, we come at first glance across the paradoxical result that the current 
density in the magneto-sphere must vanish at large distances [138]:
\begin{equation}
j_{\parallel} = 0.
\label{asy1}
\end{equation}

In fact, this simply means that at large distances almost all outflowing longitudinal 
current must be concentrated near the rotational axis. Indeed, as shown in Ref. [139], 
a cylindrical region containing almost all outflowing current must inevitably appear 
near the axis.{\footnote{Strictly speaking, this terminology corresponds to the case 
of ${\bf \Omega}{\bf B} < 0$, where $\rho_{\rm GJ} > 0$. For the opposite orientation, 
the current near the rotational axis will flow towards the central engine.}}
This behavior was later verified by numerical simulations [120]. We shall consider this 
point in more detail in Section 4.1.3.

In the relativistic case, the electric force  $\rho_{\rm e}{\bf E}$ and the component of the 
Ampere force related to the longitudinal current ${\bf j}_{\parallel} \times {\bf B_{\varphi}}$ 
will mostly act on the outflowing plasma. However, as follows from equation (60), these forces 
almost mutually balance each other. Therefore, in addition to the bulk force{\footnote
{The corresponding formula (4.227) from monograph [9] has the incorrect sign}}
\begin{equation}
{\cal F}_{\rm jpol} \approx \rho_{\rm e}{\bf E} -
\nabla \left(\frac{B_{\varphi}^2}{8\pi}\right),
\label{clF2'}
\end{equation}
it is necessary to take into account the bulk centrifugal force 
\begin{equation}
{\cal F}_{\rm cent} \approx \frac{nmc^2\gamma+S/c}{R_{\rm c}}
\label{clF1}
\end{equation}
($S \approx c B_{\varphi}^2/4\pi$ is the Poynting vector) and the Ampere force 
${\cal F}_{\rm jtor} \approx {\bf j}_{\varphi} \times {\bf B_{\rm p}}/c$ pertaining 
to toroidal current. As a consequence, the Grad-Shafranov equation in the limit 
$r \gg r_{\rm F}$ can be conveniently rewritten in the form [136, 138, 140]
\begin{eqnarray}
\frac{B_{\varphi}^2 + 4 \pi n m_{\rm p} c^2\gamma}{R_{\rm c}}
+ \frac{1}{2} \, \hat{\bf n} \cdot \nabla ({\bf B}_{\rm p}^2) \nonumber \\
+ \frac{1}{2} \, \hat{\bf n} \cdot \nabla (B_{\varphi}^2 - {\bf E}^2) 
- \frac{B_{\varphi}^2 - {\bf E}^2}{\varpi} \, (\hat{\bf n} \cdot {\bf e}_{\varpi}) = 0, 
\label{balance}
\end{eqnarray}
where $\hat{\bf n} = \nabla \Psi/|\nabla \Psi|$. Notice that the Poynting vector contributes, 
in addition to the contribution from particles, to the centrifugal force. This is related to 
the fact that, as noted above, both particles and the electromagnetic energy propagate along 
the magnetic surfaces.

\vspace*{0.2cm}

{\bf 4.1.2 The collimation mechanism.}
Thus, the form of the magnetic surfaces close to the rotational axis depends on the balance 
between the collimating Ampere force arising from the longitudinal electric current (parallel 
currents are attracted) and the decollimating Ampere force related to toroidal currents. Thus, 
the question as to whether the collimation will be effective depends on the magnitude of the 
longitudinal current. A series of exact solutions [110, 111], which were possible to obtain 
by analyzing small deviations from the monopole magnetic field, showed that for nonrelativistic 
jets the collimation is large even close to the fast magnetosonic surface. This property is also 
confirmed by numerical modeling [120, 141] (see also Fig. 13).

In the relativistic case, where, as we remember, the longitudinal current is close to the 
Goldreich current ($i_0 \approx 1$), an almost full compensation of these two forces takes
place. In particular, the balance is met exactly in the force-free approximation and in the 
split monopole magnetic field [142], so the vacuum monopole solution remains exact up to 
infinity for the magnetosphere filled with the plasma as well (Fig. 14). The current $I$ here 
takes the form $I(\theta)=I_{\rm M}^{({\rm A})}\sin^2\theta$, where
\begin{equation}
I_{\rm M}^{({\rm A})} = \frac{\Omega_{\rm F}\Psi_0}{4\pi},
\label{IMA}
\end{equation}
which  exactly  coincides with the Goldreich  current \mbox{($j_{\parallel} = \rho_{\rm GJ} c$).}

\begin{figure}
\begin{center}
\includegraphics[width=0.65\columnwidth]{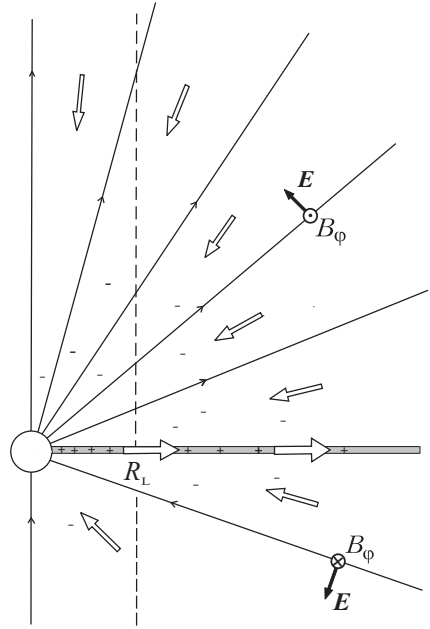}
\end{center}
\caption{The force-free monopole solution found by Michel [142], in which the electric 
field $E_{\theta}$ exactly equals the toroidal magnetic field $B_{\varphi}$. The contour 
arrows show the direction of the poloidal current.
   }
\label{fig3_02}
\end{figure}

For massive particles (and again for the split monopole magnetic field), the longitudinal 
current determined from the condition of smoothly crossing the fast magnetosonic surface 
will differ from the Goldreich current only by a factor of the order $\sigma^{-4/3}$ [143]. 
As a result, the perturbation of the magnetic flux function $\delta\Psi/\Psi$ in the 
asymptotically remote region $r \gg r_{\rm F}$ will increase logarithmically slowly [143,144]:
\begin{equation}
\frac{\delta\Psi}{\Psi} \sim \sigma^{-2/3}
\ln^{1/3}\left(\frac{r}{r_{\rm F}}\right).
\label{l119'}
\end{equation}
In other words, the current turns out to be only insignificantly larger than the critical one, 
which leads to a vanishingly small collimation. Correspondingly, the particle energy also 
increases very slowly:
\begin{equation}
\gamma \approx  \sigma^{1/3} \ln^{1/3}\left(\frac{r}{r_{\rm F}}\right).
\label{l119g}
\end{equation}

\begin{figure}
\begin{center}
\includegraphics[width=0.65\columnwidth]{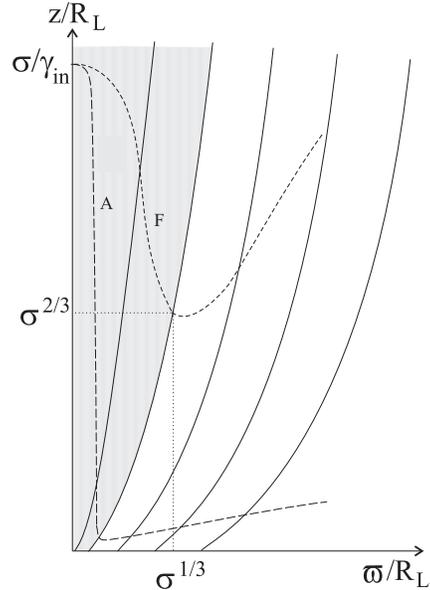}
\end{center}
\caption{The location of the Alfv\'enic (A) and fast magnetosonic (F) surfaces in a parabolic 
magnetic field. Such a field can also be generated in the accretion disc [145].
   }
\label{fig3_03}
\end{figure}

We stress that above we have considered the proper collimation, i.e., that due to bulk 
currents. However, the collimation, generally speaking, can be produced in the source itself. 
In Fig. 15, the flow obtained as a small perturbation of the force-free solution is shown, 
but for a parabolic field [145]. Such a field may be generated in the accretion disc as well 
[5]. For not too small ($\theta \gg \gamma_{\rm in}^2/\sigma$) and not too large 
($\theta \ll \sigma^{-1/3}$) angles, the location of the fast magnetosonic surface $r_{\rm F}$ 
and the value of $\gamma(r_{\rm F})$ are given by the expressions
\begin{eqnarray}
r_{\rm F} & \approx & \left(\frac{\sigma}{\theta}\right)^{1/2} R_{\rm L},
\label{r-f-par}  \\
\gamma(r_{\rm F}) & \approx & \sigma^{1/2}\theta^{1/2},
\end{eqnarray}
respectively, which again correspond to the estimates of the particle energy,  
$\gamma \sim \sigma^{1/3}$, and the axial distance, 
$r_{\rm F}\sin\theta \sim \sigma^{1/3}R_{\rm L}$ obtained above (see Ref. [145] 
for more detail). But in this case the account for nonzero particle masses also only 
slightly perturbs the force-free solution. On the fast magnetosonic surface, 
for example, one finds
\begin{equation}
\left(\frac{\delta\Psi}{\Psi}\right)_{r = r_{\rm F}} \approx \frac{1}{\sigma\theta} \ll 1.
\label{lena2}
\end{equation}

Thus, we can conclude that the proper collimation is impossible in the relativistic case. 
That is, the bulk collimation due to longitudinal currents flowing in the magnetosphere is 
impossible. Therefore, the magnetic surfaces can be collimated either by a special choice 
of currents in the source itself, or due to the interaction of the supersonic wind with the 
surrounding medium. Clearly, a sufficiently extended dense disc is required in the first case. 
Because of this, for radio pulsars where there is definitely no such disc, one usually assumes 
that the flow at distances $r \gg R_{\rm L}$ must be radial.

Unfortunately, it is not clear at present which mechanism is actually responsible for the 
collimation of relativistic jets. However, if one assumes that the collimation is indeed 
caused by the presence of the surrounding medium, it becomes possible to estimate the transverse 
size $r_{\rm jet}$ of the jets. Indeed, by assuming that the pressure of the poloidal magnetic 
field in the jet is close to the ambient pressure $P_{\rm ext}$ (this estimate is valid by 
assuming the total electric current in the jet to be zero), we obtain from the condition of 
the magnetic flux conservation that
\begin{equation}
r_{\rm jet} \sim R\left(\frac{B_{\rm in}^2}{8\pi P_{\rm ext}}\right)^{1/4},
\label{2}
\end{equation}
where $R$ and $B_{\rm in}$ are the radius and the magnetic field of the compact object, 
respectively. For example, for active galactic nuclei ($B_{\rm in} \sim 10^{4}$ G, 
\mbox{$R \sim 10^{13}$ cm)} we have
\begin{equation}
r_{\rm j} \sim 1 \, {\rm pc},
\label{3}
\end{equation}
which exactly corresponds to the observed transverse sizes of jets [12]. Correspondingly, for 
young stellar objects ($B_{\rm in} \sim 10^{3}$ G, $R \sim 10^{11}$ cm) we have 
$r_{\rm j} \sim 10^{16}$ cm, which again is in agreement with observations. Thus, 
the external medium apparently must significantly affect the collimation of jets.

\vspace*{0.2cm}

{\bf 4.1.3 The dense core.}
As we have seen, near the axis the bulk force ${\cal F}_{\rm jpol}$ related to the 
poloidal current is always directed toward the rotational axis, and in the region 
of the current closure, away from the axis. Then, for example, close to the cylindrical 
core where the curvature of the magnetic surfaces is small and, hence, the centrifugal 
force ${\cal F}_{\rm cent}$ (78) can be neglected, the equilibrium can be established 
only if the force ${\cal F}_{\rm jtor}$ is directed away from the rotational axis. But 
the poloidal magnetic field in this case must decrease with axial distance. Then, the density 
of the outflowing plasma will decrease along with the poloidal field as well. Exactly such a 
behavior of the solution was illustrated in Fig. 13.

It turned out that the dense core for the nonrelativistic supersonic flow will exist 
for both strongly and weakly magnetized flows close to the central engine, 
irrespective of the ambient pressure [146]. The radius of the core will correspond 
to such a distance from the axis, at which the toroidal field matches the poloidal one. 
A straightforward calculation shows that in both cases the longitudinal current in the 
core region will be $j = (c/v_{\rm in})j_{\rm GJ}$, i.e., $i_0 = c/v_{\rm in}$, and so 
one arrives at
\begin{equation}
r_{\rm core} = \frac{v_{\rm in}}{\Omega}.
\label{core}
\end{equation} 
The magnetic flux in such a core must be a significant fraction of the total magnetic flux. 
Such a configuration is none other than the $z$-pinch well-known in plasma physics [147] 
(the question of stability will be briefly discussed below).

As for the relativistic flow, the appearance of the dense core can be balanced here by the 
electric force $\rho_{\rm e}{\bf E}$ which, as we have seen, significantly weakens the 
force ${\cal F}_{\rm jpol}$. As a result, the answer is significantly dependent on the 
ambient pressure [9]. If the relativistic jet is surrounded by a medium with the total 
gas and magnetic pressure $P_{\rm ext}$ above some limiting value $P_{\rm min}$, the dense 
core does not form at all. In this case, the poloidal magnetic field in the jet will be 
approximately constant: $B_{\rm p}^2/8\pi \approx P_{\rm ext}$. For convenience sake, 
we shall express below the limiting value $P_{\rm ext}$ through the equivalent magnetic field 
($B_{\rm min}^2/8\pi = P_{\rm min}$). Here, one obtains the relationship
\begin{equation}
B_{\rm min} = \frac{1}{\sigma\gamma_{\rm in}}B(R_{\rm L}),
\label{38}
\end{equation} 
where $B(R_{\rm L}) = \Omega^2\Psi_{\rm tot}/\pi c^2$ is the characteristic magnetic field 
on the light cylinder. Correspondingly, the density of the outflowing plasma will be constant, 
too.

If the ambient pressure is sufficiently low, so that $P_{\rm ext} < P_{\rm min}$, then, as in the 
nonrelativistic case, the dense core is formed in the center of the outflow; the radius of 
the core must exceed that of the light cylinder:
\begin{equation}
r_{\rm core} = \gamma_{\rm in}R_{\rm L}.
\label{rcore}
\end{equation} 
It is easy to check that on the core boundary the energy flux density of the electromagnetic
field matches the particle energy density. The appearance of such a cylindrical jet was 
predicted already many years ago in many papers [138, 148, 149], but the magnetic flux in this 
core was determined only quite recently [145,146]. The magnetic field near the axis was found 
to only very weakly (logarithmically) depend on the ambient pressure, so with good accuracy 
one can set $B_{\rm core} = B_{\rm min}$. As a result, the magnetic flux 
$\Psi_{\rm core} = \pi r_{\rm core}^2B_{\rm min}$ within the core turns out 
to be much smaller than the total flux:
\begin{equation}
\frac{\Psi_{\rm core}}{\Psi_{\rm tot}} \approx \frac{\gamma_{\rm in}}{\sigma} \ll 1.
\label{39new}
\end{equation}
Here, the poloidal magnetic field and the density of matter for $x_r = \varpi/r_{\rm core} > 1$
must behave as power functions
\begin{equation}
B_{\rm p} \propto  x_r^{-k_1},\
\label{Bzpower}
\end{equation}
\begin{equation}
\rho^{({\rm lab})} \propto x_r^{-k_2}.
\label{npower}
\end{equation}
respectively. As the ambient pressure decreases, the exponents $k_1$ and $k_2$ 
gradually increase; however, their difference remains approximately constant, viz.
\begin{equation}
k_1 - k_2 \approx 0.
\end{equation}
As a result, if we have $k_1 \approx k_2 \approx 0$  for  $P_{\rm ext} > P_{\rm min}$
(i.e., the poloidal magnetic field and the outflowing plasma density are constant in the 
cross section), for ambient pressures corresponding to the magnetic field 
$B_{\rm ext} \equiv B_{\rm eq}$, where
\begin{equation}
B_{\rm eq} = \frac{1}{\sigma^2} \, B(R_{\rm L}),
\label{exep1}
\end{equation}
we have, in contrast, $k_1 \approx k_2 \approx 1$. Under such ambient pressures, as we 
shall see, the contribution from particles to the energy flux becomes dominant in the 
entire volume of the jet.

\begin{figure}
\begin{center}
\includegraphics[width=0.7\columnwidth]{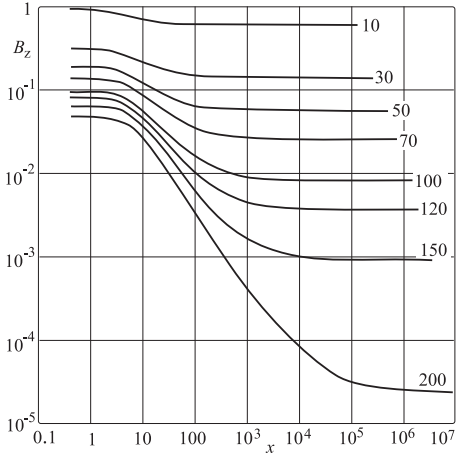}
\end{center}
\caption{Longitudinal magnetic field $B_z$ (G) as a function of the axial 
distance $x=\Omega \varpi/c$ for $\sigma=10^3$ and different Mach numbers 
${\cal M}$ on the jet axis, which was obtained as a solution of the one-dimensional 
problem [146]. The axial magnetic field only slightly deviates from $B_{\rm min}$ 
when the ambient pressure changes by several orders of magnitude.
   }
\label{fig3_04}
\end{figure}

\begin{figure}
\begin{center}
\includegraphics[width=0.8\columnwidth]{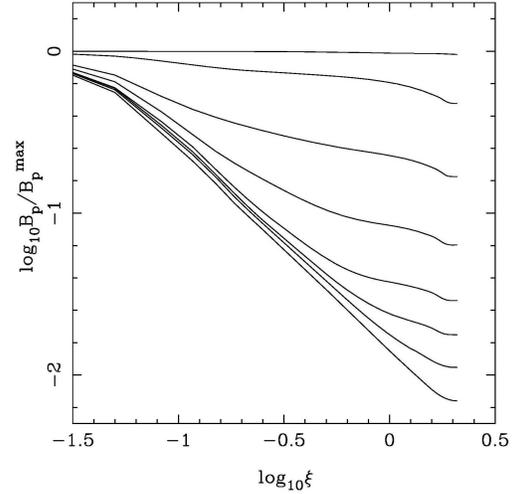}
\end{center}
\caption{Longitudinal magnetic field profile as a function of the axial 
distance $\xi \propto \varpi$ for different values of the ambient pressure, 
which was obtained as a solution of the two-dimensional time relaxation problem 
[126]. The power-law dependence corresponds to the analytical estimate (92).
   }
\label{fig3_05}
\end{figure}

In Fig. 16, the behavior of the poloidal magnetic field is shown in the double-logarithmic 
scale as a function of the distance to the axis [146] (see also Ref. [150]). It was 
obtained as the solution to the system of two ordinary differential equations ($A.14$) and 
($A.27$) to which, as shown in the Appendix, the Grad-Shafranov equation can be 
conveniently reduced in the cylindrical case. The Mach number on the rotational axis served 
as the parameter for different curves. It is evident that the poloidal magnetic field at 
distances exceeding $r_{\rm core}$ indeed starts decreasing as a power law. At the boundary 
of the jet it changes by several orders of magnitude, while the magnetic field on the axis 
changes only several-fold times. This behavior was recently confirmed by numerical two-dimensional 
calculations, too [126] (Fig. 17). It is important here that in the last case the very 
formulation of the problem was principally different (the time relaxation problem was solved 
rather than a stationary problem).

In relativistic jets there can be one more regime which is impossible in nonrelativistic 
jets [151]. In Fig. 15, it corresponds to cross sections located sufficiently close to the 
equator, when the flow in the jet center must be subsonic. By rewriting the condition 
$z < (\sigma/\gamma_{\rm in})R_{\rm L}$ in terms of an ambient pressure, we obtain the 
inequality $P_{\rm ext} > B_{\rm cr}^2/8\pi$, where
\begin{equation}
B_{\rm cr} = \frac{\gamma_{\rm in}}{\sigma}B(R_{\rm L}).
\label{55a}
\end{equation}
In this case, a subsonic flow region must inevitably be formed in the inner parts 
of the jet with $\varpi < r_{\rm s}$, where
\begin{equation}
r_{\rm s} \approx
\sigma\left[\frac{8 \pi P_{\rm ext}}{B(R_{\rm L})^2}\right]^{1/2}R_{\rm L}.
\label{56a}
\end{equation}
At last, the subsonic flow is established in the entire jet volume for higher 
ambient pressures corresponding to the magnetic field pressure on the fast 
magnetosonic surface.

\vspace*{0.2cm}

{\bf 4.1.4 The current closure.} As we have understood, according to the most 
wide-spread point of view, it is the electric current flowing along collimated 
outflows, which is responsible for the main jet energy release. Here, the following 
theorem can be formulated. 

{\it A stationary cylindrical jet with finite magnetic flux 
$\Psi_{\rm tot}$ can be formed either with the nonzero full electric current
$I(\Psi_{\rm tot}) \neq 0$ or in the presence of a surrounding medium with nonzero 
pressure.}

At first glance, these two variants fully contradict each other. However, this is 
not actually the case. As we have seen, a cylindrical core (which contains a significant 
part of the outflowing current) must be formed near the jet axis both in the relativistic 
and nonrelativistic cases. If the compact object were solitary in the Universe, the reverse 
current would indeed return near the equatorial plane. The closing of the current itself 
would occur in the switching-on wave. Thus, if one ignores the current closure region and 
studies only the structure of the central region, it is indeed possible to assume that the 
current closure takes place at infinity.

In fact, when a jet is immersed into a medium with a finite pressure, the reverse current, 
as one usually assumes, must flow along the boundary of the cocoon that forms due to the 
interaction of the supersonic wind with the external medium. The boundaries of the 
corresponding cocoons can be well seen in Figs 1 and 4. It is the force balance at the 
cocoon boundary that allowed the magnetic field $B_{\rm min}$ close to the jet axis to be 
determined [145].

\vspace*{0.2cm}

{\bf 4.1.5 The stability.} In conclusion, it is necessary to briefly consider the problem 
of the stability of jets. The nonrelativistic z-pinches observed in the laboratory are known 
to be strongly unstable with respect to constrictions and screw modes [147, 152,153]. Therefore, 
the problem of jet stability has been widely discussed in the literature [154-160]. As shown 
in Fig. 4, however, jets from young stars at large distances indeed are similar to a sequence 
of flying blobs rather than to a regular flow. Only at small distances can the flow be 
considered quite regular (see Fig. 5).

In recent years, special laboratory experiments have been carried out under conditions 
maximally similar to those in the nonrelativistic jets [161, 162]. In particular, the 
plasma velocity was as high as $100$--$400$ km s$^{-1}$ (which is comparable to the plasma 
flow velocity in jets from young stars), and the total current was about \mbox{$1$ MA.} In these 
experiments, a strong instability leading to rapid fragmentation of the flow into individual 
blobs was also observed.

Nevertheless, it is not obvious that the instability of laboratory pinches can be considered 
as an undisputable evidence for the immanent strong instability of astrophysical jets. The 
point is that astrophysical jets are always 'specially prepared', since they come out from 
the quasispherical subsonic flow region. As a result, the plasma density and the longitudinal 
magnetic field profiles near the jet base turn out to be close to the equilibrium ones. In 
laboratory experiments, in contrast, the initial plasma density is usually very different 
from the equilibrium value [153].

As for relativistic jets (which at large distances also frequently show an irregular 
structure), they proved to be more stable [163 165]. The recent numerical simulations 
[166] (where the jet was found to be stable after more than 1000 rotations of the central 
engine) confirmed this conclusion. Thus, there are no doubts now that the nonrelativistic 
Kruskal-Shafranov stability criterion [153]
\begin{equation}
\frac{r_{\rm jet}}{L} \, \frac{B_{\rm p}}{B_{\varphi}} > 1,
\label{ksh}
\end{equation}
where $L$ is the length of the jet, cannot be applied to relativistic flows. 
Unfortunately, the limits of the present review do not allow us to discuss this most 
important point in more detail.

\subsection{Acceleration}

{\bf 4.2.1 The acceleration mechanism.} First of all, let us consider
the acceleration mechanism itself. It is convenient to start
from expression (43) for the toroidal velocity of the non-
relativistic flow. It is evident that in the subsonic region
${\cal M}^2 < 1$  the flow velocity corresponds to the precise corotation:
\begin{equation}
v_{\varphi} \approx \Omega_{\rm F} \varpi.
\label{uphsmnon}
\end{equation}
In other words, particles can be considered as beads on a wire that determines their 
angular velocity of rotation. This situation is quite understandable because, as can 
be easily checked, the energy density of the magnetic field within the Alfv\'enic surface 
exceeds the plasma energy density, so it is the magnetic field that controls the motion 
of particles. In fact, the magnetic field plays the role of a slingshot that provides the 
constant angular velocity of the plasma rotation. Therefore, the velocity of particles 
linearly increases with increasing axial distance. Then, the following theorem can be 
formulated.

{\it In the nonrelativistic limit, the smooth crossing of the fast magnetosonic surface 
is possible only if the particle energy flux on this surface is at least one-third of the 
total energy losses. In other words, the transonic nonrelativistic flow in the supersonic 
region must already be effectively accelerated ($q \sim 1$).}

Indeed, we can set $E_{\rm n} \approx \Omega_{\rm F}I/2\pi c \eta_{\rm n}$ for strongly 
magnetized flows near the surface of the central engine, which immediately yields
$E_{\rm n} \approx \Omega_{\rm F}^2r_{\rm F}^2$ for $I \approx i_0 I_{\rm GJ}$. Moreover, 
it is easy to show that the poloidal velocity $v_{\rm p}$  near the singular surfaces also 
becomes on the order of $\Omega_{\rm F}r_{\rm F}$, too.

By this means the particle acceleration mechanism in the strongly magnetized wind is similar 
to that by which a slingshot accelerates a stone. This becomes possible exactly due to the 
dominant effect of the poloidal magnetic field. However, this acceleration stops at distances 
$r_{\rm F} \sim r_{\rm A}$, since there the magnetic field energy density becomes smaller than the 
plasma particle energy density. Additionally, the toroidal magnetic field beyond the 
Alfv\'enic surface becomes stronger than the poloidal one, so particles start sliding relative 
to the magnetic field lines. As a result, the flow passes to another asymptotic behaviour: 
$v_{\varphi} \approx \Omega_{\rm F}\varpi_{\rm A}^2/\varpi$.. The plasma kinetic energy in 
this region will be mainly due to their poloidal velocity component.

As for the particle acceleration in the ultrarelativistic limit, near the fast magnetosonic 
surface, as we have seen, the particle energy flux must be much smaller than the electromagnetic 
energy flux. So the question arises as to whether it is possible to effectively accelerate 
particles beyond the sonic surface. Surprisingly, the force balance across the flow should be 
considered once again to answer this question.

\vspace*{0.2cm}

{\bf 4.2.2 Efficiency. } So, let us come back to equation (79). We have already mentioned 
that the particle energy will be completely determined by the drift motion beyond the light 
cylinder $\varpi \gg R_{\rm L}$. Therefore, using Eqn (59) we can write out the 
expression for the Lorentz factor of particles in the form
\begin{equation}
\frac{1}{\gamma^2} \approx \frac{{\bf B}_{\rm p}^2}{B^2} 
+ \frac{B_{\varphi}^2 - {\bf E}^2}{B^2}. 
\label{balance1}
\end{equation}
Now, making use of the relation $B_{\varphi} \approx |{\bf E}| \approx x_r B_{\rm p}$, where 
again $x_r = \Omega \varpi/c$, we immediately come to the conclusion that when the second term 
in expression (100) can be neglected, the particle Lorentz factor must approach the following 
asymptotic solution:
\begin{equation}
\gamma = x_r,
\label{gamma!}
\end{equation} 
which, as we shall see, is universal enough. If the curvature of the magnetic field lines 
is appreciable, then, in contrast, we can neglect in formula (79) the second term corresponding 
to the bulk force $j_{\varphi} B_{\rm p}/c$. As a result, by comparing the corresponding terms 
in the force balance equation (79) for the strongly magnetized (i.e., Poynting dominated) flow, 
we arrive at another asymptotic solution [167]:
\begin{equation}
\gamma \approx {\cal C} \sqrt{\frac{R_{\rm c}}{\varpi}}.
\label{g2rc}
\end{equation}
where ${\cal C} \sim 1$. Moreover, making use of Eqn (59) and the equilibrium condition 
(79), we can write in the general case the relationship [109]
\begin{equation}
\frac{1}{\gamma^2} \approx \frac{1}{x_r^2} + \frac{\varpi}{{\cal C}^2 R_{\rm c}}.
\label{gammahar}
\end{equation}
Here, the value of ${\cal C}$ can be exactly determined for strongly collimated flows 
and $\Omega_{\rm F} =$ const [109,150]:
\begin{equation}
{\cal C} = \sqrt{3}.
\label{gammahar'}
\end{equation}
Simply speaking, the Lorentz factor will be determined by the least of the two values 
giving by expressions (101) and (102). 

\begin{figure}
\begin{center}
\includegraphics[width=0.8\columnwidth]{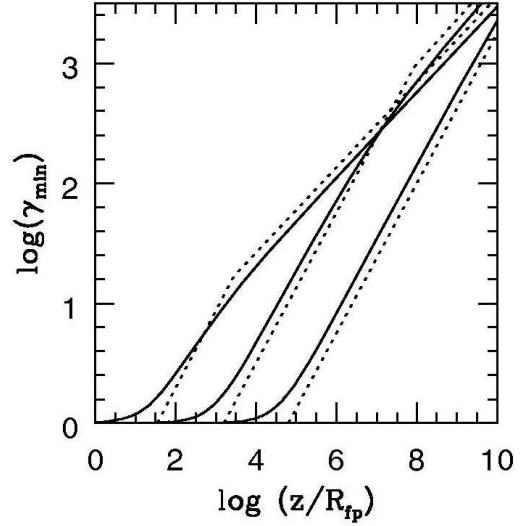}
\end{center}
\caption{The growth of a particle's Lorentz factor $\gamma$ with distance from the equatorial 
plane $z$ [168]. The right curve corresponds to strong collimation ($k > 2$). For weaker 
collimation (left curves), the particle acceleration at large distances becomes less effective.
   }
\label{fig3_06}
\end{figure}

Thus, the choice between asymptotic solutions (101) and (102) must be determined by how bent the 
magnetic surfaces are. It is easy to show that the parabolic magnetic field in which the 
field line at a large distance from the central source is given by the equation 
$z(\varpi) \propto \varpi^2$ corresponds to the terminating case [126,146]. Indeed, as 
the curvature radius can be defined as  $R_{\rm c} = [(z')^2+1]^{3/2}/z''$, in the limit 
$z' = {\rm d}z/{\rm d}\varpi \gg 1$ for the magnetic surfaces specified by the relationship 
$z(\varpi) \propto \varpi^k$, Eqn (102) gives the energy of particles moving along the 
magnetic field line:
\begin{equation}
\gamma \propto \varpi^{k-1},   
\label{asym}
\end{equation} 
where $\varpi$ is, in this case, the current axial distance of a particle. At $k = 2$,
the acceleration efficiency determined from expressions (101) and (102) is the same. Thus, 
if the magnetic surfaces are collimated more strongly than those for the parabolic field 
(i.e., $k > 2$), the curvature of the magnetic surfaces can be neglected at large distances, 
and the energy of the particles will be determined by expression (101). If the flow is poorly 
collimated (i.e., $1 < k <  2$), the particle acceleration will be less effective and one 
should use expression (102). The numerical simulations [124, 168] fully confirm the picture 
presented here. As shown in Fig. 18, the acceleration of a particle moving along the magnetic 
field line indeed follows the law $\gamma(z) \propto z^{1/k}$ for strongly collimated flows, 
in correspondence with asymptotic behaviour (101), while the acceleration for poorly collimated 
outflows at large distances, in full agreement with expression (105), becomes less effective. 

We thus arrive at the most important conclusion that the acceleration efficiency of particles 
in a supersonic ultrarelativistic wind is determined by the degree of collimation of the magnetic 
surfaces. The plasma can be effectively accelerated only in the case where the magnetic 
surfaces are collimated more strongly than those for the parabolic field. In this event, one 
can use asymptotic solution (101) which shows that the acceleration mechanism, in fact, again is due 
to the slingshot effect (the larger the axial distance, the higher the energy). But in this 
case, too, the total transformation of the electromagnetic energy into the particle flux 
energy, $\gamma \approx \sigma$, can take place only if the jet transverse size exceeds the 
value
\begin{equation}
r_{\rm eff} =  \sigma R_{\rm L}.   
\label{rsigma}
\end{equation} 
In particular, the fraction of energy carried by particles for $P_{\rm ext} < B_{\rm eq}^2/8\pi$
can be determined from the simple relationship [10,151]
\begin{equation}
\frac{W_{\rm part}}{W_{\rm tot}} \sim
\frac{1}{\sigma}\left[\frac{B^2(R_{\rm L})}{8 \pi P_{\rm ext}}\right]^{1/4}.
\label{46b}
\end{equation}
If the collimation is poor, the acceleration will be ineffective, since particles begin 
sliding from the magnetic field lines. In this case, a much higher transverse size of the jet 
is required for accelerating particles to limiting energies:
\begin{equation}
r_{\rm eff} =  \sigma^{1/(k-1)} R_{\rm L}.   
\label{rsigma'}
\end{equation}
This dependence was also numerically confirmed [169]. In particular, the acceleration becomes 
actually impossible for the split monopole magnetic field $k = 1$ (which, as we see, is a 
special case that requires separate consideration), since, as follows from estimate (82), 
the condition $\gamma \approx \sigma$ is reached only at exponentially large distances from 
the compact object. On the other hand, it easy to check that for $k > 1$ both for effective 
and for ineffective acceleration mechanisms the condition $\gamma \vartheta \sim 1$ will be 
satisfied, where $\vartheta \approx \varpi/z$ is the characteristic opening angle of the 
acceleration region.

Here, it is important to emphasize that the self-consistent analysis, in which the magnetic 
surfaces for the split monopole field were not assumed to be exactly conical, allowed the 
determination of the correct structure of the flow [143]. For example, the expression (82) 
for the particle's Lorentz factor discussed above is determined precisely by the small 
curvature of the magnetic surfaces, and so exactly corresponds to expression (102). It should 
be remembered that for a long time the flow in the split monopole magnetic field had been, 
in fact, the only example in which analytical results could be obtained. Therefore, a 
strong opinion was that the effective particle acceleration beyond the fast magnetosonic 
surface is completely impossible. As wee see, this conclusion proved to be incorrect in the 
general case.

\subsection{Subsonic flows}

As discussed above, all observed jets must be transonic. It is this property that 
allowed us to find the longitudinal current flowing in the magnetosphere and, hence, 
to estimate the central engine power. However, until recently many models considered 
subsonic flows, and the magnitude of the current was determined from other considerations 
[170, 171]. We shall briefly discuss below two the most known examples of such subsonic flows.

\vspace*{0.2cm}

{\bf 4.3.1 Pulsar magnetospheres.} Starting in the early 1970s, pulsar magnetospheres have 
been discussed mostly in the force-free approximation [107, 142, 172-174]. This was based 
on the fact that the plasma filling the neutron star magnetosphere is secondary with respect 
to the magnetic field, and so (at least inside the light cylinder) the particle energy density 
can be neglected. The Grad-Shafranov equation ($A.28$) in the force-free approximation (which 
in this case is simply called the pulsar equation) becomes elliptical. Therefore, for numerical 
modeling of the axisymmetric magnetospheres one has to impose an additional condition at the 
external boundary of the integration region [98, 100 104]. Usually, one chooses the condition 
of radiality of the magnetic field lines (Fig. 19). In this case it is this additional condition 
that fixes the longitudinal current in the magnetosphere. Thus, it is not surprising that this 
current turns out to be close to the longitudinal current $I_{\rm M}(\theta)$  (80) obtained by 
F C Michel for the monopole solution shown in Fig. 14. If the absence of the reverse current 
flowing along the equator (so that the current closing was done by bulk currents only) was 
chosen as the additional condition, the magnetic field structure beyond the light cylinder 
turned out to be significantly different [175].

\begin{figure}
\begin{center}
\includegraphics[width=0.8\columnwidth]{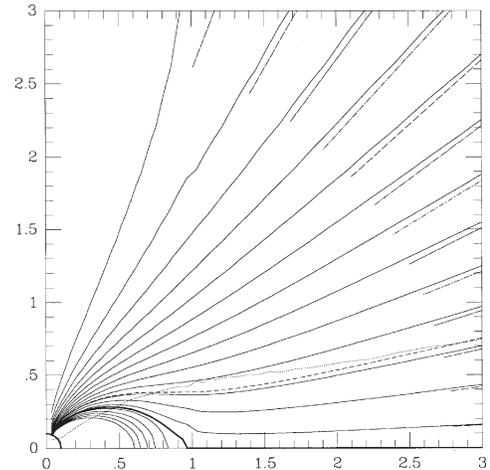}
\end{center}
\caption{The structure of the axisymmetric radio-pulsar magnetosphere in the model [98]. 
The flow is assumed to be radial at large distances.
   }
\label{fig3_07}
\end{figure}

\begin{figure}
\begin{center}
\includegraphics[width=0.8\columnwidth]{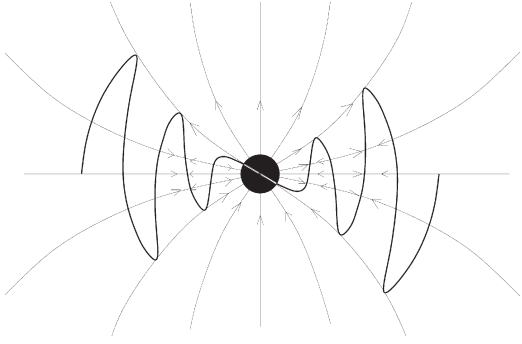}
\end{center}
\caption{The magnetic field structure in the 'rotating split monopole' 
model [176]. In the force-free case, particles moving radially at the speed 
of light can provide the formation of the current sheet (the wavy curve) which 
separates magnetic fluxes in the equatorial region.
   }
\label{fig3_08}
\end{figure}

Interestingly, a similar structure also appears in different models of the inclined rotator. 
Here, first of all, we should highlight the model of the 'rotating split monopole' (Fig. 20). 
In the force-free approximation (when massless particles move with the speed of light), the 
monopole magnetic field, namely
\begin{eqnarray}
&&\Psi(r,\theta, \varphi, t)  =  \Psi_{0}(1-\cos\theta), \nonumber \\
&& \hspace*{1cm} \theta < \pi/2 - \chi \cos(\varphi - \Omega t + \Omega r/c), \\
&&\Psi(r,\theta, \varphi, t)  =  \Psi_{0}(1+\cos\theta), \nonumber \\
&& \hspace*{1cm} \theta > \pi/2 - \chi \cos(\varphi - \Omega t + \Omega r/c), 
\label{e38''}
\end{eqnarray}
was found to be also the solution of the problem [176]. In this case, the electromagnetic 
fields inside cones with angles \mbox{$\theta < \pi/2 - \chi$}  and 
\mbox{$\pi - \theta < \pi/2 - \chi$} near the rotational axis are independent of time and 
the angle $\varphi$ and coincide with the fields of an axisymmetric rotator, while 
in the equatorial plane the field directions reverse sign at the moment when 
the current sheet intersects the given point. A similar structure beyond the light cylinder 
was obtained for the rotating dipole, too [177]. True enough, in contrast to the 'rotating 
magnetic monopole', the total longitudinal current here depends on the inclination angle 
$\chi$, although not very strongly.

The last property can be easily explained. As we have seen, the very existence of the MHD 
wind far away from the light cylinder is possible only if the toroidal magnetic field is 
comparable in magnitude to the electric field. But this is possible only if a sufficiently 
large longitudinal current $I \approx I_{\rm M}^{({\rm A})}$ flows in the magnetosphere. 
Let us keep in mind that this value of the current can also be found from the 
condition of the smooth crossing of the singular surfaces in the full MHD version. In none 
of the numerical calculations mentioned above were bounds on the longitudinal currents 
flowing from the neutron star surface set. Thus, it is not surprising that the longitudinal 
current obtained from the solution of the problem considered turned out to be on the order 
of $I_{\rm M}(\theta)$. As a result, the energy losses in model [176] were found to be 
independent of the angle $\chi$. Energy losses in model [177] were also found to be weakly
dependent on the inclination angle:
\begin{equation}
W_{\rm tot}  \approx \frac{1}{4} \, \frac{B_0^2\Omega^4R^6}{c^3}\left(1 + \sin^2\chi\right).
\label{Wspit}
\end{equation} 

However, the following problem emerges here. All the theories of particle generation near 
the neutron star magnetic poles [178-180] unambiguously suggests that the longitudinal 
current density cannot be higher than that of the local Goldreich current which, as follows 
from definition (9), depends on the inclination angle $\chi$
\begin{equation}
j_{\rm GJ} \approx \frac{ \Omega B}{2\pi}\cos\chi.
\label{GJforj}
\end{equation}
Indeed, the local Goldreich charge density near the magnetic poles for the orthogonal 
rotator must be $(\Omega R/c)^{1/2}$ times smaller than in the axisymmetric magnetosphere. 
Thus, the longitudinal current flowing along open magnetic field lines should be correspondingly 
smaller (for ordinary pulsars this factor can be as high as $10^{2}$). In the 'rotating split
monopole' model considered above, this problem does not arise, because at any inclination angle 
in the polar magnetospheric regions the current is always the same as in the axisymmetric case. Just 
this current provided the necessary toroidal magnetic field. For the inclined dipole, in contrast, 
it is necessary to additionally assume that the longitudinal current in the polar cap regions can 
be significantly higher than the local Goldreich current (A Spitkovsky, private communication).

There is one more problem related to the decrease in the longitudinal current density as 
$\chi \rightarrow 90^{\circ}$ . The point is that the current losses for the local Goldreich 
current must decrease as the inclination angle $\chi$ increases [56,107]:
\begin{equation}
W_{\rm tot} = \frac{f_*^2}{4} \, \frac{B_0^2\Omega^{4}R^6}{c^3}i_0\cos\chi.
\label{e15}
\end{equation}
Here, the coefficient $f_* = 1.59$--$1.96$ depends only on the inclination angle $\chi$.
It is necessary to stress that in addition to the $\cos \chi$ factor (which is related to the 
scalar product $W_{\rm tot} = -{\bf \Omega} \cdot {\bf K}$, where ${\bf K}$  is the braking 
torque), the significant dependence of the current losses $W_{\rm tot}$ on the inclination 
angle is also contained in the quantity $i_0$. The matter is that in the definition of the 
dimensionless current $i_0 = I/I_{\rm GJ}$ in expression (113), the denominator contains the 
Goldreich current for the axisymmetric case, whereas at nonzero angles  $\chi$ the 
Goldreich charge density itself depends on the angle  $\chi$ near the magnetic poles. It is 
logical to expect that the dimensionless current $i_0$ for the inclined rotator will be bounded 
from above as $i_0^{(\rm max)}(\chi) \sim \cos\chi$. As a result, the current losses, in 
contrast to relationship (111), must decrease with angle $\chi$ at least as $\cos^2\chi$.
In particular, for $\chi = 90^{\circ}$ (when $\cos^2\chi$ must be replaced by its characteristic 
value within the polar cap region $\cos^2\chi$, we obtain [56]
\begin{equation}
W_{\rm tot} = c_{\perp}\frac{B_0^2\Omega^{4}R^6}{c^3}
\left(\frac{\Omega R}{c}\right) i_{\rm A}.
\label{e17'}
\end{equation}
Here, $i_{\rm A} = 1$  for the local Goldreich current, and the coefficient $c_{\perp} \sim 1$
now depends not only on the profile of the asymmetric longitudinal current, but also on the 
polar cap shape.

Usually, when discussing this point, the following counter-argument against the decrease 
in current losses with increasing $\chi$ is invoked. In the expression for the braking
torque, namely 
\begin{equation}
{\bf K} =
\frac{1}{c}\int[{\bf r}\times[{\bf J}_{\rm s}\times{\bf B}]]{\rm d}S,
\label{e3}
\end{equation}
The surface current ${\bf J}_{\rm s}$ indeed must decrease as  $\cos \chi$ as angle $\chi$ 
increases. But then the characteristic distance from the axis to the polar cap, in contrast, 
will increase as $\sin \chi$, so ultimately the losses will be weakly dependent on the
inclination angle $\chi$ even for the local Goldreich current.

\begin{figure}
\begin{center}
\includegraphics[width=0.7\columnwidth]{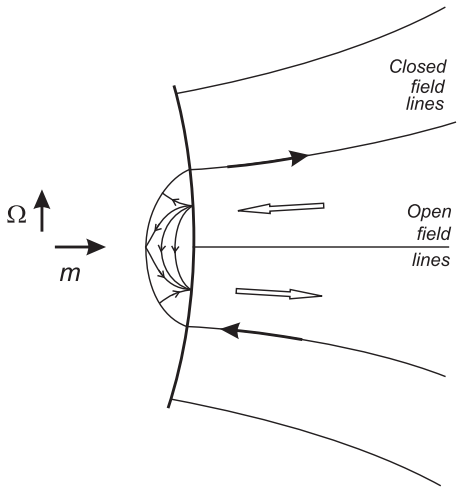}
\end{center}
\caption{The structure of electric currents flowing near the magnetic poles of an 
orthogonal rotator. Currents flowing along the sepratrix (thin arrows) that separates 
the regions of the open and closed magnetic field lines are adjusted to bulk currents 
(contour arrows) in such a way that the closing surface current is fully concentrated 
within the polar cap.
   }
\label{fig3_09}
\end{figure}

As follows from a more precise analysis [107], however, the real structure of the surface 
currents in the polar cap region was ignored in this obvious, at first glance, consideration. 
Referring to Fig. 21, the closing currents in fact must be such that the current averaged over 
the polar cap region is zero. As a result, one needs to consider higher-order effects (like, 
for example, the neutron star surface curvature effect), when determining the radio pulsar 
spin-down rate. But if the surface current averaged over the polar cap region is indeed zero, 
then, as shown in Fig. 21, a surface current comparable in amplitude to the bulk current in 
the magnetosphere must flow along the separatrix separating the regions of open and closed 
magnetic field lines. Remarkably, in numerical modeling of the inclined rotator [181], 
reverse currents flowing along the separatrix were indeed discovered. Finally, it worth 
noting that there is no contradiction between relations (111) and (113), either. As we 
specially emphasized, the approximation expression (111) was obtained in Ref. [177] for a 
flow in which the longitudinal current was larger than the local Goldreich current, which 
corresponds to the condition $i_0 > 1$ ($i_{\rm A} > 1$, respectively).

In any case, studies of the last decade, in our opinion, have at last formulated the problem 
whose solution will provide significant progress in the understanding of the structure of 
radio pulsar magnetospheres. The problem is whether the plasma generation region in the neutron 
star magnetosphere can provide a sufficiently large longitudinal current which 
is necessary to launch the MHD wind from the inclined rotator. If the necessary current can be 
produced, nothing will prohibit the formation of the MHD wind in which the main part of the 
energy will be carried by the electromagnetic field. If the generation of a current which is
significantly larger than the local Goldreich current is impossible, then the toroidal magnetic 
field near the light cylinder will be smaller than the poloidal magnetic field. In that case, a 
light surface will inevitably be formed near the light cylinder, on which the current closing 
and particle acceleration up to energies $\gamma \sim \sigma$ will occur [56]. Thus, the problem 
of the effective particle acceleration in pulsar winds, which we mentioned in Section 2.4, can be 
solved.

Interestingly, the possibility of answering this question has apparently emerged a short 
time ago. This test is related to the unusual properties of the radio pulsar
PSR $B$1931$+$24 [182]. It differs from other pulsars in that it stays in the active 
state for $5$--$10$ days, and then its radio emission switches off in less than 10 s, 
and the source is not observed during the next $25$--$35$ days. It is important that 
the spin-down rate in these two states be different:
\begin{eqnarray}
\dot \Omega_{\rm on} & = & -1.02 \times 10^{-14} \, {\rm s}^{-2}, \\
\dot \Omega_{\rm off} & = & -0.68 \times 10^{-14} \, {\rm s}^{-2},
\end{eqnarray} 
so that
\begin{equation}
\frac{\dot \Omega_{\rm on}}{\dot \Omega_{\rm off}} \approx 1.5.
\end{equation}
A similar behavior was later observed in the pulsar PSR $J$1832$+$0031 
($t_{\rm on} \sim 300$ days, $t_{\rm off} \sim 700$ days), with the ratio 
$\dot \Omega_{\rm on}/\dot \Omega_{\rm off} \approx 1.5$ again.

It is logical to assume that the difference in the spin-down rates in these pulsars 
is simply associated with the fact that the spin-down in the switch-on state is due 
to the current losses, and in the switch-off state, when the magnetosphere is not 
filled with plasma, is due to magnetodipole radiation [183, 184]. Then, making use 
of equations (3) and (113) we obtain
\begin{equation}
\frac{\dot\Omega_{\rm on}}{\dot\Omega_{\rm off}} 
= \frac{3f_{\ast}^2}{2} \, {\rm cot}^2\chi,
\end{equation}
which yields the reasonable inclination angle $\chi \approx 60^{\circ}$. On the other hand, 
using expression (111) [177] for the switch-on state, we arrive at the relationship
\begin{equation}
\frac{\dot\Omega_{\rm on}}{\dot\Omega_{\rm off}} = \frac{3}{2} \,
\frac{(1+ \sin^2\chi)}{\sin^2\chi}.
\end{equation}
Clearly, this quantity cannot be equal to 1.5 for any $\chi$. If this interpretation of observations 
holds true, this implies that the longitudinal current in the magnetosphere indeed does not exceed 
the local Goldreich current.

It should be emphasized that we have assumed above that the magnetospheric plasma 
fully screens the magnetodipole radiation of the neutron star. This conclusion, 
which was formulated for the first time in paper [107], seems now to be directly 
confirmed, since both in the model [176] and in the numerical calculations [107] 
there are no alternating electromagnetic fields decaying as $1/r$.

\vspace*{0.2cm}

{\bf 4.3.2 'The magnetic tower'.} 'The magnetic tower' is the model of collimated jets 
proposed by D Lynden-Bell in 
1996 [128] (Fig. 22) and later on widely discussed in relation to both relativistic 
and nonrelativistic sources [171, 185-187]. It is based on the assumption that there 
exists an intensive wind outflowing perpendicular to the accretion disc and stretching 
the magnetic field lines along the rotational axis. Here, one usually assumes that 
the initial magnetic field was quasi-dipole, i.e., it consisted of magnetic field 
lines with one end frozen in the central star (or in the inner regions of the accretion 
disc) and another end frozen in the outer parts of the disc. Such a cylindric flow cannot 
intersect singular surfaces since the magnetic field lines remain at a constant distance 
from the axis of rotation. That is why, the longitudinal current in this model will be 
determined exactly by the differential rotation that leads to the magnetic field line 
twisting. Because of this, as in the case shown in Fig. 11, the magnetic tower top will 
propagate upwards by gradually increasing the volume occupied by the twisted magnetic 
field, while a stationary configuration restricted by the ambient pressure will be formed 
at smaller distances.

\begin{figure}
\begin{center}
\includegraphics[width=0.55\columnwidth]{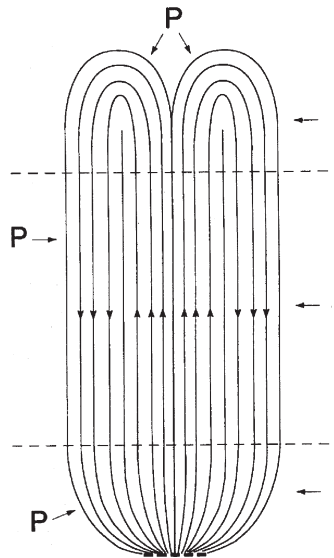}
\end{center}
\caption{The magnetic field structure in the model of the 'magnetic tower' that can 
exist under sufficiently strong ambient pressure $P$ [128]. The energy is transferred 
from the central engine along the rotational axis, and back to the accretion disc along 
the jet periphery.
   }
\label{fig3_10}
\end{figure}

There are two additional important properties that distinguish the magnetic tower model 
from the transonic flows we are considering. First, if the magnetic field lines are anchored 
in the accretion disc, a configuration with almost zero total magnetic flux will form during 
the outflow (see Fig. 22). In other words, the direction of the poloidal magnetic field on the 
periphery of the jet will be different from that near the jet axis. Second, when the magnetic 
field loops do not extend beyond the light cylinder and, hence, do not open, the energy will 
continue being transferred along the magnetic fields lines. But this means that the energy 
will be carried away from the central engine only along the rotational axis, while on the 
jet periphery the energy flux will be directed back to the accretion disc [186,187].
 
If such a configuration were stationary, the reverse energy flux would be exactly the same 
as the energy flux outgoing from the central engine closely to the rotational axis. However, 
we have seen that the equipotenital condition is violated in the switching-on wave, so the 
reverse energy flux turns to be smaller than the escaping flux.

Thus, in the very setup the problem of the magnetic tower formation is different from 
that describing transonic flows. The longitudinal current determining the energy losses 
in no way relates to the critical conditions on the singular surfaces which, as we specially 
stressed above, should unavoidably appear in all real compact sources (see also Section 2). 
Hence, apparently, the magnetic tower model cannot correspond to the reality. The results 
of numerical simulations also support this conclusion. When the flow intersected the singular 
surfaces, the magnetic field lines became open and the energy flux was directed outward from 
the central engine in the regions of both outgoing and incoming magnetic field lines [188]. 
On the other hand, in the case where the flow remained subsonic in numerical simulations, the
magnetic tower formation was indeed observed [50,189].

\section{Estimation of parameters}

Thus, we have seen that at present it has turned out to be possible to understand many 
key points related to the formation and the internal structure of collimated outflows. 
In doing so, we have managed to find several key parameters that determine the basic 
physical properties of ejected matter. First and foremost, these include the magnetization 
parameter $\sigma$, the multiplicity parameter $\lambda$, and the initial velocity 
$v_{\rm in}$ (the Lorentz factor $\gamma_{\rm in}$) of the outflow. We shall try below 
to understand how precisely these parameters can be estimated for the observed jets.

It should be emphasized from the very beginning that each time it is necessary to clearly 
separate which collimated outflows are being considered. Indeed, the properties of jets 
from active galactic nuclei can be quite different on the scale of the host galaxy 
(see Fig. 1) and in the formation region (see Fig. 2), where their transverse size is 
close to 1 pc. However, as we mentioned above, the relativistic jets can be stable on 
all scales. On the other hand, in the majority of cases nonrelativistic jets at large 
distances from the star can indeed be unstable (see Fig. 4), so for them the treatment 
considered here, strictly speaking, can be applied only in the innermost parts.

\subsection{Active galactic nuclei}

Despite longstanding efforts, we know very little about the internal structure of jets 
from active galactic nuclei. In particular, we still do not have an answer to the key 
question of whether it is the black hole, and not the inner parts of the accretion disc, 
that is the central engine which is responsible for the black hole power [190,191]. 
In the practical sense, the main uncertainty appears in the determination of the particle
production multiplicity $\lambda$. Indeed, as already noted, the plasma on the magnetic 
field lines threading the black hole horizon (which is needed both to screen the 
longitudinal electric field and to produce the longitudinal electric current) must 
be generated in the magnetosphere itself between two families of singular surfaces. 
Some fraction of the plasma will escape the magnetosphere, while another part will
accrete onto the black hole. Correspondingly, it is still unclear which quantities 
determine the density of matter flowing out of the accretion disc surface (see, for 
example, Refs [192, 193]). Moreover, it cannot be ruled out that the jet at large 
distances from the central engine will be additionally 'loaded' due to interaction 
with stellar winds from surrounding stars [194], or, for example, due to the 'photon 
breeding' effect (creation of the secondary electron-positron plasma by hard gamma-ray 
quanta generated in the interaction of the relativistic outflow with the ambient medium) 
[195,196]. That is why, the properties of jets on the scales of several kiloparsecs can 
be significantly different from those in the jet formation region.

In the region of field lines threading the black hole horizon several plasma generation 
mechanisms are currently being discussed, in which, however, plasma is ultimately generated 
always due to two-photon pair creation. The one-photon conversion, which plays the leading 
role in radio pulsar magnetospheres, here turns out to be ineffective, since the probability 
of pair creation in magnetic fields $B \sim B_{\rm Edd} \sim 10^4$ G is vanishingly small.

First and foremost, secondary plasma generation can be related to the direct two-photon 
process $\gamma + \gamma \rightarrow e^{+} + e^{-}$ (see, for example, Ref. [105]), where the necessary 
gamma-quanta are emitted from the inner regions of an accretion disc. This mechanism, with 
a high value of the parameter $\lambda_{\rm AGN_1} \sim 10^{10}$--$10^{12}$, was discussed 
in the pioneering paper by Blandford and Znajek [33]. However, sufficiently high temperatures 
providing the necessary number of hard gamma-ray photons with energies above the pair creation 
threshold ${\cal E}_{\rm min} = m_{\rm e}c^2$, and small free path lengths of photons are required 
for this mechanism to be effective. Presently, the accuracy of the compactness parameter estimate 
$l_{a,{\rm AGN}} \sim 1$--$100$ does not allow one to make definitive conclusions on the 
efficiency of this mechanism of particle creation.

On the other hand, such a high particle density must be typical for a wind outflowing from 
the accretion disc surface. Let us keep in mind that even if the energy release related to 
such a wind is insignificant, it can play the decisive role in the matter outflow collimation 
[197]. Here, the energy of the jet core observed at high radio frequencies and in gamma-rays 
will be associated with ultrarelativistic particles extracting energy from the rotating black 
hole.

There is another mechanism capable of bringing particles into the region of magnetic field 
lines threading the black hole horizon even in the absence of hard gamma-ray quanta. This 
mechanism is similar to the particle creation process in the outer gap of the pulsar 
magnetosphere [198]. Indeed, the exact relativistic expression for the Goldreich charge density 
$\rho_{\rm GJ}$  takes the form [9]
\begin{equation}
\rho_{\rm GJ} = -\frac{1}{8\pi^2}\nabla_k
\left(\frac{\Omega_{\rm F}-\omega}{\alpha}\nabla^k\Psi\right).
\label{k51}
\end{equation}
In particular, near the rotational axis we simply have
\begin{equation}
\rho_{\rm GJ} \approx  -\frac{(\Omega_{\rm F} - \omega)B}{2\pi \alpha}.
\label{k52}
\end{equation}
As a result, the general relativity effects cause the Goldreich density to 
vanish at $\omega \approx \Omega_{\rm F}$. Therefore, a region quite similar 
to the outer gap in pulsar magnetospheres appears in the black hole magnetospheres. 
The formation of longitudinal electric fields is also possible in this region, since 
the charge-separated plasma flow cannot provide the fullfilment of the condition 
$\rho_{\rm e} = \rho_{\rm GJ}$. As a result, it turned out that under real 
conditions the size of the acceleration region is much smaller than the system's size, 
so that the acceleration region does not affect the global structure of the magnetosphere 
[34, 35]. In this model, the particle production multiplicity is rather small:
\begin{equation}
\lambda_{\rm AGN_2} \sim 10 - 100.
\label{lmbagn2}
\end{equation}
Hence, we shall consider below both large and small values of the parameter $\lambda$.

Next, it should be remembered that to explain the high efficiency of the energy release 
from the central engine we need to assume that the rotation parameter $\Omega R/c$ must be 
not too much smaller than unity. In other words, the light cylinder radius must not exceed 
the central engine size too much. As a result, the observed transverse size $r_{\rm jet}$ of 
relativistic jets will be three-five orders of magnitude greater than the light cylinder
radius $R_{\rm L}$. That is why, far from the central engine most magnetic field lines 
must be far beyond the light cylinder.

However, according to the relationship
\begin{equation}
\frac{B_{\varphi}}{B_{\rm p}} = x_r,
\label{bphi'}
\end{equation}
where again $x_r = \Omega_{\rm F}\varpi/c$, which follows from the definition of the magnetic 
field components at $I \approx I_{\rm GJ}$, this implies that the toroidal magnetic field will 
be the same three-five orders of magnitude stronger than the poloidal magnetic field. Therefore, 
the magnetic field must have a strongly pronounced spiral structure. Correspondingly, the 
electric field must also be three-five orders of magnitude larger than the poloidal magnetic 
field. At present, VLBI (Very Large Baseline Interferometry) methods provide a lot of data on 
the polarization of the innermost parts of jets [199 201]; however, so far it is impossible 
to unambiguously determine the magnetic field structure from the observations.

Notice, finally, that strong twisting of magnetic field lines does not imply that the plasma 
motion will also occur along strongly twisted trajectories. As stressed above, an almost 
poloidal drift motion in the crossed electric and magnetic fields will be the principal motion 
of particles beyond the light cylinder. Thus, the toroidal velocity for $x_r > 1$ is given by 
the simple relationship
\begin{equation}
v_{\varphi}(x_r) = \frac{c}{x_r}.
\label{44a}
\end{equation}

Further, if the magnetization parameter $\sigma$ exceeds the ratio $r_{\rm jet}/R_{\rm L}$,
the plasma Lorentz factor can again be estimated from the asymptotic solution (101), so that 
$\gamma \approx r_{\rm jet}/R_{\rm L}\sim 10^{3}$--$10^{5}$. If $\sigma < r_{\rm jet}/R_{\rm L}$, 
the effects of the finite mass of particles will limit their energy growth at large axial 
distances. To determine the magnetization parameter $\sigma$, as was shown, it is necessary 
to know the particle production multiplicity $\lambda$.

As noted above, it is impossible at present to determine the basic parameters of the 
outflowing plasma from observations. Nevertheless, some estimates can still be made. 
For example, one of the methods to determine the value of $\lambda$ is based on the 
assumption that the synchrotron radiation self-absorption occurs at the base of the 
jet [202]. This assumption allows one to estimate the particle number density [203]. 
On one-parsec scales, the characteristic particle number densities thus were found to 
be $10^2$--$10^4$ cm$^{-3}$, which gives
\begin{equation}
\lambda_{\rm AGN_1} \sim 10^{10} - 10^{12}.
\label{lmbagn1}
\end{equation}
If this indeed is the case, intensive secondary particle creation near the black hole 
horizon should have occurred. Here, according to relation (56) the value of $\sigma$ cannot 
exceed one hundred:
\begin{equation}
\sigma_{\rm AGN_1} \sim 10^{2} - 10^{3},
\label{sgmagn1}
\end{equation}
which is much smaller than the ratio $r_{\rm jet}/R_{\rm L} \sim 10^5$ corresponding 
to the maximum possible Lorentz factor derived from the asymptotic solution (101). Therefore, 
an almost complete transformation of the electromagnetic energy into particles' 
energy must occur in the process of collimation. Notice that in this case, although 
$\gamma_{\rm max} = \sigma \sim 10^2$--$10^3$ exceeds the particle energies that are 
required to explain the apparent superluminal motion effect, it is still insignificant. 
Here, almost all the energy flux in the jet will be related to the flux of accelerated 
particles. For $\sigma \sim 10^2$--$10^3$, the radius \mbox{$r_{\rm F} \sim \sigma^{1/3}R_{\rm L}$} 
(66) of the fast magnetosonic surface must be smaller than one hundred radii from the 
central engine, which is \mbox{$10^{14}$--$10^{16}$ cm.} Thus, the flow in the jet must indeed 
be supersonic.

Now making use of expressions (89), (95), and (96), we arrive at the conclusion that 
in this case all critical magnetic fields for the reasonable value of $\gamma_{\rm in} \sim 10$
are larger than $B_{\rm ext} \sim 10^{-6}$ G corresponding to the ambient pressure 
(for convenience, all parameters discussed here are listed below in Table 2). But this 
means that a denser core must exist in the center of the jet, and the subsonic flow near 
the axis will not be formed. At last, the ejection rate of electron-positron pairs, 
${\dot N} = \pi  R_0^2 \cdot \lambda n_{\rm GJ} c$, can be estimated, which, as can be 
easily checked, is determined using the simple relation
\begin{equation}
{\dot N} \sim \lambda \left(\frac{W_{\rm tot}c}{e^2}\right)^{1/2}.
\end{equation}
As a result, we have ${\dot N}_{\rm AGN_1} \sim 10^{49}$ particles s$^{-1}$ 
(hereinafter we set $M = 10^9 M_{\odot}$). In other words, about $10^{63}$ 
electron-positron pairs will be injected into host galaxy over the time 
of the active life of a galactic nucleus, $\tau \sim 10^7$ years. This number, 
incidentally, is quite sufficient to explain the intensity of the annihilation 
line emitted from the Galactic center, which, as is well known, requires about 
$10^{43}$ annihilations per second [204].

If the multiplicity factor of the secondary particle creation is small, 
$\lambda_{\rm AGN_2} \sim 10$--$100$, the inner structure of the jet must 
be significantly different, since now all critical fields are below the 
value of $10^{-6}$ G corresponding to the ambient pressure. Here, one obtains
\begin{equation}
\sigma_{\rm AGN_2} \sim 10^{10} - 10^{12},
\label{sgmagn2}
\end{equation}
so that the plasma Lorentz factor, according to the asymptotic solution (101), can be as high 
as approximately $10^4$--$10^5$, and a subsonic flow region must be formed in the center 
of the jet. In this case, the energy flux is Poynting-dominated. Correspondingly, the 
dense core will not be formed, so that both the poloidal magnetic field and the plasma 
density are weakly dependent on the axial distance. The electron-positron pair injection 
rate will be much smaller: ${\dot N}_{\rm AGN_2} \sim 10^{39}$ particles s$^{-1}$. However, 
in this case, too, the fast magnetosonic surface radius $r_{\rm F} \sim 10^{17}$--$10^{18}$ cm 
will be smaller than the jet transverse size. Notice that here there is no direct contradiction 
with observations, since, as has been noted, the drift motion of particles will be directed 
almost along the poloidal magnetic field. This particle motion does not produce synchrotron
radiation. Hence, one should be cautious in using the standard synchrotron radiation formulas 
to estimate the value of the magnetic field and the lifetime of relativistic particles.

We should keep in mind that, when estimating radiation from relativistic jets, one usually 
assumes that an approximate equipartition between the energy densities of particles and the magnetic 
field takes place ($B^{({\rm com})2}/8\pi \sim \gamma^{({\rm com})} n^{({\rm com})} m c^2$) in 
the rest frame of plasma. The parameters we introduced allow us to obtain simple relationships 
for all quantities in this reference frame. In particular, it is easy to show that the 
characteristic Lorentz factor of particles in the plasma rest frame is simply the following:
\begin{equation}
\gamma^{({\rm com})} \approx \frac{\sigma}{\gamma},
\label{gcom}
\end{equation}
where $\gamma$ is the bulk Lorentz factor of the hydrodynamic flow. On the other hand, 
one finds $B^{({\rm com})} \approx (x_r/\gamma)B_{\rm p}$. Consequently, we have 
$B^{({\rm com})} \approx B_{\rm p}$ in the effective acceleration region ($\gamma = x_r$), 
while for the asymptotic solution (102) we obtain $B^{({\rm com})} \gg B_{\rm p}$.

\subsection {Microquasars}

If the operation of the central engine in microquasars indeed can be described by the 
MHD model considered here, it is again possible to assume that the observed subrelativistic 
jet velocities must correspond to $\sigma$ on the order of $3$--$10$. Then, according to 
relation (56), we should conclude that the particle production multiplicity in the 
microquasar magnetosphere must be fairly large:
\begin{equation}
\lambda_{\mu{\rm QSO}} \sim 10^{10}.
\label{lmbmQSO}
\end{equation}
Such a high value is also supported by the compactness parameter $l_{a,\mu{\rm QSO}} \sim 10^4$. 
Then, the electron-positron pair ejection rate should be ${\dot N}_{\mu{\rm QSO}} \sim 10^{43}$ 
particles s$^{-1}$. Finally, the large values of the critical fields \mbox{($B_{\rm min} \sim 10^{4}$ G,}
$B_{\rm eq} \sim 10^{3}$ G, $B_{\rm cr} \sim 10^{6}$ G) indicate that a dense core must exist 
in the jet center, and the subsonic flow near the rotational axis does not form.

On the other hand, if $\sigma$  is indeed not too high, the fast magnetosonic surface 
($r_{\rm F} \sim \sigma^{1/3}R_{\rm L}$) must lie close to the light cylinder, i.e., 
at distances of order \mbox{$10^{7}$--$10^{8}$ cm.} This scale is much smaller than the distance 
from the central engine to the companion star in binaries involving a microquasar. Hence, 
we can conclude that the interaction of the stellar wind and matter ejected from the 
microquasar occurs in the supersonic regime.

\subsection {Sources of cosmological gamma-ray bursts}

Now let us see which parameters can characterize jets outflowing cosmological gamma-ray 
bursts. It should be recalled that, according to one of the most popular models, a rapidly 
rotating central engine (magnetar, black hole) is immersed in the progenitor massive star 
[40]. That is why, the ambient pressure for the jet in its formation region is provided 
not by the surrounding medium with very small pressure, but by the matter of the massive 
star itself (the equivalent magnetic field $B_{\rm ext} \sim 10^{6}$--$10^{8}$ G). Notice 
also that in cosmological gamma-ray bursts there can be one more mechanism of the 
electron-positron pair creation, related to neutrino annihilation. Such neutrinos can be 
copiously created during supercritical accretion onto the collapsing stellar core [205].

The starting point that can shed light on the physical conditions inside the central engine 
can be the characteristic particle Lorentz factor $\gamma \sim 300$, which in this case can 
be naturally related to the magnetization parameter $\sigma$:
\begin{equation}
\sigma_{\rm GRB} \sim 10^{2} - 10^{3}.
\label{sigmaGRB}
\end{equation}
If the condition $\gamma \ll \sigma$ is satisfied, the total energy release from 
the central engine would be unrealistically high. Now, using Eqn (56) to estimate the 
plasma generation multiplicity, we obtain
\begin{equation}
\lambda_{\rm GRB} \sim 10^{13} - 10^{14}.
\label{lambdaGRB}
\end{equation}
Such a huge value unambiguously evidences that the particle creation efficiency must be high 
enough. Indeed, formula (57) gives very large compactness parameter $l_{a} \sim 10^{15}$. 
Correspondingly, for the electron-positron pair ejection rate we find 
${\dot N}_{\rm GRB} \sim 10^{53}$ particles s$^{-1}$.

Next, the very small sizes of the central engine together with the moderate value of the 
magnetization parameter $\sigma$ shows that the fast magnetosonic surface radius $r_{\rm F}$ 
(66) should not exceed $10^{7}$--$10^{8}$ cm, which is significantly smaller 
than the size of the progenitor star. Consequently, the matter outflow in the jet becomes 
supersonic before it exits the star. Finally, expressions (89) and (95) for the characteristic 
magnetic fields $B_{\rm min}$ and $B_{\rm eq}$ indicate that they are in the range of 
\mbox{$10^{6}$--$10^{8}$ G,} i.e., their pressure is comparable with that inside the progenitor star. 
Therefore, the jet transverse size will indeed be sufficient to accelerate particles up to 
energies $\gamma \sim \sigma$.

Notice at last that the condition $\gamma \vartheta \sim 1$ is certainly not satisfied for 
gamma-ray bursts, since in that case the jet spread angles would only be $0.1^{\circ}$, while 
observations indicate that $\vartheta \sim 1$--$10^{\circ}$ [206]. Such a flow can also be 
realized. For example, it was shown in paper [207] (see also Ref. [208]) that in the model 
of an infinitely extended progenitor star, where the ambient pressure decreases graduately
according to a power law, the acceleration turns out to be not very effective in comparison 
with a more realistic model in which the ambient pressure beyond the star is assumed to be 
low. As it has turned out, in both cases the flow corresponds to a weakly collimated flux 
with $1 < k < 2$, where the particle energy follows the asymptotic behaviour
$\gamma \approx (R_{\rm c}/\varpi)^{1/2}$ (102). In the former case, however, the radius 
of curvature $R_{\rm c}$ of the magnetic surfaces, which is determined by the pressure 
decrease law inside the progenitor star, turns out to be sufficiently small, which precludes 
plasma from being effectively accelerated. Beyond the star, magnetic field lines straighten 
up (and hence the radius of curvature increases), which leads to a more effective acceleration. 
As we see, the simple analytical asymptotic solutions obtained above allow easy interpretation 
of the numerical experiment.

\subsection{Radio pulsars}

Radio pulsars, undoubtedly, only indirectly relate to the topic considered here, 
since their magnetospheres are certainly not axisymmetric and stationary. 
It is not then surprising that jets, as we have noted, are observed only from two 
energetic pulsars. Nevertheless, many points, which were possible to clarify in 
the theory of the neutron star magnetosphere, undoubtedly allow us to shed light 
on the nature of other compact objects, too.

First and foremost, it should be noted that basic parameters characterizing the pulsar 
wind are well known inside the light cylinder. This is due to our good knowledge of 
the process of plasma creation near the magnetic polar caps of a neutron star. Numerous 
calculations have shown that the general properties of the secondary electron-positron 
plasma flowing out from the magnetosphere turned out to be only a little sensitive to 
the details of the acceleration region structure. For most models [178, 180, 209, 210], 
both the density and the energy spectrum of the plasma flowing out are universal enough. 
Therefore, with certainty we can say that the plasma flowing along open field lines in 
the pulsar magnetosphere includes both the primary particle beam with energy 
${\cal E} \approx 10^7$ MeV and density close to the Goldreich density and the secondary 
electron-positron component. Its energy spectrum with a good accuracy exhibits a power 
law dependence
\begin{equation}
{\cal E} \propto {\cal E}^{-2},
\end{equation}
ranging from ${\cal E}_{\rm min} \sim 10$--$100$ MeV to 
${\cal E}_{\rm max} \sim 10^4$ MeV. The total density of the secondary plasma for ordinary 
pulsars exceeds the Goldreich density by $10^3$--$10^4$ times. And only for the most energetic 
pulsars can the multiplicity factor reach $10^5$. 

Thus, the parameter $\lambda$ for radio pulsars is determined quite reliably:
\begin{equation}
\lambda_{\rm PSR} \sim 10^3 - 10^4.
\label{lmbpsr}
\end{equation}
Now, making use again of relation (56) we obtain 
\begin{equation}
\sigma_{\rm PSR} \sim 10^3 - 10^4.
\label{sgmpsr}
\end{equation}
And only for the most energetic pulsars do we find $\sigma_{\rm PSR} \sim 10^5$--$10^6$. 
Thus, the condition \mbox{$\gamma_{\rm in} \gg \sigma^{1/3}$} is satisfied in 
the vast majority of pulsars, which corresponds to a slow rotation of the central object 
[99]. Indeed, the equality $\gamma_{\rm in} = \sigma^{1/3}$ can be written out in the 
form $P = P_{\rm cr}$, where
\begin{eqnarray}
P_{\rm cr} = \pi \frac{R}{c}\,
\left[ \frac{2}{\lambda \gamma_{\rm in}^3}
\, \left(\frac{\omega_{B}R}{c}\right)\right]^{1/2}
\nonumber \\ 
\sim 10^{-3} \,  \left(\frac{\lambda}{10^4}\right)^{-1/2}
\left(\frac{\gamma}{10^{2}}\right)^{-3/2}
\left(\frac{B_0}{10^{12} \, {\rm G}}\right)^{1/2} \, {\rm s}.
\label{l116'}
\end{eqnarray}
For fast rotation ($P \ll P_{\rm cr}$), the particle energy significantly increases when 
particles approach the fast magnetosonic surface, whereas for slowly rotating objects 
($P \gg P_{\rm cr}$ the particle energy remains practically unchanged. Their further fate, 
as we have shown above, depends on whether the flow intersects the light surface or not.

\subsection{Young stars}

In conclusion, we discuss the main parameters which characterize nonrelativistic jets from 
young stars. It should be recalled that in this case the nonrelativistic magnetization 
parameter an $\sigma_{\rm n} \approx (\Omega_{\rm F}/\Omega_{\rm cr})^2$ (53) plays the key 
role. Under the condition $\sigma_{\rm n} \gg 1$ ($\Omega_{\rm F} \gg \Omega_{\rm cr}$), the 
electromagnetic energy flux near the central engine will be much greater than the particle 
energy flux; beyond singular surfaces, as stressed above, particles must carry a significant
fraction of the total energy.

As the critical period $P_{\rm cr} = 2 \pi/\Omega_{\rm cr}$, namely
\begin{eqnarray}
P_{\rm cr} \approx 10 
\left(\frac{R}{10^{11}{\rm cm}}\right)^{2}
\left(\frac{B_0}{10^{3}\,{\rm G}}\right)  
\left(\frac{r_{\rm d}/R}{30}\right)^{-1}
\nonumber \\
\left(\frac{v_{\rm in}}{100 \,{\rm km \, s^{-1}}}\right)^{-3/2}
\left(\frac{\dot M}{10^{-9}M_{\odot} \, {\rm yr}^{-1}}\right)^{-1/2}
 \, {\rm d}
\label{pcr}
\end{eqnarray}
($r_{\rm d}$ is the inner radius of the accretion disc) is close to the spin periods 
of young stars, in the region of magnetic field lines coming out of the surface of 
the star we have $\sigma_{\rm n} \sim 1$. On the other hand, the period $P_{\rm cr}$ 
is one two orders of magnitude larger than the rotation periods in the inner regions of 
accretion discs, $P_{\rm b} = 2 \pi (GM/r^3)^{1/2}$, so that for the corresponding magnetic 
field lines one finds $\sigma_{\rm n} \sim 10$--$1000$. That is why, the inner parts of the 
accretion disc, not the central star, must play the role of the central engine rotor. 
As noted above, there are observational evidences of this being the case [76].

Next, from relation (66) we obtain
\begin{equation}
r_{\rm A} \sim r_{\rm F} \sim \frac{v_{\rm in}}{\Omega} \, \sigma_{\rm n}^{1/3},
\label{aqa}
\end{equation}
so that $r_{\rm F} \sim 10$--$30 \, R$. Thus, the distance to singular surfaces exceeds 
the size of the star by $10$--$30$ times, but it is the same $10$--$30$ times smaller 
than the transverse size of jets. Consequently, the flow inside the jets must be supersonic 
and, hence, the longitudinal current for these objects again must be derived from the 
critical conditions on the singular surfaces. Correspondingly, the radius of the jet core 
must be on the order of $r_{\rm core} = v_{\rm in}/{\Omega} \sim 0.1$ a.u., the jet magnetic
field should be as high as 
$B_{\rm core} \sim 0.1 (\Omega r_{\rm core}/v_{\rm in})^2 B_{\rm in} \sim 10^{-2}$ G, and the 
particle number density must range from $10^{8}$ to \mbox{$10^{9}$ cm$^{-3}$.}

Finally, we note that the existence of the integrals of motion allows us to obtain direct 
information about the plasma outflow region. For example, if the radial and longitudinal 
velocities of the flow in the jet are known at the axial distance $r_{\perp}$ (and such 
observations, as noted above, have already been performed for several young stars [72, 73]), 
it is possible to estimate the distance $r_{\rm st}$ from the central star at which the 
corresponding force line is 'anchored' in the accretion disc [211, 212]:
\begin{eqnarray}
r_{\rm st} \approx 0.7 \,  \left(\frac{r_{\perp}}{10 \, {\rm a.u.}}\right)^{2/3}
\left(\frac{{v}_{\varphi}}{10 \, {\rm} \, {\rm km} \,  {\rm s}^{-1}}\right)^{2/3} \nonumber \\
\left(\frac{{v}_{\rm p}}{100 \, {\rm km} \, {\rm s}^{-1}}\right)^{-4/3}
\left(\frac{M}{M_{\odot}}\right)^{1/3} \, {\rm a.u.}
\end{eqnarray}
It is obvious that this scale is much larger than the size of the star, so at present it has 
been possible to resolve only outermost regions of the outflow.

\section{Conclusions}

Thus, significant progress has indeed been recently achieved in the understanding 
of the nature of jets observed in different classes of astrophysical sources. This 
became possible because the analytical approach allowed sufficiently simple relationships 
between physical parameters characterizing the outflows to be found, and the numerical 
modeling (in which, we should keep in mind, the setup of the problem itself has been 
significantly different from the analysis of stationary equations) has confirmed these 
relationships.

The most important result of the analytical theory includes the understanding of the role 
of key dimensionless parameters. For clearness, they are listed in Table 2 (for active 
galactic nuclei we set $M = 10^9 \, M_{\odot}$). It turned out that the knowledge of these 
parameters allows us to estimate many characteristics of jets, including the fraction of 
energy carried by particles, the plasma Lorentz factor, the electron positron pair injection 
rate ${\dot N}$, and the compactness parameter $l_{a}$, as well as to determine the main 
parameters of the internal structure of jets. The determination of these parameters from 
observations would be a significant breakthrough in our understanding of physical processes 
which are underway in active astrophysical sources.

Next, we have shown that many properties of relativistic and nonrelativistic jets are 
significantly different from each other. For convenience, we also collect them together 
in Table 3. As can be seen, one should be very cautious when applying the results, which 
were obtained for nonrelativistic jets, to ultrarelativistic flows. Moreover, the asymptotic 
relations formulated above clarify many results obtained by means of numerical simulations.

The limited space of the present review did not allow us to discuss in detail many 
important issues. In particular, we only briefly discussed the stability of jets. 
Finally, here we have no space at all to discuss the nonstationary performance of 
the central engine (most papers have recently started focusing exactly on this topic) 
or the proper radiation of jets. Nevertheless, we would like to hope that questions 
addressed in this review will be useful for future studies of relativistic and 
nonrelativistic outflows observed in many astrophysical objects.

\begin{table}[ht]
\caption{
Parameters of jets outflowing from relativistic compact objects. All values are given to 
an order of magnitude.
} 
\centering
\begin{tabular}{|l|c|c|c|c|c|}
\hline
                           & AGN$_1$        & AGN$_2$          & $\mu$QSO   & GRB       & PSR        \\
\hline
$\sigma$                   & $100$          & $10^{12}$        & $10$       & $10^3$      & $10^{4}$        \\
$\lambda$                  & $10^{12}$      & $100$            & $10^{10}$  & $10^{14}$   & $10^{3}$ \\
$\gamma_{\rm in}$          & $10$           & $10$             & $10$       & $10$        & $100$     \\
\hline
$l_{a}$                    & $1$--$100$     & $1$--$100$       & $10^{4}$   & $10^{15}$   & $10^{-5}$ \\
$W_{\rm part}/W_{\rm tot}$ & $1$            &  $10^{-9}$       & $ 10^{-5}$ & $1$         & $10^{-2}$ \\
$\gamma$                   & $10$--$100$    & $10^4$--$10^5$   & $10^3$     & $300$       & $10^3$\\
${\dot N}$, s$^{-1}$ & $10^{49}$      & $10^{39}$        & $10^{33}$  & $10^{53}$   & $10^{32}$   \\
\hline
$B_{\rm min}$, G          & $10^{-2}$      & $10^{-12}$       & $10^{4}$   & $10^{8}$    & $10^{-6}$  \\
$B_{\rm eq}$, G         & $10^{-4}$      & $10^{-24}$       & $10^{3}$   & $10^{6}$    & $10^{-8}$  \\
$B_{\rm cr}$, G         & $10^{-1}$      & $10^{-11}$       & $10^{6}$   & $10^{10}$   & $10^{-2}$  \\
% & & &  &  &   \\
\hline
\end{tabular}
\label{table5_01} 
\end{table}

\begin{table*}[ht]
\caption{Main differences between relativistic and nonrelativistic jets.} 
\vspace{0.3cm}
\centering
\begin{tabular}{|l|l|}
\hline
{\bf Relativistic flow}                         & {\bf Nonrelativistic flow} \\
\hline 
Longitudinal current is close  to the  Goldreich current   & Longitudinal current is much larger than the\\
                             &  Goldreich current   \\
\hline
For strongly magnetized flow ($\sigma \gg 1$), fast  & Fast magnetosonic surface is located near Alfv\'enic \\  
magnetosonic surface near equatorial plane is located       &   surface \\
 $\sigma^{1/3}$ times farther than Alfv\'enic surface&   \\
\hline                                 
On fast magnetosonic surface, the particle energy flux   & On fast magnetosonic surface, the particle energy flux\\
is much smaller than the electromagnetic energy flux      &   is close to the electromagnetic energy flux \\
\hline
Beyond Alfv\'enic surface, the electric field  is close   & Beyond Alfv\'enic surface, the electric field is much\\
 in magnitude to the magnetic field     &   smaller than the magnetic field   \\
\hline
Proper collimation is impossible for both strongly and   & Proper collimation becomes effective for strongly \\ 
 weakly magnetized outflows         &  magnetized outflows\\   
\hline
The dense core in jets can be formed only under & The dense core is always formed in jets. Magnetic flux \\
sufficiently small ambient pressures. Magnetic flux &  in the core is a significant fraction of the total \\
in the core is much smaller than the total flux   &  magnetic flux  \\
\hline
Cylindrical flow with subsonic core is possible   &   Cylindrical flow with subsonic core is impossible  \\
\hline
\end{tabular}
\label{table5_02} 
\end{table*}

\noindent
{\bf Acknowledgments}

\vspace{0.3cm}

I would like to acknowledge M Barkov, S Komissarov, M Romanova, and A D Tchekhovskoy 
for carefully reading the manuscript and many notes, and also A V Gurevich, Ya N Istomin, 
Yu Yu Kovalev, R Lovelace, Yu Lyubarsky, and A Spitkovsky for useful discussions. The work 
was supported by RFBR grant 09-02-00749 and the Federal Agency of Science and Innovations 
(contract No. 02.740.11.0250).
 
\section{Appendix}

In the Appendix, we give for reference the complete set of equations of the Grad-Shafranov 
method written out in themost general case, i.e., for axisymmetric stationary flows in the 
vicinity of a rotating black hole. First of all, we keep in mind the basic relations for 
the Kerr metric of a rotating black hole. In the Boyer-Lindquist coordinates $t$, $r$, 
$\theta$, and $\varphi$, it assumes the form
$$
{\rm d}s^{2}=-\alpha^{2}{\rm d}t^{2}
+g_{ik}({\rm d}x^{i}+\beta^{i}{\rm d}t)({\rm d}x^{k}+\beta^{k}{\rm d}t),
\label{c1}
\eqno{(A.1)}
$$
where the quantity
$$
\alpha=\frac{\rho}{\Sigma}\sqrt{\Delta}
\label{c2}
\eqno{(A.2)}
$$
is the gravitational redshift, and the vector ${\bf \beta}$  is toroidal:
$$
\beta^{r} = \beta^{\theta} = 0, \qquad
\beta^{\varphi}=-\omega.
\label{c3}
\eqno{(A.3)}
$$
Here
$$
\omega = \frac{2aMr}{\Sigma^{2}}
\label{c2**'}
\eqno{(A.4)}
$$
is the so-called Lense-Thirring angular velocity. Finally, $M$ and $a$ are the mass and 
specific angular momentum of the black hole ($a = J/M$), respectively. In addition, we 
introduced the standard notations
$$
 \Delta=r^{2}+a^{2}-2Mr, \qquad
 \rho^{2}=r^{2}+a^{2}\cos^{2}\theta, \nonumber\\
$$
$$
\Sigma^{2}=(r^{2}+a^{2})^{2}-a^{2}\Delta\sin^{2}\theta, \qquad
 \varpi=\frac{\Sigma}{\rho}\sin\theta.
\label{c5}
\eqno{(A.5)}
$$
Here, in all relativistic expressions we use the units in which $c = G = 1$. Finally, it is 
important that the three-dimensional metric $g_{ik}$ in formula ($A.1$) be diagonal:
$$
g_{rr}=\frac{\rho^{2}}{\Delta},\qquad
g_{\theta \theta}=\rho^{2}, \qquad
g_{\varphi \varphi}=\varpi^{2}.
\label{c4}
\eqno{(A.6)}
$$
As for the flat space limit, it can be easily obtained by passing to the limit 
$\alpha \rightarrow 1$ and $\omega \rightarrow 0$.

As is well known, for calculations it is convenient to introduce a special reference 
frame, the so-called ZAMO (Zero Angular Momentum Observers) [131], which has the 
following properties:
\begin{itemize}
\item
ZAMO observers are located at constant radius \mbox{$r =$ const,} $\theta =$ const but rotate 
with the Lense Thirring angular velocity ${\rm d}\varphi/{\rm d}t=\omega$;
\item
for ZAMO, the four-dimensional metric $g_{\alpha\beta}$ is diagonal, with its three-dimensional 
part $g_{ik}$ coinciding with Eqn ($A.6$).
\end{itemize}
Below, all vectors will be written out in this reference frame. In particular, 
the operator $\nabla_{i}$, means the covariant derivative in the three-dimensional 
metric ($A.6$).

As a result, the electric and magnetic fields can be conveniently written as
$$
{\bf B}  =  \frac{{\bf\nabla}\Psi \times {\bf e}_{\hat \varphi}}{2\pi\varpi}
-\frac{2I}{\alpha\varpi}{\bf e}_{\hat \varphi},
\label{k21'P} 
\eqno{(A.7)}
$$
$$
{\bf E}  =  -\frac{\Omega_{\rm F}-\omega}{2\pi\alpha}{\bf\nabla}\Psi,
\label{k24''P}
\eqno{(A.8)}
$$
respectively, and the four-velocity of matter is written as
$$
{\bf u} = \frac{\eta}{\alpha n}{\bf B} + \gamma(\Omega_{\rm F}-\omega)\frac{\varpi}
{\alpha}{\bf e}_{\hat\varphi},
\label{p3'}
\eqno{(A.9)}
$$
where $\gamma = 1/\sqrt{1-v^2}$ is the Lorentz factor of matter, and the subscripts with a 
cap over them correspond to physical components of vectors. The quantity $\Omega_{\rm F}$ 
remains the integral of motion. In turn, integrals of motions $E$ and $L$ will be written 
out as
$$
 E=E(\Psi) = \frac{\Omega_{\rm F}I}{2\pi}
+ \mu\eta(\alpha\gamma + \omega \varpi u_{\hat\varphi}),
\eqno{(A.10)}
\label{p31P} 
$$
$$
 L=L(\Psi) = \frac{I}{2\pi} + \mu\eta \varpi u_{\hat\varphi}.
\label{p32P}
\eqno{(A.11)}
$$
Next, the relativistic Bernoulli equation $\gamma^{2} - u_{\hat\varphi}^{2} = u_{\rm p}^2 +1$
takes the form 
$$
\frac{K}{\varpi^{2}A^{2}}=\frac{1}{64\pi^{4}}\frac{{\cal M}^{4}({\bf\nabla}
\Psi)^{2}}{\varpi^{2}}+\alpha^{2}\eta^{2}\mu^{2},
\label{p38}
\eqno{(A.12)}
$$
where the Alfv\'en factor is 
$$
A=\alpha^{2}-(\Omega_{\rm F}-\omega)^{2}\varpi^{2}-{\cal M}^{2},
\label{p39app}
\eqno{(A.13)}
$$
and
$$
K=\alpha^{2}\varpi^{2}(E-\Omega_{\rm F}L)^{2}\left[\alpha^{2}-(\Omega_{\rm F}-
\omega)^{2}\varpi^{2}-2{\cal M}^{2}\right] \nonumber 
$$
$$
+{\cal M}^{4}\left[\varpi^{2}(E-\omega L)^{2}-\alpha^{2}L^{2}\right].
\eqno{(A.14)}
\label{p40}
$$
This equation defines the Alfv\'enic Mach number {\cal M}, where 
$$
{\cal M}^{2}=\frac{4\pi\eta^{2}\mu}{n}.
\label{p36}
\eqno{(A.15)}
$$
Now, making use of relations ($A. 12$)-($A. 14$), which can be recast in the form
$({\bf\nabla}\Psi)^{2}=F({\cal M}^{2},E,L,\eta,\Omega_{\rm F},\mu)$,
where
$$
F=\frac{64\pi^{4}}{{\cal M}^{4}}\frac{K}{A^{2}}-\frac{64\pi^{4}}{{\cal M}^{4}}\alpha^{2}
\varpi^{2}\eta^{2}\mu^{2},
\label{p45}
\eqno{(A.16)}
$$
we obtain
$$
\nabla_{k}{\cal M}^{2}=\frac{N_k}{D},
\label{p46}
\eqno{(A.17)}
$$
where
$$
N_k = -\frac{A}{({\bf\nabla}\Psi)^{2}}\nabla^{i}\Psi \cdot \nabla_{i}
\nabla_{k}\Psi+\frac{A}{2}\frac{\nabla_{k}'F}{({\bf\nabla}\Psi)^{2}}.
\label{p47}
\eqno{(A.18)}
$$
Here, the operator $\nabla'_k$ acts on all quantities but ${\cal M}^{2}$. The
quantity $\nabla'_k\mu$ must be determined from the relation [97]
$$
\nabla'_k\mu
=\frac{2 c^{2}_{\rm s}}{1-c^{2}_{\rm s}}\mu \, \frac{\nabla_k\eta}{\eta}
+\frac{1}{1-c^{2}_{\rm s}}\left[\frac{1}{n}\left(\frac{\partial
P}{\partial s}\right)_{n}+T\right]\nabla_k s,
\label{p49}
\eqno{(A.19)}
$$
where $c^{2}_{\rm s} = 1/\mu(\partial P/\partial n)_{s}$ is the speed of sound, and $s$ 
is the entropy. In turn, the denominator $D$ can be rewritten in the form
$$
D=\frac{A}{{\cal M}^{2}}+\frac{\alpha^{2}}{{\cal M}^{2}}\frac{B^{2}_{\hat\varphi}}{B^{2}_
{\rm p}}-\frac{1}{u^{2}_{\rm p}}\frac{A}{{\cal M}^{2}}
\frac{c^{2}_{\rm s}}{1-c^{2}_{\rm s}}.
\label{p48}
\eqno{(A.20)}
$$

As for the Grad-Shafranov equation, in the compact form it can be written out as [96, 97]
$$
\frac{1}{\alpha}\nabla_{k}\left\{\frac{1}{\alpha\varpi^2}
[\alpha^{2}-(\Omega_{\rm F}-\omega)^{2}\varpi^{2}-{\cal M}^{2}]
\nabla^{k}\Psi\right\}  \nonumber
$$
$$
+\frac{\Omega_{\rm F}
-\omega}{\alpha^{2}}({\bf\nabla}\Psi)^{2}\frac{{\rm d}
\Omega_{\rm F}}{{\rm d}\Psi}
\nonumber \\
\eqno{(A.21)}
\label{p64}
$$
$$
+\frac{64\pi^{4}}{\alpha^{2}\varpi^{2}}\frac{1}{2{\cal M}^{2}}
\frac{\partial}{\partial\Psi}\left(\frac{G}{A}\right)
-16\pi^{3}\mu n\frac{1}{\eta}\frac{{\rm d}\eta}{{\rm d}\Psi}
-16\pi^{3}nT\frac{{\rm d}s}{{\rm d}\Psi}=0,  \nonumber
$$
where
$$
G=\alpha^{2}\varpi^{2}(E-\Omega_{\rm F}L)^{2}+\alpha^{2}{\cal M}^{2}L^{2}
-{\cal M}^{2} \varpi^{2}(E-\omega L)^{2}.
\label{p65}
\eqno{(A.22)}
$$
Now, expanding terms $\nabla_{k}{\cal M}^{2}$ in Eqn ($A.21$) according to definitions 
($A.17$)--($A.19$), we finally arrive at
$$
 A\left[\frac{1}{\alpha}\nabla_{k}\left(\frac{1}{\alpha\varpi^{2}}
 \nabla^{k}\Psi\right)+\frac{1}{\alpha^{2}\varpi^{2}({\bf\nabla}\Psi)^{2}}
 \frac{\nabla^{i} \Psi \cdot \nabla^{k} \Psi \cdot \nabla_{i}\nabla_{k}\Psi}{D}\right]
 \nonumber
$$
$$ 
 +\frac{1}{\alpha^{2}\varpi^{2}}\nabla_{k}'A \cdot \nabla^{k}\Psi 
 -\frac{A}{\alpha^{2}\varpi^{2}({\bf\nabla}\Psi)^{2}}
 \frac{1}{2D}\nabla_{k}'F \cdot \nabla^{k}\Psi
\nonumber \\
 \nonumber 
$$
$$
  +\frac{\Omega_{\rm F}-\omega}{\alpha^{2}}
 ({\bf\nabla}\Psi)^{2} \frac{{\rm d}\Omega_{\rm F}}{{\rm d}\Psi}
+\frac{64\pi^{4}}{\alpha^{2}\varpi^{2}}\frac{1}{2{\cal M}^{2}}
 \frac{\partial}{\partial\Psi}\left(\frac{G}{A}\right)
\nonumber \\
$$
$$
  -16\pi^{3}\mu n\frac{1}{\eta}
 \frac{{\rm d}\eta}{{\rm d}\Psi}-16\pi^{3}nT\frac{{\rm d}s}{{\rm d}\Psi}=0,
\label{p66} 
\eqno{(A.23)}
$$
where again the gradient $\nabla_{k}'$ acts on all quantities but
${\cal M}^2$, and the derivative $\partial/\partial\Psi$
acts only onh the integrals of motion. Formula ($A27$) determines 
in the most general form the equilibrium equation for the magnetic 
surfaces. Finally algebraic relations have the form
$$
\frac{I}{2\pi} =  \frac{\alpha^{2}L-(\Omega_{\rm F}-\omega)\varpi^{2}
(E-\omega L)}{\alpha^{2}-(\Omega_{\rm F}-\omega)^{2}\varpi^{2}-{\cal M}^{2}},
\label{p33} \\
\eqno{(A.24)}
$$
$$
\gamma  =  \frac{1}{\alpha\mu\eta}\frac{\alpha^{2}(E-\Omega_{\rm F}L)-{\cal M}^{2}
(E-\omega L)}{\alpha^{2}-(\Omega_{\rm F}-\omega)^{2}\varpi^{2}-{\cal M}^{2}},
\label{p34} 
\eqno{(A.25)} 
$$
$$
u_{\hat\varphi}  =  \frac{1}{\varpi\mu\eta}\frac{(E-\Omega_{\rm F}L)
 (\Omega_{\rm F}-\omega)\varpi^{2}-L{\cal M}^{2}}{\alpha^{2}-(\Omega_{\rm F}
 -\omega)^{2}\varpi^{2}-{\cal M}^{2}}.
\label{p35}
\eqno{(A.26)}
$$

Equations ($A.12$) and ($A.24$)--($A.26$) represent algebraic bounds which allow the 
determination, albeit in an indirect form, of all characteristics of the flow from the 
given poloidal field ${\bf B}_{\rm p}$ (i.e., from the known potential $\Psi$) and five 
integrals of motion. It should be emphasized that for a nonzero temperature they are 
extremely lengthy, mainly due to the need to resolve equation ($A.19$). In the case of 
cold plasma ($s$ = 0, i.e., \mbox{$\mu = $ const),} Bernoulli equation ($A.12$) becomes a 
fourth-order algebraic equation with respect to ${\cal M}^2$. As shown above, this fact 
often allows analytical asymptotics to be found.

In the cylindrical case, the second-order Grad-Shafranov equation can be conveniently 
reduced to the system of two ordinary differential equations of the first order for 
the magnetic flux $\Psi(\varpi)$ and the Mach number ${\cal M}^2$. The equation for 
the Mach number has therewith the form [146]
$$
\left[\frac{(e')^2}{\mu^2\eta^2}-1+\frac{\Omega_{\rm F}^2 r^2}{c^2}
-A\frac{c_{\rm s}^2}{c^2}\right]
\frac{{\rm d}{\cal M}^2}{{\rm d} r } = 
\frac{{\cal M}^6L^2}{A r ^3 \mu^2\eta^2c^2}
$$
$$
+\frac{\Omega_{\rm F}^2 r {\cal M}^2}{c^{2}}\left[2 - \frac{(e')^2}{A\mu^2\eta^2c^4}\right]
+{\cal M}^2 \frac{e'}{\mu^2\eta^2c^4}\frac{{\rm d}\Psi}{{\rm d}r}\frac{{\rm d}e'}{{\rm d}\Psi}
$$
$$
+{\cal M}^2\frac{ r ^2}{c^2}\Omega_{\rm F}\frac{{\rm d}\Psi}{{\rm d}r}
\frac{{\rm d}\Omega_{\rm F}}{{\rm d}\Psi}
-{\cal M}^2 \left(1-\frac{\Omega_{\rm F}^2 r ^2}{c^2} + 2A\frac{c_{\rm s}^2}{c^2}\right)
\frac{{\rm d}\Psi}{{\rm d} r }\frac{1}{\eta}\frac{{\rm d}\eta}{{\rm d}\Psi}
$$
$$
-\left[\frac{A}{n}\left(\frac{\partial P}{\partial s}\right)_n
+\left(1-\frac{\Omega_{\rm F}^2 r ^2}{c^2}\right)T\right]
\frac{{\cal M}^2}{\mu}\frac{{\rm d}\Psi}{{\rm d}r}
\frac{{\rm d}s}{{\rm d}\Psi},
\label{p36new1}
\eqno{(A.27)}
$$
where $e' = E - \Omega_{\rm F}L$. The equation for the magnetic flux 
$\Psi$ will coincide with Bernoulli equation ($A.14$).
Finally, the force-free pulsar equation takes on the form
$$
-\left(1-\frac{\Omega_{\rm F}^2\varpi^2}{c^2}\right)\nabla^2\Psi
+\frac{2}{\varpi}\frac{\partial \Psi}{\partial\varpi}
$$
$$
-\frac{16\pi^{2}}{c^2}I\frac{{\rm d}I}{{\rm d}\Psi}
+\frac{\varpi^{2}}{c^2}\left(\nabla\Psi\right)^{2}
\Omega_{\rm F}\frac{{\rm d}\Omega_{\rm F}}{{\rm d}\Psi}=0,
\label{d39}
\eqno{(A.28)}
$$
where $\nabla^2$ is the Laplace operator. Its generalization to the force-free 
black hole magnetosphere is written as [131]
$$
\frac{1}{\alpha}{\bf\nabla}_k\left\{\frac{\alpha}{\varpi^{2}}
\left[1-\frac{(\Omega_{\rm F}-\omega)^{2}\varpi^{2}}{\alpha^{2}}\right]
{\bf\nabla}^k\Psi \right\} 
$$
$$
+\frac{\Omega_{\rm F}-\omega}{\alpha^{2}}
({\bf\nabla}\Psi)^{2} \frac{{\rm d} \Omega_{\rm F}} {{\rm d}\Psi}
+\frac{16\pi^{2}}{\alpha^{2} \varpi^{2}} I\frac{{\rm d}I}{{\rm d}\Psi} = 0.
\eqno{(A.29)}
$$
These equations are elliptical in all the space, and so they require boundary conditions on the 
integration region boundary or on the black hole horizon.

\newpage

%\vspace{0.5cm}

\noindent
{\bf References}

\vspace{0.3cm}

{\small

\noindent
1.     Kardashev N S {\it Astron. Zh.} {\bf 41} 807 (1964) [ {\it Sov. Astron.} {\bf 8} 643 (1965)] 

\noindent
2.     Pacini V {\it Nature} {\bf 216} 567 (1967) 

\noindent
3.     Goldreich P, Julian W H {\it Astrophys. J.} {\bf 157} 869 (1969)

\noindent
4.      Michel F C {\it Astrophys. J.} {\bf 158} 727 (1969)\

\noindent
5.      Blandford R D {\it Mon. Not. R. Astron. Soc.} {\bf 176} 465 (1976)

\noindent
6.      Lovelace R W E{\it Nature} {\bf 262} 649 (1976)

\noindent
7.     Bisnovatyi-Kogan G S, Popov Yu P, Samokhin A A {\it Astrophys. Space Sci.} {\bf 41} 321 (1976)

\noindent
8.     Moiseenko S G, Bisnovatyi-Kogan G S, Ardeljan N V {\it Astrophys. J.} {\bf 370} 501 (2006)

\noindent
9.      Beskin V S {\it Osesimmetrichnye Statsionamye Techeniya v Astrofizike} ({\it MHD Flows in Compact Astrophysical Objects}) (Moscow: Fizmatlit, 2005) [Translated into English (Berlin: Springer, 2009)]

\noindent
10.     Beskin V S {\it Usp. Fiz. Nauk} {\bf 167} 689 (1997) [Phys. Usp. {\bf 40} 659 (1997)]

\noindent
11.     Shapiro S, Teukolsky S {\it Black Holes, White Dwarfs, and Neutron Stars} (New York: Wiley, 1983) [Translated into Russian (Moscow: Mir, 1985)]

\noindent
12.     Begelman M C, Blandford R D, Rees M J Rev. Mod. Phys. {\bf 56} 255 (1984) [Translated into Russian: in {\it Fizika Vnegalakticheskikh Istochnikov Radioizlucheniya} (Ed. R D Dagkesamanskii) (Moscow: Mir, 1987) p. 1

\noindent
13.     Rees M J Annu. Rev. Astron. Astrophys. {\bf 22} 471 (1984)

\noindent
14.     Zeldovich Ya B, Novikov I D {\it Relyativistskaya Astrofizika} ({\it Relativistic Astrophysics}) (Moscow: Nauka, 1967) [Translated into English (Chicago: Univ. of Chicago Press, 1971,1983)]

\noindent
15.     Lynden-Bell D {\it Nature} {\bf 223} 690 (1969)

\noindent
16.     Bisnovatyi-Kogan G S, Ruzmaikin A A {\it Astrophys. Space Sci.} {\bf 28} 45 (1974)

\noindent
17.     Bisnovatyi-Kogan G S, Ruzmaikin A A {\it Astrophys. Space Sci.} {\bf 42} 401(1976)

\noindent
18.     Novikov I D, Frolov V P {\it Fizika Chernykh Dyr} ({\it Physics of Black Holes}) (Moscow: Nauka, 1986) [Translated into English (Dordrecht: Kluwer Acad., 1989)]

\noindent
19.     Junor W, Biretta J A, Livio M {\it Nature} {\bf 401} 891 (1999)

\noindent
20.     Perley R A, Dreher J W, Cowan J J {\it Astrophys. J.} {\bf 285} L35 (1984)

\noindent
21.     Lobanov A, Hardee P, Eilek J {\it New Astron. Rev.} {\bf 47} 629 (2003)

\noindent
22.     Kovalev Y Y et al. {\it Astrophys. J.} {\bf 668} L27 (2007)

\noindent
23.     Junor W, Biretta J A Astron. J. {\bf 109} 500 (1995)

\noindent
24.     Reynolds C S et al. {\it Astrophys. J.} {\bf 283} 873 (1996)

\noindent
25.     Sikora M, Madejski G {\it Astrophys. J.} {\bf 534} 109 (2000)

\noindent
26.     Blandford R D, Rees M J {\it Astrophys. J.} {\bf 169} 395 (1974)

\noindent
27.     Fabian A C, Rees M J {\it Astrophys. J.} {\bf 277} L55 (1995)

\noindent
28.     Feretti L et al. Astron. Astrophys. {\bf 298} 699 (1995)

\noindent
29.     Cheng A Y S, O'Dell S L {\it Astrophys. J.} {\bf 251} L49 (1981)

\noindent
30.     Proga D, Stone J M, Kallman T R {\it Astrophys. J.} {\bf 543} 686 (2000)

\noindent
31.     Ghisellini G et al. {\it Astrophys. J.} {\bf 362} L1 (1990)

\noindent
32.     Konigl A, Kartje J F {\it Astrophys. J.} {\bf 434} 446 (1994)

\noindent
33.     Blandford R D, Znajek R L {\it Astrophys. J.} {\bf 179} 433 (1977)

\noindent
34.     Beskin V S, Istomin Y N, Parev V I {\it Astron. Zh.} {\bf 69} 1258 (1992) [{\it Sov. Astron.} {\bf 36} 642 (1992)]

\noindent
35.     Hirotani K, Okamoto I {\it Astrophys. J.} 497 563 (1998)

\noindent
36.     Fender R P, in {\it Compact Stellar X-ray Sources} (Eds W Lewin, M van der Klis) (Cambridge: Cambridge Univ. Press, 2006) p. 381

\noindent
37.     Spencer R E {\it Nature} {\bf 282} 483 (1979)

\noindent
38.     Mirabel I F, Rodriguez L F {\it Nature} {\bf 371} 46 (1994)

\noindent
39.     Akerlof C et al. {\it Nature} {\bf 398} 400 (1999)

\noindent
40.     Postnov K A {\it Usp. Fiz. Nauk} {\bf 169} 545 (1999) [{\it Phys. Usp.} {\bf 42} 469 (1999)]

\noindent
41.     Ruderman M {\it Ann. New York Acad. Sci.} {\bf 262} 164 (1975)

\noindent
42.     Panaitescu A, Kumar P {\it Astrophys. J.} {\bf 571} 779 (2002)

\noindent
43.     Blinnikov S I et al. {\it Pis'ma Astron. Zh.} {\bf 10} 422 (1984) [{\it Sov. Astron.} Lett. {\bf 10} 177 (1984)]

\noindent
44.     Eichler D et al. {\it Nature} {\bf 340} 126 (1989)

\noindent
45.     Paczynski B {\it Acta Astron.} {\bf 41} 257 (1991)

\noindent
46.     Woosley S E {\it Astrophys. J.} {\bf 405} 273 (1993)

\noindent
47.     Paczynski B {\it Astrophys. J.} {\bf 494} L45 (1998)

\noindent
48.     Meszaros P, Rees M J {\it Astrophys. J.} {\bf 482} L29 (1997)

\noindent
49.     van Putten M H P M, Levinson A {\it Astrophys. J.} {\bf 584} 937 (2003)

\noindent
50.     Komissarov S S, Barkov M V {\it Astrophys. J.} {\bf 382} 1029 (2007)

\noindent
51.     Usov V V {\it Nature} {\bf 357} 472 (1992)

\noindent
52.     Thompson C, Duncan R C {\it Astrophys. J.} {\bf 275} 255 (1995)

\noindent
53.     Hewish A et al. {\it Nature} {\bf 217} 708 (1968)

\noindent
54.     Baade W, Zwicky F {\it Proc. Natl. Acad. Sci.} {\bf 20} 254 (1934)

\noindent
55.     Landau L D, Lifshitz E M {\it Teoriya Polya} ({\it The Classical Theory of Fields}) (Moscow: Nauka, 1973) [Translated into English (Oxford: Pergamon Press, 1975)]

\noindent
56.     Beskin V S, Gurevich A V, Istomin Ya N {\it Physics of the Pulsar Magnetosphere} (Cambridge: Cambridge Univ. Press, 1993)

\noindent
57.     Mestel L, Panagi P, Shibata S {\it Astrophys. J.} {\bf 309} 388 (1999)

\noindent
58.     Weisskopf M C et al. {\it Astrophys. J.} {\bf 536} L81 (2000)

\noindent
59.     Helfand D J, Gotthelf E V, Halpern J P {\it Astrophys. J.} {\bf 556} 380 (2001)

\noindent
60.     Chiueh T, Li Z-Y, Begelman M C {\it Astrophys. J.} {\bf 505} 835 (1998)

\noindent
61.     Lyubarsky Y, Kirk J G {\it Astrophys. J.} {\bf 547} 437 (2001)

\noindent
62.     Petri J, Lyubarsky Y {\it Astron. Astrophys.} {\bf 473} 683 (2007)

\noindent
63.     Kennel C F, Coroniti F V {\it Astrophys. J.} {\bf 283} 694 (1984)

\noindent
64.     Komissarov S S, Lyubarsky Y E {\it Astrophys. J.} {\bf 349} 779 (2004)

\noindent
65.     Bogovalov S V et al. {\it Astrophys. J.} {\bf 358} 705 (2005)

\noindent
66.     Del Zanna L, Amato E, Bucciantini N {\it Astron. Astrophys.} {\bf 421} 1063 (2004)

\noindent
67.     Djorgovsky S, Evans C R {\it Astrophys. J. Lett.} {\bf 335} L61 (1988)

\noindent
68.     Herbig G H {\it Astrophys. J.} {\bf 111} 11 (1950)

\noindent
69.     Haro G {\it Astron. J.} {\bf 55} 72 (1950)

\noindent
70.     Reipurth B, Bally J {\it Annu. Rev. Astron. Astrophys.} {\bf 39} 403 (2001)

\noindent
71.     Surdin V G {\it Rozhdenie Zvezd} ({\it The Birth of Stars}) (Moscow: URSS, 2001)

\noindent
72.     Bacciotti F et al. {\it Astrophys. J.} {\bf 576} 222 (2002)

\noindent
73.     Coffey D et al. {\it Astrophys. J.} {\bf 663} 350 (2007)

\noindent
74.     Chrysostomou A, Lucas P W, Hough J H {\it Nature} {\bf 450} 71 (2007)

\noindent
75.     Pelletier G, Pudritz R E {\it Astrophys. J.} {\bf 394} 117 (1992)

\noindent
76.     Edwards S {\it Proc. Int. Astron. Union} {\bf 3} 171 (2007)

\noindent
77.     Shu F H et al. {\it Astrophys. J.} {\bf 429} 797 (1994)

\noindent
78.     Pudritz R E, Norman C A {\it Astrophys. J.} {\bf 301} 571 (1986)

\noindent
79.     Brenneman L W, Reynolds C S {\it Astrophys. J.} {\bf 652} 1028 (2006)

\noindent
80.     Daly R A {\it Astrophys. J. Lett.} {\bf 696} L32 (2009)

\noindent
81.     Lyne A G, Graham-Smith F {\it Pulsar Astronomy} 2nd ed. (Cambridge: Cambridge Univ. Press, 1998)

\noindent
82.     Bisnovatyi-Kogan G S, Lovelace R V E {\it New Astron. Rev.} {\bf} 45 663 (2001)

\noindent
83.     Alfven H, Falthammar C-G {\it Cosmical Electrodynamics} (Oxford: Clarendon Press, 1963) [Translated into Russian (Moscow: Mir, 1967)]

\noindent
84.     Weber E J, Davis L (Jr.) {\it Astrophys. J.} {\bf 148} 217 (1967)

\noindent
85.     Solov'ev L S, in {\it Voprosy Teorii Plazmy} ({\it Reviews of Plasma Physics}) Vol. 3 (Ed. M A Leontovich) (Moscow: Gosatomizdat, 1963) p. 245 [Translated into English (New York: Consultants Bureau, 1967) p. 277

\noindent
86.     Heinemann M, Olbert S J {\it Geophys. Res.} {\bf 83} 2457 (1978)

\noindent
87.     Okamoto I {\it Astrophys. J.} {\bf 173} 357 (1975)

\noindent
88.     Heyvaerts J, in {\it Plasma Astrophysics} (Lecture Notes Phys., Vol. 468, Eds C Chiuderi, G Einaudi) (Berlin: Springer-Verlag, 1996) p. 31

\noindent
89.     Shafranov V D {\it Zh. Eksp. Tear. Fiz.} {\bf 33} 710 (1957) [{\it Sov. Phys. JETP} {\bf 6} 545 (1958)]

\noindent
90.     Grad H {\it Rev. Mod. Phys.} {\bf 32} 830 (1960)

\noindent
91.     Von Mises R {\it Mathematical Theory of Compressible Fluid Flow} (New York: Academic Press, 1958) [Translated into Russian (Moscow: IL, 1961)]

\noindent
92.     Guderley K G {\it Theorie Schallnaher Strbmungen} ({\it Theory of Transonic Flow}) (Berlin: Springer-Verlag, 1957) [Translated into English (Oxford: Pergamon Press, 1962); Translated into Russian (Moscow: IL, I960)]

\noindent
93.     Frankl F I {\it Izbrannye Trudy po Gazovoi Dinamike} ({\it Selected Papers on Gasdynamics}) (Moscow: Nauka, 1973)

\noindent
94.     Ardavan H {\it Astrophys. J.} {\bf 204} 889 (1976)

\noindent
95.     Lovelace R V E et al. {\it Astrophys. J. Suppl.} {\bf 62} 1 (1986)

\noindent
96.     Nitta S, Takahashi M, Tomimatsu A {\it Phys. Rev. D} {\bf 44} 2295 (1991)

\noindent
97.     Beskin V S, Par'ev V I {\it Usp. Fiz. Nauk} {\bf 163} 95 (1993) [{\it Phys. Usp.} {\bf} 36 529 (1993)]

\noindent
98.     Contopoulos I, Kazanas D, Fendt C {\it Astrophys. J.} {\bf 511} 351 (1999)

\noindent
99.     Bogovalov S V {\it Astron. Astrophys.} {\bf 371} 1155 (2001)

\noindent
100.    Goodwin S P et al. {\it Astrophys. J.} {\bf 349} 213 (2004)

\noindent
101.    Gruzinov A {\it Phys. Rev. Lett.} {\bf 94} 021101 (2005)

\noindent
102.    Komissarov S S {\it Astrophys. J.} {\bf 367} 19 (2006)

\noindent
103.    McKinney J C {\it Astrophys. J.} {\bf 368} L30 (2006)

\noindent
104.    Timokhin A N {\it Astrophys. J.} {\bf 368} 1055 (2006)

\noindent
105.    Svensson R {\it Astrophys. J.} {\bf 209} 175 (1984)

\noindent
106.    Zakamska N L, Begelman M C, Blandford R D {\it Astrophys. J.} {\bf 679} 990 (2008)

\noindent
107.    Beskin V S, Gurevich A V, Istomin Ya N {\it Zh. Eksp. Teor. Fiz.} {\bf 85} 401 (1983) [{\it Sov. Phys. JETP} {\bf 58} 235 (1983)]

\noindent
108.    Beskin V S, Rafikov R R {\it Astrophys. J.} {\bf 313} 433 (2000)

\noindent
109.    Tchekhovskoy A, McKinney J C, Narayan R {\it Astrophys. J.} {\bf 388} 551 (2008)

\noindent
110.    Bogovalov S V {\it Pis'ma Astron. Zh.} {\bf 18} 832 (1992) [{\it Sov. Astron.} Lett. {\bf 18} 337 (1992)]

\noindent
111.    Beskin V S, Okamoto I {\it Astrophys. J.} {\bf 313} 445 (2000)

\noindent
112.    Toropina O D et al. {\it Mem. Soc. Astron. Ital.} {\bf 76} 508 (2005)

\noindent
113.    Beskin V S, Kuznetsova I V {\it Nuovo Cimento B} {\bf 115} 795 (2000)

\noindent
114.    Kazhdan Y M, Murzina M {\it Astrophys. J.} {\bf 270} 351 (1994)

\noindent
115.    Uchida Y, Shibata K {\it Astron. Soc. Jpn.} {\bf 36} 105 (1984)

\noindent
116.    Hawley J F, Smarr L L, Wilson J R {\it Astrophys. J.} {\bf 277} 296 (1984)

\noindent
117.    Shima E et al. {\it Astrophys. J.} {\bf 217} 367 (1985)

\noindent
118.    Petrich L I et al. {\it Astrophys. J.} {\bf 336} 313 (1989)

\noindent
119.    Ustyugova G V et al. {\it Astrophys. J.} {\bf 439} L39 (1995)

\noindent
120.    Bogovalov S, Tsinganos K {\it Astrophys. J.} {\bf 305} 211 (1999)

\noindent
121.    Ustyugova G V et al. {\it Astrophys. J.} {\bf 541} L21 (2000)

\noindent
122.    Komissarov S S {\it Astrophys. J.} {\bf 326} L41 (2001)

\noindent
123.    Komissarov S S et al. {\it Astrophys. J.} {\bf 380} 51 (2007)

\noindent
124.    Barkov M V, Komissarov S S {\it Astrophys. J.} {\bf 385} L28 (2008)

\noindent
125.    Tchekhovskoy A, McKinney J C, Narayan R {\it Astrophys. J.} {\bf 699} 1789 (2009)

\noindent
126.    Komissarov S S et al. {\it Astrophys. J.} {\bf 394} 1182 (2009)

\noindent
127.    Tchekhovskoy A, Narayan R, McKinney J C {\it Astrophys. J.} {\bf 711} 50 (2010)

\noindent
128.    Lynden-Bell D {\it Astrophys. J.} {\bf 279} 389 (1996)

\noindent
129.    Lynden-Bell D {\it Astrophys. J.} {\bf 341} 1360 (2003)

\noindent
130.    Landau L D, Lifshitz E M {\it Elektrodinamika Sploshnykh Sred} ({\it Electrodynamics of Continuous Media}) (Moscow: Nauka, 1982) [Translated into English (Oxford: Pergamon Press, 1984)]

\noindent
131.    Thorne K S, Price R H, Macdonald D A {\it Black Holes: the Membrane Paradigm} (New Haven: Yale Univ. Press, 1986) [Translated into Russian (Moscow: Mir, 1988)]

\noindent
132.    Punsly B {\it Black Hole Gravitohydromagnetics} (Berlin: Springer, 2001)

\noindent
133.    Al'pert Ya L, Gurevich A V, Pitaevskii L P {\it Iskusstvennye Sputniki v Razrezhennoi Plazme} ({\it Space Physics with Artifical Satellites}) (Moscow: Nauka, 1964) [Translated into English (New York: Consultants Bureau, 1965)]

\noindent
134.    Komissarov S S {\it Astrophys. J.} {\bf 350} 1431 (2004)

\noindent
135.    Camenzind M {\it Compact Objects in Astrophysics} (Heidelberg: Springer, 2007)

\noindent
136.    Okamoto I {\it Astrophys. J.} {\bf 307} 253 (1999)

\noindent
137.    Heyvaerts J, Norman C {\it Astrophys. J.} {\bf 596} 1240 (2003)

\noindent
138.    Heyvaerts J, Norman C {\it Astrophys. J.} {\bf 347} 1055 (1989)

\noindent
139.    Bogovalov S V Pis'ma {\it Astron. Zh.} {\bf 21} 633 (1995) [{\it Astron. Lett.} {\bf 21} 565 (1995)]

\noindent
140.    Bogovalov S V Pis'ma {\it Astron. Zh.} {\bf 24} 381 (1998) [{\it Astron. Lett.} {\bf 24} 321 (1998)]

\noindent
141.    Sakurai T {\it Astron. Astrophys.} {\bf 152} 121 (1985)

\noindent
142.    Michel F C {\it Astrophys. J.} {\bf 180} L133 (1973)

\noindent
143.    Beskin V S, Kuznetsova IV, Rafikov R R {\it Astrophys. J.} {\bf 299} 341 (1998)

\noindent
144.    Tomimatsu A {\it Publ. Astron. Soc. Jpn.} {\bf 46} 123 (1994)

\noindent
145.    Beskin V S, Nokhrina E E {\it Astrophys. J.} {\bf 367} 375 (2006)

\noindent
146.    Beskin V S, Nokhrina E E {\it Astrophys. J.} {\bf 397} 1486 (2009)

\noindent
147.    Trubnikov B A {\it Teoriya Plazmy} ({\it Plasma Theory}) (Moscow: Energoatomizdat, 1996)

\noindent
148.    Bogovalov S V {\it Astrophys. J.} {\bf 280} 39 (1996)

\noindent
149.    Bogovalov S V Pis'ma {\it Astron. Zh.} {\bf 16} 844 (1990) [{\it Sov. Astron. Lett.} {\bf 16} 362 (1990)]

\noindent
150.    Lyubarsky Yu {\it Astrophys. J.} {\bf 698} 1570 (2009)

\noindent
151.    Beskin V S, Malyshkin L M {\it Pis'ma Astron. Zh.} {\bf 26} 253 (2000) [{\it Astron. Lett.} {\bf 26} 208 (2000)]

\noindent
152.    Kadomtsev B B {\it Kollektivnye Yavleniya v Plazme} ({\it Cooperative Effects in Plasmas}) 2nd ed. (Moscow: Nauka, 1988) [Translateed into English, in Reviews of Plasma Physics Vol. 22 (Ed. V D Shafranov) (New York: Kluwer Acad./Plenum Publ., 2001) p. 1

\noindent
153.    Ryutov D D, Derzon M S, Matzen M K {\it Rev. Mod. Phys.} {\bf 72} 167 (2000)

\noindent
154.    Bisnovatyi-Kogan G S, Komberg B V, Fridman A M {\it Astron. Zh.} {\bf 46} 465 (1969) [{\it Sov. Astron.} {\bf 13} 369 (1969)]

\noindent
155.    Benford G {\it Astrophys. J.} {\bf 247} 792 (1981)

\noindent
156.    Hardee P E {\it Astrophys. J.} {\bf 313} 607 (1987)

\noindent
157.    Hardee P E, Norman M L {\it Astrophys. J.} {\bf 334} 70 (1988)

\noindent
158.    Appl S, Camenzind M {\it Astron. Astrophys.} {\bf 256} 354 (1992)

\noindent
159.    Lyubarskii Yu E {\it Astrophys. J.} {\bf 308} 1006 (1999)

\noindent
160.    Bisnovatyi-Kogan G {\it Astrophys. Space Sci.} {\bf 311} 287 (2007)

\noindent
161.    Ciardi A et al. {\it Astrophys. J.} {\bf 691} L147 (2009)

\noindent
162.    Bellan P M et al. {\it Phys. Plasmas} {\bf 16} 041005 (2009)

\noindent
163.    Istomin Y N, Pariev VI {\it Astrophys. J.} {\bf 267} 629 (1994)

\noindent
164.    Meliani Z, Keppens R {\it Astron. Astrophys.} {\bf 475} 785 (2007)

\noindent
165.    Narayan R, Li J, Tchekhovskoy A {\it Astrophys. J.} {\bf 697} 1681 (2009)

\noindent
166.    McKinney J C, Blandford R D {\it Astrophys. J.} {\bf 394} L126 (2009)

\noindent
167.    Beskin V S, Zakamska N L, Sol H {\it Astrophys. J.} {\bf 347} 587 (2004)

\noindent
168.    Narayan R, McKinney J C, Farmer A J {\it Astrophys. J.} {\bf 375} 548 (2006)

\noindent
169.    Barkov M V, Komissarov S S {\it Int. J. Mod. Phys. D} {\bf 17} 1669 (2008)

\noindent
170.    Bardou A, Heyvaerts J {\it Astron. Astrophys.} {\bf 307} 1009 (1996)

\noindent
171.    Lovelace R V E, Romanova M M {\it Astrophys. J.} {\bf 596} L159 (2003)

\noindent
172.    Michel F C {\it Astrophys. J.} {\bf 180} 207 (1973)

\noindent
173.    Mestel L, Wang Y-M {\it Astrophys. J.} {\bf 188} 799 (1979)

\noindent
174.    Lyubarskii Y E Pis'ma {\it Astron. Zh.} {\bf 16} 34 (1990) [{\it Sov. Astron. Lett.} {\bf 16} 16 (1990)]

\noindent
175.    Lovelace R V E, Turner L, Romanova M M {\it Astrophys. J.} {\bf 652} 1494 (2006)

\noindent
176.    Bogovalov S V {\it Astron. Astrophys.} {\bf 349} 1017 (1999)

\noindent
177.    Spitkovsky A {\it Astrophys. J.} {\bf 648} L51 (2006)

\noindent
178.    Ruderman M A, Sutherland P G {\it Astrophys. J.} {\bf 196} 51 (1975)

\noindent
179.    Arons J {\it Astrophys. J.} {\bf 248} 1099 (1981)

\noindent
180.    Gurevich A V, Istomin Ya N {\it Zh. Eksp. Teor. Fiz.} {\bf 89} 3 (1985) [{\it Sov. Phys.JETP} {\bf 62} 1(1985)]

\noindent
181.    Bai X-N, Spitkovsky A {\it Astrophys. J.} {\bf 715} 1282 (2010)

\noindent
182.    Kramer M et al. {\it Science} {\bf 312} 549 (2006)

\noindent
183.    Beskin V S, Nokhrina E E {\it Astrophys. Space Sci.} {\bf 308} 569 (2007)

\noindent
184.    Gurevich A V, Istomin Ya N {\it Astrophys. J.} {\bf 377} 1663 (2007)

\noindent
185.    Kato Y, Mineshige S, Shibata K {\it Astrophys. J.} {\bf 605} 307 (2004)

\noindent
186.    Sherwin B D, Lynden-Bell D {\it Astrophys. J.} {\bf 378} 409 (2007)

\noindent
187.    Uzdensky D A, MacFadyen A I {\it Astrophys. J.} {\bf 669} 546 (2007)

\noindent
188.    Romanova M M et al. {\it Astrophys. J.} {\bf 399} 1802 (2009)

\noindent
189.    Lovelace R V E et al. {\it Astrophys. J.} {\bf 572} 445 (2002)

\noindent
190.    Ghosh P, Abramowicz M A {\it Astrophys. J.} {\bf 292} 887 (1997)

\noindent
191.    Livio M, Ogilvie G I, Pringle J E {\it Astrophys. J.} {\bf 512} 100 (1999)

\noindent
192.    Konigl A {\it Astrophys. J.} {\bf 342} 208 (1989)

\noindent
193.    Ferreira J, Pelletier G {\it Astron. Astrophys.} {\bf 295} 807 (1995)

\noindent
194.    Krolik J H {\it Active Galactic Nuclei, from the Central Black Hole to the Galactic Environment} (Princeton, NJ: Princeton Univ. Press, 1999)

\noindent
195.    Derishev E V et al. {\it Astrophys. Space Sci.} {\bf 297} 21 (2005)

\noindent
196.    Stern B E, Poutanen J {\it Astrophys. J.} {\bf 383} 1695 (2008)

\noindent
197.    Sol H, Pelletier G, Asseo E {\it Astrophys. J.} {\bf 237} 411 (1989)

\noindent
198.    Cheng K S, Ho C, Ruderman M {\it Astrophys. J.} {\bf 300} 500 (1986)

\noindent
199.    Zavala R T, Taylor G B {\it Astrophys. J.} {\bf 612} 749 (2004)

\noindent
200.    Lister M L, Homan D C {\it Astron. J.} {\bf 130} 1389 (2005)

\noindent
201.    Gabuzda D C et al. {\it Astrophys. J.} {\bf 384} 1003 (2008)

\noindent
202.    Blandford R D, Konigl A {\it Astrophys. J.} {\bf 232} 34 (1979)

\noindent
203.    Lobanov A P {\it Astron. Astrophys.} {\bf 330} 79 (1998)

\noindent
204.    Churazov E M et al. {\it Usp. Fiz. Nauk} {\bf 176} 334 (2006) [{\it Phys. Usp.} {\bf 49} 319 (2006)]

\noindent
205.    MacFadyen A I, Woosley S E {\it Astrophys. J.} {\bf 524} 262 (1999)

\noindent
206.    Piran T {\it Rev. Mod. Phys.} {\bf 76} 1143 (2004)

\noindent
207.    Tchekhovskoy A, Narayan R, McKinney J C {\it New Astron.} {\bf 15} 749 (2010)

\noindent
208.    Komissarov S S, Vlahakis N, Konigl A {\it Astrophys. J.} {\bf 407} 17 (2010)

\noindent
209.    Daugherty J K, Harding A K {\it Astrophys. J.} {\bf 252} 337 (1982)

\noindent
210.    Medin Z, Lai D {\it Astrophys. J.} {\bf 406} 1379 (2010)

\noindent
211.    Anderson J M et al. {\it Astrophys. J.} {\bf 590} L107 (2003)

\noindent
212.    Ferreira J, Dougados K, Cabrit S {\it Astron. Astrophys.} {\bf 453} 785 (2006)

}

\end{document}